\documentclass[fleqn,usenatbib]{mnras}

\usepackage[T1]{fontenc}

\DeclareRobustCommand{\DE}[3]{#2}
\let\DEthebibliography\thebibliography
\def\thebibliography{\DeclareRobustCommand{\DE}[3]{##3}\DEthebibliography}

\usepackage{graphicx}	
\usepackage{amsmath}	
\usepackage{amssymb}	
\usepackage{subfig}

\usepackage{newtxtext,newtxmath}

\title[Discovery and Characterisation of TOI-1064]{A pair of Sub-Neptunes transiting the bright K-dwarf TOI-1064 characterised with {\em CHEOPS}\thanks{Based on observations made with ESO Telescopes at the La Silla Observatory under program ID 1102.C-0923.}}

\author[T. G. Wilson et al.]{
Thomas G. Wilson$^{1}$\thanks{E-mail: tgw1@st-andrews.ac.uk},
Elisa Goffo$^{2,3}$, 
Yann Alibert$^{4}$, 
Davide Gandolfi$^{2}$, 
Andrea Bonfanti$^{5}$, 
\newauthor Carina M. Persson$^{6}$, 
Andrew Collier Cameron$^{1}$, 
Malcolm Fridlund$^{6,7}$, 
Luca Fossati$^{5}$, 
Judith Korth$^{8}$
% ...
\\ \\
\textit{(Affiliations can be found after the references)}
}

\date{Accepted XXX. Received YYY; in original form ZZZ}

\pubyear{2021}

\begin{document}
\label{firstpage}
\pagerange{\pageref{firstpage}--\pageref{lastpage}}
\maketitle

\clearpage

\begin{abstract}
We report the discovery and characterisation of a pair of sub-Neptunes transiting the bright K-dwarf TOI-1064 (TIC\,79748331), initially detected in {\it TESS} photometry. To characterise the system, we performed and retrieved {\it CHEOPS}, {\it TESS}, and ground-based photometry, HARPS high-resolution spectroscopy, and Gemini speckle imaging. We characterise the host star and determine $T_{\rm eff, \star}=4734\pm67\,$K, $R_{\star}=0.726\pm0.007\,R_{\odot}$, and $M_{\star}=0.748\pm0.032\,M_{\odot}$. We present a novel detrending method based on PSF shape-change modelling and demonstrate its suitability to correct flux variations in {\em CHEOPS} data. We confirm the planetary nature of both bodies and find that TOI-1064\,b has an orbital period of $P_{\rm b} = 6.44387\pm0.00003$\,d, a radius of $R_{\rm b} = 2.59\pm0.04$\,$R_{\oplus}$, and a mass of $M_{\rm b} = 13.5_{-1.8}^{+1.7}$\,$M_{\oplus}$, whilst TOI-1064\,c has an orbital period of $P_{\rm c} = 12.22657^{+0.00005}_{-0.00004}$\,d, a radius of $R_{\rm c} = 2.65\pm0.04$\,$R_{\oplus}$, and a 3$\sigma$ upper mass limit of 8.5\,$\mathrm{M_{\oplus}}$. From the high-precision photometry we obtain radius uncertainties of $\sim$1.6\%, allowing us to conduct internal structure and atmospheric escape modelling. TOI-1064\,b is one of the densest, well-characterised sub-Neptunes, with a tenuous atmosphere that can be explained by the loss of a primordial envelope following migration through the protoplanetary disc. It is likely that TOI-1064\,c has an extended atmosphere due to the tentative low density, however further RVs are needed to confirm this scenario and the similar radii, different masses nature of this system. The high-precision data and modelling of TOI-1064\,b is important for planets in this region of mass-radius space, and it allows us to identify a trend in bulk density-stellar metallicity for massive sub-Neptunes that may hint at the formation of this population of planets.

\end{abstract}




\section{Introduction}
\label{sec:intro} 

The study of exoplanets yields improvements in our understanding of planet formation and evolution, and has done so since the discovery of the first hot Jupiter \citep{Mayor1995} challenged planet formation paradigms and compelled the development of planet migration theories. Since this initial discovery and subsequent pioneering works \citep{Charbonneau2000,Charbonneau2009,Rivera2005,Leger2009,Queloz2009}, thousands of planets have been detected using ground-based and space-based photometric instruments such as; HATNet, SuperWASP, TrES, {\em CoRoT}, {\em Kepler}, {\em K2}, and {\em TESS} \citep{Bakos2002,Alonso2004,Baglin2006,Pollacco2006,Borucki2010,Howell2014,Ricker2015}, and radial velocity (RV) surveys, as in; HIRES, HARPS, HARPS-N, SOPHIE, and ESPRESSO \citep{Vogt1994,Mayor2003,Bouchy2006,Cosentino2012,Pepe2021}, that have subsequently paved the way for follow-up characterisation of these systems and yielded further insight into planet formation and evolution.

One such advancement has come from the precise measurements of the radii and masses of transiting exoplanets orbiting bright host stars, as knowledge of these key properties permits the determination of both empirical mass-radius curves \citep{Weiss2013,Weiss2014,Chen2017,Otegi2020} and bulk densities. Planetary densities can be converted into bulk compositional information using internal structure models (\citealt{Fortney2007,Seager2007,Sotin2007,Valencia2007,Zeng2013,Howe2014,Dorn2015,Dorn2017,Brugger2017,Zeng2019,Mousis2020,Aguichine2021}). However, to break degeneracies in the modelling, both elemental stellar abundances \citep{Santos2015,Adibekyan2021} and low uncertainties on planetary radii and masses are needed. The former can be resolved via spectroscopic follow-up to produce a combined high-resolution spectrum that can be used to extract stellar information. High-precision photometry and RVs are needed to constrain radii and masses.

\begin{figure*}
  \includegraphics[width = \textwidth]{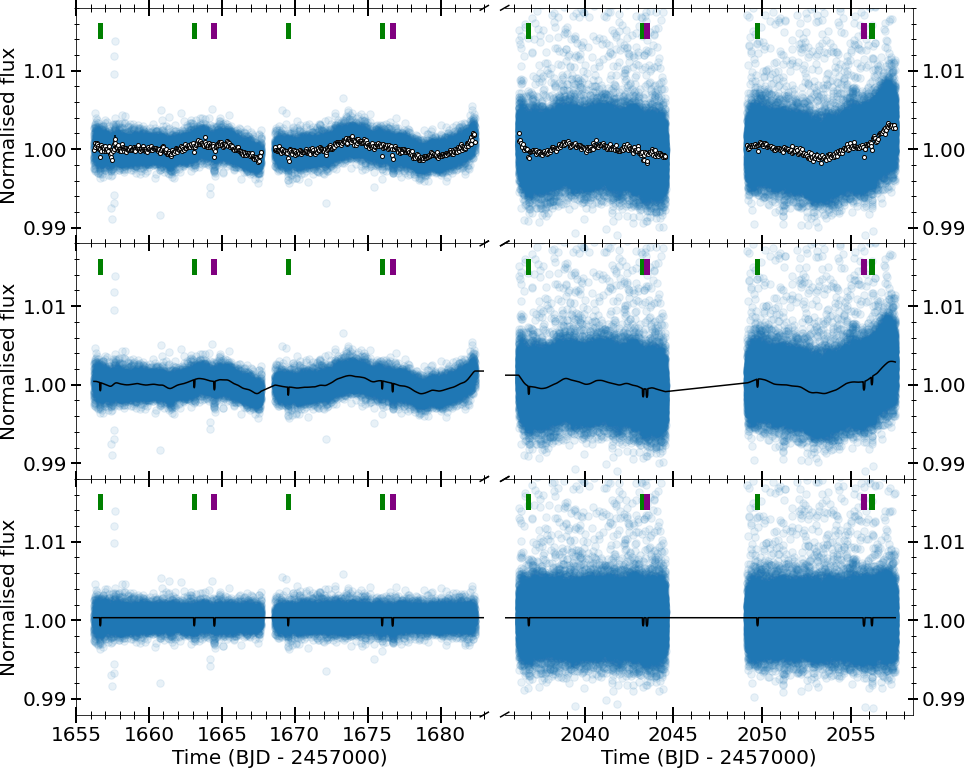}
  \caption{The Sector 13 ({\em left}) and 27 ({\em right}) {\it TESS} light curves (with cadences of 2\,min and 20\,s, respectively) of TOI-1064 with the locations of the transits of planet b (green) and planet c (purple) shown. {\it Top panel}: The retrieved PDCSAP fluxes in blue with data binned every 1\,hr given as white circles. {\it Middle panel}: The photometry with transit and Mat\'{e}rn-3/2 kernel GP models used in the global analysis, as detailed in Section~\ref{sec:analysis}. {\it Bottom panel}: The photometry and transit model with trends modelling by the Mat\'{e}rn-3/2 kernel GP removed.}
  \label{fig:tess_lcs}
\end{figure*}

Studying multi-planet systems offers a further, unique opportunity for planetary system characterisation via comparative planetology, as they have formed within the same proto-planetary disc. Therefore, observations can constrain formation and evolution models from the analysis of mutual inclinations \citep{Steffen2010,Fang2012,Fabrycky2014} and eccentricities \citep{VanEylen2015,VanEylen2019,Limbach2015,Mills2019}. Additionally, determination of orbital spacings between pairs of planets \citep{Weiss2018,Jiang2020} and the assessment of correlations or discrepancies between the spacings and planetary sizes \citep{Ciardi2013} and masses \citep{Lissauer2011,Millholland2017,Adams2020,Weiss2020} can lead to a better understanding on a demographic scale. Moreover, should individual planets be well characterised, the bulk internal and atmospheric compositions could be compared and, when combined the fluxes of stellar radiation at the positions of the planets, could inform formation and evolution modelling.

In this paper, we report the discovery of the two-planet system TOI-1064 using {\it TESS}, {\it CHEOPS}, LCOGT, NGTS, ASTEP, ASAS-SN, and WASP photometry (Sections~\ref{sec:tess}-\ref{sec:astep},~\ref{sec:asas_sn}, and~\ref{sec:wasp}), HARPS radial velocities (Section~\ref{sec:harps}), and Gemini speckle imaging (Section~\ref{sec:gemini}). We determine the stellar parameters of the host star (Section~\ref{sec:hoststar}), and through photometric and radial velocity analyses we precisely characterise the planetary properties of TOI-1064\,b and\,c (Section~\ref{sec:analysis}). Utilising these results, we present internal structure and atmospheric modelling of the planets (Section~\ref{sec:charac}), discuss important aspects of the system (Section~\ref{sec:disc}), and summarise our conclusions (Section~\ref{sec:concs}).

\section{Observations}
\label{sec:obs} 

To fully characterise the TOI-1064 system, we collate photometric, spectroscopic, and imaging data from multiple sources detailed below.

\subsection{{\em TESS}}
\label{sec:tess}

\begin{table*}
\caption{The {\it CHEOPS} observing log of TOI-1064.}             
\label{tab:obs_log}      
\centering   
\small
\begin{tabular}{c c c c c c c c}
\hline\hline
 visit & Planets & Start date & Duration & Data points & File key & Efficiency & Exp. Time  \\ 
 \# &  & [UTC] & [h] & [\#] & & [\%] & [s] \\
\hline                    
 1 & b,c & 2020-07-12T10:11:36 & 13.64 & 473 & CH\_PR100031\_TG027501\_V0200 & 58 & 60 \\%
 2 & b & 2020-07-31T23:57:36 & 7.50 & 290 & CH\_PR100031\_TG029601\_V0200 & 64 & 60 \\%
 3 & c & 2020-08-18T12:12:36 & 7.52 & 293 & CH\_PR100031\_TG029701\_V0200 & 65 & 60 \\%
 4 & b & 2020-08-20T07:04:36 & 7.50 & 270 & CH\_PR100031\_TG029101\_V0200 & 60 & 60 \\%
 5 & b & 2020-08-26T18:28:17 & 7.99 & 263 & CH\_PR100031\_TG029501\_V0200 & 55 & 60 \\%
 6 & c & 2020-08-30T13:04:57 & 24.23 & 830 & CH\_PR100031\_TG027401\_V0200 & 57 & 60 \\%
\hline\hline                  
\end{tabular}
\end{table*}

The {\it TESS} spacecraft has been conducting survey-mode observations of stars in the {\it TESS} Input Catalogue (TIC; \citealt{Stassun2018,Stassun2019}) since launch in 2018, covering 26 sectors in the two-year primary mission \citep{Ricker2015}. Subsequently, {\it TESS} has entered its extended mission with a main goal to provide additional, shorter-cadence photometry of targets observed during the primary mission. Initially listed as TIC 79748331, TOI-1064 was observed by {\it TESS} in Sector 13, camera 1, CCD 2, from 2019-June-19 to 2019-July-17. A gap occurred at the midpoint of the sector due to telemetry operations around spacecraft periastron passage, yielding 27.51\,d of science observations. Individual frames were processed into 2\,min cadence calibrated pixel files and reduced into light curves by the Science Processing Operations Center (SPOC; \citealt{Jenkins2016}), at NASA Ames Research Center. A transit search was conducted on the Presearch Data Conditioning simple aperture photometric (PDCSAP) light curves \citep{Smith2012,Stumpe2012,Stumpe2014} using a wavelet-based, noise-compensating matched filter \citep{Jenkins2002,Jenkins2010} with two candidate planetary signals passing all diagnostic tests \citep{Twicken2018,Li2019}. The {\it TESS} Science Operations Center (TSO) announced these as {\it TESS} Objects of Interest \citep{Guerrero2021}. The two candidates were detected at SNRs of 9.4 and 8.1, respectively, and no additional transiting planet signatures were identified in the residual light curves. Following the commencement of the extended mission, {\it TESS} observed TOI-1064 in Sector 27, camera 1, CCD 1, from 2020-July-05 to 2020-July-30. Allowing for the midsector observation gap, this results in 23.35\,d of science observations that were subsequently processed and reduced by the SPOC into 20\,s cadence photometry (in GI Cycle 3 program 3278; P.I. Andrew Mayo). In total, eight transits of TOI-1064{\bf\,b} and four transits of TOI-1064{\bf\,c} were identified with the {\it TESS}-reported periods of the two candidates of 6.4422$\pm$0.0015\,d and 12.2371$\pm$0.0071\,d, respectively.

We retrieved both sectors of photometry (denoted {\tt LC} and {\tt FAST\_LC} for the 2\,min and 20\,s cadence data respectively) from the Mikulski Archive for Space Telescopes (MAST), by selecting the systematics-corrected PDCSAP light curves \citep{Smith2012,Stumpe2012,Stumpe2014}, and using the default quality bitmask. Finally, for the photometric analysis detailed below, we rejected data points flagged by the SPOC as being of bad quality ({\tt QUALITY} > 0) and those with Not-a-Number fluxes or flux errors. This quality control yields a total of 89\,642 data points from both {\it TESS} sectors, with the resulting light curves from Sectors 13 and 27 shown in Fig.~\ref{fig:tess_lcs}.

\subsection{{\em CHEOPS}}
\label{sec:cheops} 

The {\em CHEOPS} spacecraft \citep{Benz2021}, launched on 2019-December-18 from Kourou, French Guiana, is an ESA small-class mission with the prime aim of observing bright (V $<$ 12\, mag), exoplanet-hosting stars to obtain ultra high-precision photometry \citep{Broeg2013}. Since launch, {\em CHEOPS} successfully passed In-Orbit Commissioning and was verified to achieve a photometric noise of 15\,ppm per 6\,hr for a V $\sim$ 9\,mag star \citep{Benz2021}. A study of the photometric precision of {\it CHEOPS} has shown that the depth uncertainty of a 500\,ppm transit from one {\it CHEOPS} observation is comparable to eight transits from {\it TESS} \citep{Bonfanti2021}.

The first scientific results from {\em CHEOPS} show the range of science that can be achieved with such a high-precision instrument (\citealt{Lendl2020,Borsato2021,Delrez2021,Leleu2021,Morris2021,Szabo2021,VanGrootel2021}; Maxted et al. {\it accepted}). One such study reports on the improvement in precision of exoplanet sizes in the HD\,108236 system \citep{Bonfanti2021}, that highlights a key scientific goal of the {\it CHEOPS} mission: {\it the refinement of exoplanet radii to decrease bulk density uncertainties, and thereby allowing internal structure and atmospheric evolution modelling}.

\begin{figure}
  \includegraphics[width = 8.5cm]{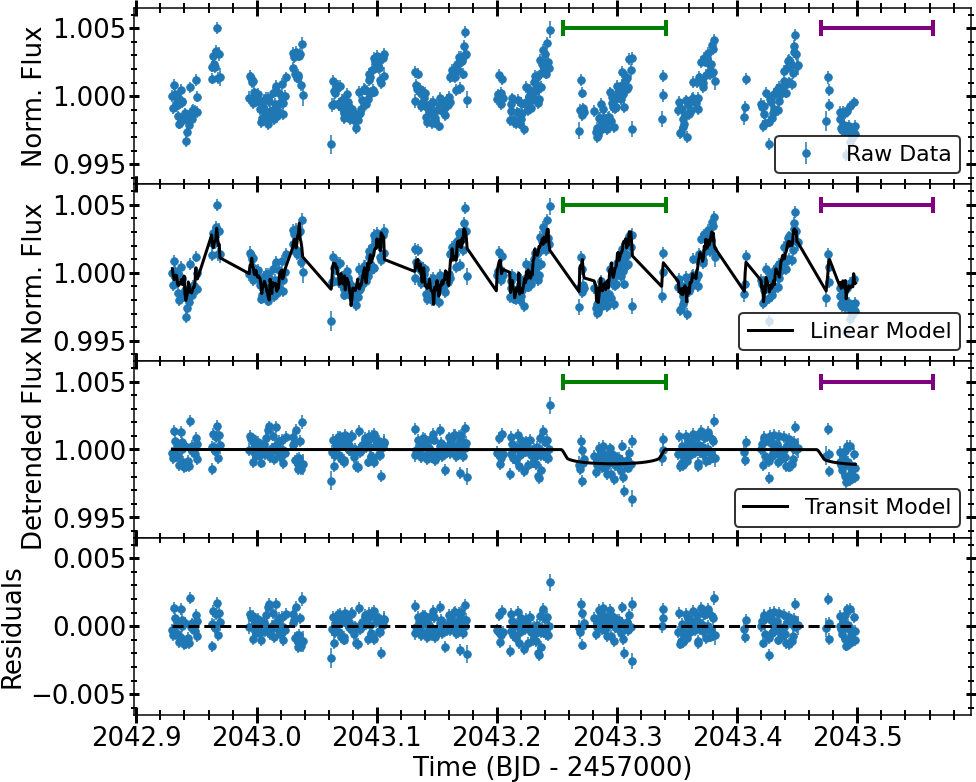}
  \caption{The normalised light curve of the first {\em CHEOPS} visit of TOI-1064 with the locations of the transits of planet b (green) and planet c (purple) shown. {\em Top panel}: The raw DRP-produced fluxes. {\em Second panel}: The DRP fluxes with the fitted linear model produced from the measured PSF shape changes, as detailed in Appendix~\ref{sec:scalpels_desc}. {\em Third panel}: The detrended fluxes with the nominal transit model from the global analysis (Section~\ref{sec:joint_analysis}). {\em Bottom panel}: Residuals to the fit.}
  \label{fig:cheops_raw1}
\end{figure}

To better characterise and to secure the validation of both planetary candidates, we obtained six visits of TOI-1064 with the {\it CHEOPS} spacecraft between 2020-July-12 and 2020-August-31, as a part of the Guaranteed Time Observers programme, yielding a total of 68.38\,h on target. We identify four transits of TOI-1064{\bf\,b} and three transits of TOI-1064{\bf\,c} across these runs. A breakdown of the individual visit start times and durations is detailed in Table~\ref{tab:obs_log}. For all visits, we used an exposure time of 60\,s.

As the {\it CHEOPS} spacecraft is in a low-Earth orbit, sections of observations are unobtainable due to the on-board rejection of images due the level of stray-light being higher than the accepted threshold, occultations of the target by the Earth, or passages through the South Atlantic Anomaly (SAA) during which no data is downlinked. These effects occur on orbital timescales ($\sim$98.77\,min) and lead to a decrease in the observational efficiency. As can be seen in Fig.~\ref{fig:cheops_raw1} (and Figs.~\ref{fig:cheops_raw2}-\ref{fig:cheops_raw6}), these interruptions are apparent in the {\it CHEOPS} photometric data with the efficiencies for the six visits listed in Table~\ref{tab:obs_log}. 

\begin{figure*}
  \includegraphics[width = \textwidth]{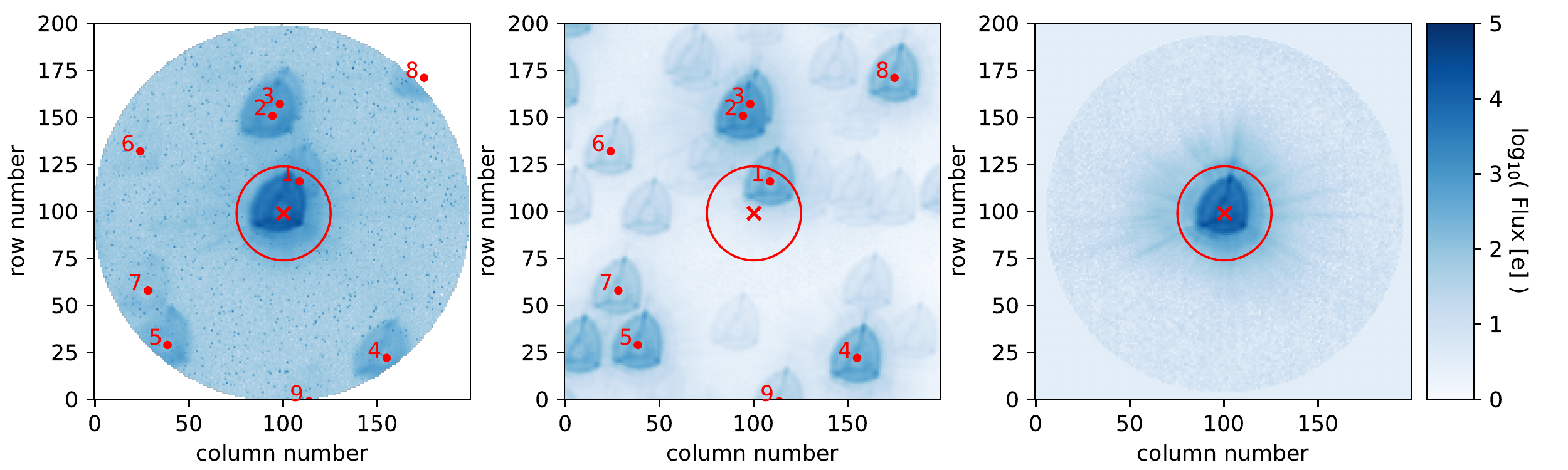}
  \caption{The 3.3' diameter FoV of TOI-1064 with the PSF centroid indicated by the red cross and the photometric aperture given by the red circle. {\em Left panel}: the FoV as observed by {\em CHEOPS}. {\em Middle panel}: DRP model of the FoV showing the contamination estimate of the background stars with the target removed. {\em Right panel}: DRP model of the FoV showing TOI-1064 with contamination estimate subtracted.}
  \label{fig:cheops_psf}
\end{figure*}

Data for all visits were automatically processed using the latest version of the {\it CHEOPS} data reduction pipeline (DRP v13; \citealt{Hoyer2020}). The DRP undertakes image calibration, such as bias, gain, non-linearity, dark current, and flat fielding corrections, and conducts rectifications of instrumental and environmental effects, for example cosmic-ray impacts, smearing trails of nearby stars, and background variations. Subsequently, it performs aperture photometry on the corrected frames using a set of defined-radius apertures; $R$ = 22.5\arcsec (RINF), 25.0\arcsec (DEFAULT), and 30.0\arcsec (RSUP), and an additional aperture that aims to optimise the radius based on contamination level and instrumental noise (ROPT). For the six observations of TOI-1064 this radius was determined to be between 15.0\arcsec and 16.0\arcsec, due to the nearby source discussed below. Furthermore, the DRP computes the contribution of nearby stars to the photometry by simulating the Field-of-View (FoV) of the {\it CHEOPS} observations of the target using the {\it Gaia} DR2 catalogue \citep{GaiaCollaboration2018} as an input source list for objects' locations and brightnesses. By conducting aperture photometry on the simulated FoV with the target removed, light curves of the contamination from nearby sources are produced, as detailed in Section~6.1 of \cite{Hoyer2020}. As can be seen in Fig.~\ref{fig:cheops_psf}, in the {\it CHEOPS} FoV there is a nearby object ({\it Gaia} EDR3 6683371813007224960, $\Delta G$ = +3.9\,mag) 24.6\arcsec away from the target that may affect the photometry of the target, and thus the contamination estimates were subtracted from the light curves of TOI-1064. The {\it Gaia} parallax and proper motion data indicate that this object and the target do not form a larger bound system. The right panel of Fig.~\ref{fig:cheops_psf} shows the inferred FoV of the target with the contamination removed. Fig.~\ref{fig:cheops_psf} also reveals additional multiple nearby sources with stars that have $G < 17$\,mag numbered. It should be noted that the sources numbered 1$-$5 are the same as those detected in the {\it TESS} FoV (Fig.~\ref{fig:tess_cens}), whereas 6$-$9 are too faint to be seen in the {\it TESS} target pixel file. The remaining objects in the FoV have $\Delta G >$ +7\,mag compared to TOI-1064 and thus, they will not contribute considerably to the photometry. For this study, we selected datasets that minimised the root mean square (RMS) of the light curves, which for all visits were obtained with the RINF aperture. This radius minimised the contribution of nearby sources whilst ensuring that the majority of the target's Point Spread Function (PSF) was within the aperture.

Due to the nature of the {\it CHEOPS} orbit and the rotating FoV, non-astrophysical, short-term photometric trends caused by a varying background, nearby contaminants, or other sources, can be seen in the data. Several studies (e.g. \citealt{Lendl2020,Bonfanti2021,Delrez2021,Leleu2021}) have found success in removing these systematics by conducting a linear decorrelation using several basis vectors, such as background, contamination, orbital roll angle, and $x$ and $y$ centroid positions. Upon inspection of the {\it CHEOPS} observations of TOI-1064, we found significant flux variations on orbital timescales. The selection of basis vectors of concern typically involves assessing the Bayesian Information Criteria (BIC) upon the detrending of the data using a combination of DRP-provided vectors. However, in this study a more data-driven approach was taken to identify which basis vectors were to be used in the detrending, as is detailed below in Appendix~\ref{sec:scalpels_desc}.

\subsection{LCOGT}
\label{sec:lco} 

We acquired ground-based time-series follow-up photometry of TOI-1064 as part of the {\it TESS} Follow-up Observing Program (TFOP)\footnote{https://tess.mit.edu/followup} using the {\it TESS} Transit Finder, which is a customised version of the {\tt Tapir} software package \citep{Jensen2013}, to schedule our transit observations.

We observed full transits of TOI-1064{\bf\,b} in the Pan-STARRS $z$-short band on 2020-June-03, 2020-June-16, and 2020-August-26 from the Las Cumbres Observatory Global Telescope (LCOGT) \citep{Brown2013} 1.0\,m network node at South African Astronomical Observatory (SAAO). We also observed full transits of TOI-1064{\bf\,c} on 2019-August-30 in Pan-STARRS $z$-short band from the LCOGT 1\,m node at Cerro Tololo Inter-American Observatory and on 2019-October-05 in B-band from the LCOGT 1\,m network node at SAAO. The 0.389$\arcsec$ pixel scale images were calibrated by the standard LCOGT {\tt BANZAI} pipeline \citep{McCully2018}, and photometric data were extracted with {\tt AstroImageJ} \citep{Collins2017}. The images were focused and have typical stellar point-spread-functions with a full-width-half-maximum (FWHM) of $\sim 2$\arcsec, and circular apertures with radius $\sim 4$\arcsec were used to extract the differential photometry. 

\subsection{NGTS}
\label{sec:ngts} 

The Next Generation Transit Survey \citep[NGTS;][]{Wheatley2018} was used to observe a partial transit egress of TOI-1064{\bf\,c} on 2019-October-17. NGTS is a photometric facility, which is located at the ESO Paranal Observatory in Chile and consists of twelve robotic telescopes, each with a 20\,cm diameter, an 8 square-degree FoV, and a pixel scale of 4.97$\arcsec$. Each NGTS telescope is operated independently, however simultaneous observations with multiple NGTS telescopes have been shown to greatly improve the photometric precision achieved \citep{Smith2020, Bryant2020}. TOI-1064 was observed using two NGTS telescopes with both telescopes observing with an exposure time of 10\,s and using the custom NGTS filter (520-890\,nm). A total of 2016 images were obtained during the observation.

The image reduction was performed using a custom photometry pipeline \citep{Bryant2020}. The source extraction and photometry in the pipeline are performed using the \texttt{SEP} Python library \citep{Bertin1996, Barbary2016}. Comparison stars which are similar to TOI-1064 in brightness, colour, and CCD position were automatically identified using \textit{Gaia} DR2 \citep{GaiaCollaboration2018}.

\subsection{ASTEP}
\label{sec:astep} 

We observed five transits of the TOI-1064 planets as part of the Antarctica Search for Transiting ExoPlanets (ASTEP) program \citep{Guillot2015,Mekarnia2016}. The $0.4$\,m ASTEP telescope is located at the French/Italian Concordia station on the East Antarctic plateau. It is equipped with an FLI Proline science camera with a KAF-16801E, $4096\times4096$ front-illuminated CCD. The camera has an image scale of 0$\farcs$93 pixel$^{-1}$ resulting in a $1^{\circ}\times1^{\circ}$ corrected field of view. The focal instrument dichroic plate splits the beam into a blue wavelength channel for guiding, and a non-filtered red science channel roughly matching an R$_{c}$ transmission curve. The telescope is automated or remotely operated when needed. Due to the extremely low data transmission rate at the Concordia Station, the data are processed on-site using an automated IDL-based pipeline, and the result is reported via email and then transferred to Europe on a server in Rome, Italy. The raw light curves of about 1,000 stars are then available for deeper analysis.  

Three full transits of planet TOI-1064{\bf\,c} were observed on 2020-June-30, 2020-August-18 and 2021-March-26. On 2021-March-14 both an egress of TOI-1064{\bf\,c} and a full transit of TOI-1064{\bf\,b} were observed during the same light curve. In all cases, the weather was fair, with temperatures ranging between $-46^\circ$C and $-65^\circ$C, a stable relative humidity around 50\% and wind speeds between 2 and 7\,$\mathrm{ms}^{-1}$. Exposure times were chosen to be 40\,s in 2020 and 50 to 60\,s in 2021, with a read-out time of about 25\,s. A 9 to 13$\farcs$ radius photometric aperture was found to give the best results.

\subsection{HARPS}
\label{sec:harps} 

\begin{table*}
\caption{A sample of the HARPS observing log of TOI-1064. This table is available in its entirety online.}             
\label{tab:HARPS_obs_log}
\begin{center}
\small
\begin{tabular}{c c c c c c c c c c}
\hline\hline
 Time & RV & $\sigma_{\rm RV}$ & BIS & FWHM & Contrast & S-index & H$\alpha$ & Na D1 & Na D2 \\ 
 {[BJD$-$2457000]} & [km~s$^{-1}$] & [km~s$^{-1}$] & [km~s$^{-1}$] & [km~s$^{-1}$] &  &  &  &  &  \\
\hline                    
  1734.5360 & 21.2169 & 0.0024 & 0.0487 & 6.4368 & 46.1189 & 0.5954 & 0.6869 & 1.2457 & 0.9962 \\ 
  1738.5247 & 21.2254 & 0.0016 & 0.0487 & 6.4494 & 45.9029 & 0.5870 & 0.6882 & 1.2341 & 0.9905 \\
  1739.5374 & 21.2245 & 0.0023 & 0.0562 & 6.4469 & 45.7285 & 0.6237 & 0.6773 & 1.2366 & 0.9907 \\
  ... & \\
\hline\hline                  
\end{tabular}
\end{center}
\end{table*}

We acquired 26 high-resolution ($R$ = 115\,000) spectra of TOI-1064 between 2019-September-08 and 2019-October-29 using the High Accuracy Radial velocity Planet Searcher (HARPS) spectrograph \citep{Mayor2003} mounted at the ESO 3.6m telescope of La Silla Observatory. The observations were carried out as part of the observing program 1102.C-0923. We set the exposure time to 1800\,s, leading to a signal-to-noise (S/N) ratio per pixel at 550\,nm ranging between 18 and 66, with a median of 51. We used the second fibre of the instrument to monitor the sky background and we reduced the data using the dedicated HARPS Data Reduction Software \citep[DRS;][]{Lovis2007}. For each spectrum, the DRS provides also the contrast, the full width at half maximum (FWHM) and the bisector inverse slope (BIS) of the cross-correlation function (CCF). We also extracted additional activity indices and spectral diagnostics, namely the Ca\,{\sc ii} H\,\&\,K lines activity indicator (S-index), H$\alpha$, Na D1 and Na D2, using the code \texttt{TERRA} \citep{AngladaEscude2012}. The 26 DRS and \texttt{TERRA} RV measurements and activity indicators are listed in Table ~\ref{tab:HARPS_obs_log}. Time stamps are given in Barycentric Julian Date in the Barycentric Dynamical Time ($BJD_{\mathrm{TDB}}$).

\subsection{Gemini-South}
\label{sec:gemini} 

Due to the 21\arcsec size of {\it TESS} pixels the {\it TESS} light curves may be contaminated by nearby sources. As mentioned above, for {\it CHEOPS} datasets contamination from stars in the {\it Gaia} DR2 catalogue is simulated and removed. However, should there be sources very near TOI-1064 that were not detected with {\it Gaia}, contamination may still occur. If an exoplanet host star has a spatially close companion, that companion (bound or line of sight) can create a false-positive transit signal if it is, for example, an eclipsing binary. ``Third-light'' flux from the companion star can lead to an underestimated planetary radius if not accounted for in the transit model \citep{Ciardi2015} and cause non-detections of small planets residing with the same exoplanetary system \citep{Lester2021}. Additionally, the discovery of close, bound companion stars, which exist in nearly one-half of FGK-type stars \citep{Matson2018} provides crucial information toward our understanding of exoplanetary formation, dynamics and evolution \citep{Howell2021}. Thus, to search for close-in bound companions unresolved in {\em TESS} or other ground-based follow-up observations, we obtained high-resolution speckle imaging observations of TOI-1064.

TOI-1064 was observed on 2019-September-12 using the Zorro speckle instrument on Gemini South \footnote {https://www.gemini.edu/sciops/instruments/alopeke-zorro/}. Zorro provides simultaneous speckle imaging with a pixel scale of 0.01$\arcsec$ in two bands (562\,nm and 832\,nm) with output data products including a reconstructed image with robust contrast limits on companion detections (e.g. \citealt{Howell2016}). Five sets of 1000$\times$0.06\,s exposures were collected and subjected to Fourier analysis in our standard reduction pipeline (see \citealt{Howell2011}). 

\subsection{ASAS-SN}
\label{sec:asas_sn} 

To further assess the host star, we obtained publicly available ASAS-SN V-band photometry \citep{Shappee2014,Kochanek2017} of TOI-1064 taken over five consecutive seasons between 2014-May-04 and 2018-September-24. Upon inspection of the data, a dimming trend of roughly 0.3\,mag was seen over the four years of observations, with an abrupt increase (roughly 0.2\,mag) in flux occurring during the final season. Therefore, as the goal of using this dataset was to study shorter-period variation, we rejected data after BJD 2458300 in order to avoid photometry taken during the brightening event, and we removed the long-term trend by modelling the dataset with a broad Savitzky–Golay smoothing filter and dividing the fluxes by this model. Lastly, we removed outliers by conducting a 5-sigma clip, which resulted in 401 data points taken on 328 epochs which can be seen in Fig.~\ref{fig:asas_sn_wasp}a). The median flux errors for this dataset are 18\,ppt and thus, whilst these data are not precise enough for transit detection of the two planetary candidates around TOI-1064, they can be used to study photometric variability in the host star.

\subsection{WASP}
\label{sec:wasp} 

Additionally, TOI-1064 was observed during the SuperWASP project \citep{Pollacco2006}, with data taken between 2008-March-26 and 2014-November-10. We retrieved photometry that had been extracted following standard procedures \citep{Pollacco2006}, and detrended for systematic effects using SysRem (see \citealt{CollierCameron2006, Mazeh2007}), that preserves stellar variability and transit-like features in the dataset. Flux and flux uncertainty outliers were rejected via a 5-sigma clipping, with the subsequent light curve shown in Fig.~\ref{fig:asas_sn_wasp}b). This yielded a total of 66\,832 data points on 387 epochs covering four seasons of observations. Similarly to the ASAS-SN dataset, the median flux error for the WASP observations (28\,ppt) means that the photometry is not precise enough to be used to detect the transits of the planetary candidates. However, due to the substantial baseline of the data, they can be used to search for photometric variability.

\section{Characterisation of TOI-1064}
\label{sec:hoststar}

\begin{table}
\caption{Stellar properties of TOI-1064.}
\label{tab:stellarParam}      
\begin{center}
\begin{tabular}{lll}        
\hline\hline                 
\multicolumn{3}{c}{TOI-1064} \\    
\hline                        
2MASS & \multicolumn{2}{l}{J19440094-4733417} \\
Gaia EDR3 & \multicolumn{2}{l}{6683371847364921088} \\
TIC & \multicolumn{2}{l}{79748331} \\
UCAC2 & \multicolumn{2}{l}{11441398} \\ 
\hline
Parameter & Value & Note \\ 
\hline
   $\alpha$ [J2000] & 19$^{\rm h}$44$^{\rm m}$00.95$^{\rm s}$ & 1 \\
   $\delta$ [J2000] & -47$^{\circ}$33$^{'}$41.75$^{\arcsec}$ & 1 \\
   $\mu_{\alpha}$ [mas/yr] & -3.543$\pm$0.015 & 1 \\
   $\mu_{\delta}$ [mas/yr] & -100.885$\pm$0.012 & 1 \\
   $\varpi$ [mas] & 14.532$\pm$0.015 & 1 \\
   $d$ [pc] & 68.81$\pm$0.07 & 5 \\
   RV [km~s$^{-1}$] & 20.7$\pm$0.7 & 1 \\
   U [km~s$^{-1}$] & 11.50$\pm$0.61 & 5$^{\rm a}$ \\
   V [km~s$^{-1}$] & -34.29$\pm$0.08 & 5$^{\rm a}$ \\
   W [km~s$^{-1}$] & -14.32$\pm$0.34 & 5$^{\rm a}$ \\
\hline
   $V$ [mag] & 10.95$\pm$0.06 & 2 \\
   $G_{\rm BP}$ [mag] & 11.207$\pm$0.003 & 1 \\
   $G$ [mag] & 10.645$\pm$0.003 & 1 \\
   $G_{\rm RP}$ [mag] & 9.939$\pm$0.004 & 1 \\
   $J$ [mag] & 9.10$\pm$0.02 & 3 \\
   $H$ [mag] & 8.63$\pm$0.04 & 3 \\
   $K$ [mag] & 8.47$\pm$0.03 & 3 \\
   $W1$ [mag] & 8.41$\pm$0.02 & 4 \\
   $W2$ [mag] & 8.48$\pm$0.02 & 4 \\
\hline
   $T_{\mathrm{eff}}$ [K] & 4734$\pm$67 & 5; spectroscopy \\
   $\log{g}$ [cm~s$^{-2}$]      & 4.60$\pm$0.06 & 5; spectroscopy \\\relax
   [Fe/H] [dex] & 0.05$\pm$0.08 & 5; spectroscopy \\\relax
   [Mg/H] [dex] & 0.03$\pm$0.06 & 5; spectroscopy \\\relax
   [Si/H] [dex] & 0.06$\pm$0.08 & 5; spectroscopy \\\relax
   [Ca/H] [dex] & 0.11$\pm$0.10 & 5; spectroscopy \\\relax
   [Na/H] [dex] & 0.17$\pm$0.12 & 5; spectroscopy \\
   $v \sin i_\star$ [km~s$^{-1}$] & 2.7$\pm$0.7 & 5; spectroscopy \\\relax
   $\log{R}_{\rm HK}^{'}$ & -4.633$\pm$0.024 & 5; spectroscopy \\
   $E(B-V)$ & 0.056$\pm$0.032 & 5; IRFM \\
   $R_{\star}$ [$R_{\odot}$] & 0.726$\pm$0.007 & 5; IRFM \\
   $M_{\star}$ [$M_{\odot}$] & 0.748$\pm$0.032 & 5; isochrones \\  
   $t_{\star}$ [Gyr]        & 9.4$\pm$3.8 & 5; isochrones \\
   $L_{\star}$ [$L_{\odot}$] & 0.238$\pm$0.014 & 5; from $R_{\star}$ and $T_{\mathrm{eff}}$\\
   $\rho_{\star}$ [$\rho_\odot$] & 1.95$\pm$0.10 & 5; from $R_{\star}$ and $M_{\star}$ \\
   $\rho_{\star}$ [$\mathrm{g\,cm^{-3}}$] & 2.76$\pm$0.14 & 5; from $R_{\star}$ and $M_{\star}$ \\
\hline\hline                                   
\end{tabular}
\end{center}
[1] \cite{GaiaCollaboration2021}, [2] \cite{Zacharias2012}, [3] \cite{Skrutskie2006}, [4] \cite{Wright2010}, [5] This work \\
$^{\rm a}$ Calculated via the right-handed, heliocentric Galactic spatial velocity formulation of \cite{Johnson1987} using the proper motions, parallax, and radial velocity from [1].
\end{table}

\subsection{Atmospheric Properties and Abundances}

\begin{figure*}
    \centering
    \subfloat[\centering ASAS-SN]{{\includegraphics[width = 8.5cm]{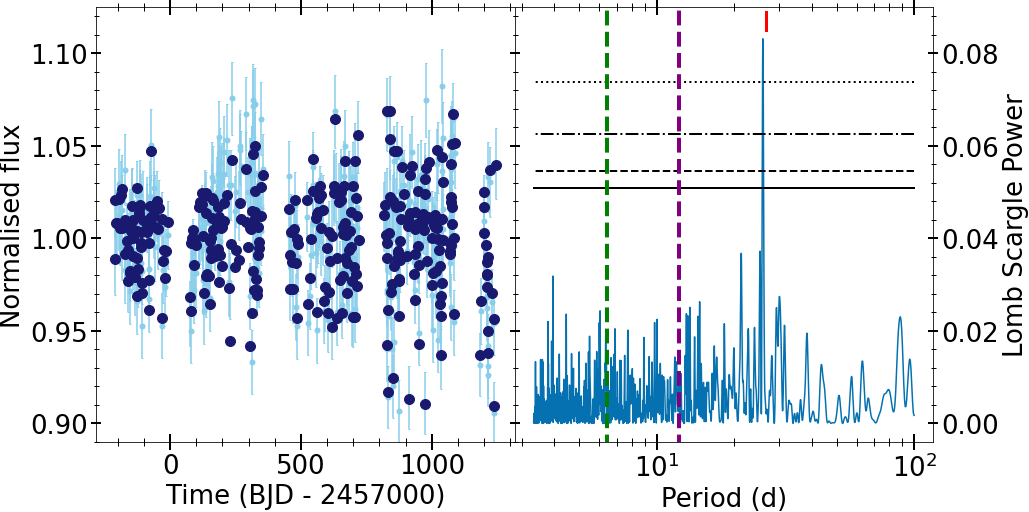} }}
    \qquad
    \subfloat[\centering WASP]{{\includegraphics[width = 8.5cm]{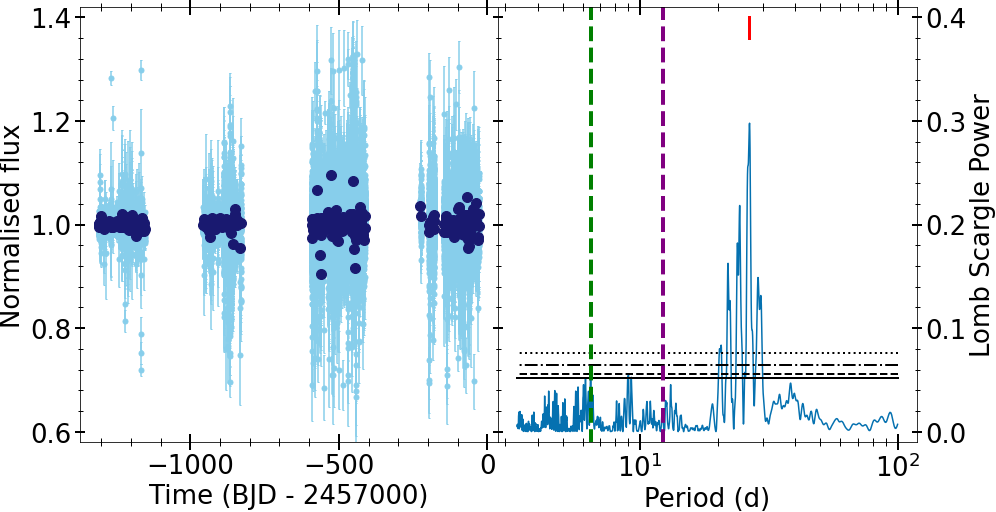} }}
    \caption{a) ASAS-SN and b) WASP data used to study the photometric variability of TOI-1064. {\em Left panels}: The full light curves (light blue) and nightly-binned data (dark blue). {\em Right panels}: Lomb-Scargle periodograms of the nightly-binned data, with the orbital periods of planet b and c shown as vertical green and purple lines, respectively. The horizontal black dotted, dash-dot, dashed, and solid lines (from top to bottom) are the 0.1\%, 1\%, 5\%, and 10\% False Alarm Probabilities, respectively. The adopted rotation period is given as red vertical marks.}
    \label{fig:asas_sn_wasp}
\end{figure*}

The spectral analyses of the host star of the TOI-1064 system were based on the co-addition of all the radial velocity observations carried out with the HARPS spectrograph, detailed in Section~\ref{sec:harps}, at a spectral resolution of 115\,000. 
We began with applying the {\tt{SpecMatch-Emp}} \citep{Yee2017} software to our data. Our observed optical spectrum is compared to a spectral library of approximately four hundred stars observed by Keck/HIRES with spectral classes M5 to F1. 
Interpolating and minimising differences, the direct output is the effective temperature of the star, $T_\mathrm{eff}$, the stellar radius $R_{\star}$, and the 
iron abundance [Fe/H] which are found to be $4780\pm110$~K, $0.77\pm 0.10$~$R_{\odot}$, and $0.05\pm0.09$ (dex) for TOI-1064, respectively.

We then calculate a completely synthesised spectrum, using the \href{http://www.stsci.edu/~valenti/sme.html}{{\tt{SME}} }\citep[Spectroscopy Made Easy;][]{Valenti1996, Piskunov2017} 
package version 5.22 in the fashion explained in more detail in e.g. \citet{Fridlund2017}. We use as starting values the $T_\mathrm{eff}$ and [Fe/H] from  {\tt{SpecMatch-Emp}}. We model the $v \sin i_\star$ by fitting large numbers of narrow and unblended metal lines between 6000 and 6500\,\AA. 
By determining the depths and profiles of Na, Ca, and Fe lines, the [Na/H], [Ca/H], and [Fe/H] abundances 
are computed, as well as the logarithm of the stellar surface gravity, $\log{g}$, from the \ion{Ca}{I} triplet around 6200\,\AA. We hold the 
turbulent velocities, $V_{\rm mic}$ and $V_{\rm mac}$ fixed to 0.47\,km~s$^{-1}$ and 1.2\,km~s$^{-1}$, respectively, 
based on $T_\mathrm{eff}$ and $\log{g}$ \citep{Adibekyan2012b, Doyle2014}. As a final step, our model was checked with the \ion{Na}{I} doublet 5589 and 5896\,\AA, sensitive to both $T_\mathrm{eff}$ and $\log{g}$. Through these steps we arrive at the values listed in Table~\ref{tab:stellarParam} which are in excellent agreement with the {\tt{SpecMatch-Emp}} model, and confirm that TOI-1064 is a K-dwarf.

We also derive stellar parameters using the ARES+MOOG tools \citep{Sousa2014}. In particular, the equivalent widths of iron lines are measured using the ARES code\footnote{The last version of ARES code (ARES v2) can be downloaded at \url{http://www.astro.up.pt/$\sim$sousasag/ares}} \citep{Sousa2007,Sousa2015} and the iron abundance are then computed using Kurucz model atmospheres \citep{Kurucz1993a} and the radiative transfer code MOOG \citep{Sneden1973}. The values reported here are obtained via convergence of both ionisation and excitation equilibria. We obtain $T_\mathrm{eff} = 4636\pm102\,$K, $\log{g} = 4.52\pm0.24$, and ${\rm [Fe/H]} = -0.04\pm0.05$, which are consistent with the results presented above. Due to the smaller uncertainties we adopt the {\tt{SME}} values.

Using the stellar parameters listed in Table~\ref{tab:stellarParam}, we determine the abundances of [Mg/H] and [Si/H] to be $0.03\pm0.06$ and $0.06\pm0.08$, respectively, using the classical curve-of-growth analysis method assuming local thermodynamic equilibrium. We use the ARES v2 code \citep{Sousa2015} to measure the equivalent widths (EW) of the spectral lines of these elements. Then we use a grid of Kurucz model atmospheres \citep{Kurucz1993a} and the radiative transfer code MOOG \citep{Sneden1973} to convert the EWs into abundances. When doing so, we closely follow the methods described in e.g. \citet{Adibekyan2012a, Adibekyan2015}.
In Table~\ref{tab:stellarParam} we also include the right-handed, heliocentric galactic spatial velocities that, along with the derived [Fe/H], indicate that TOI-1064 is a member of the galactic thin disc population.

\subsection{Radius, Mass, and Age}

In order to determine the stellar radius of TOI-1064 we use a modified version of the infrared flux method (IRFM; \citealt{Blackwell1977}) that allows the derivation of stellar angular diameters and effective temperatures through known relationships between these properties, and an estimate of the apparent bolometric flux, recently detailed in \citet{Schanche2020}. We perform the IRFM in a Markov-Chain Monte Carlo (MCMC) approach in which the stellar parameters, derived via the spectral analysis detailed above, are used as priors in the construction of spectral energy distributions (SEDs) from stellar atmospheric models. The SEDs are subsequently attenuated to account for reddening and used for synthetic photometry. This is conducted by convolving the SED with the broadband response functions for the chosen bandpasses, with the fluxes compared to the observed data to compute the apparent bolometric flux. For this study, we retrieve broadband fluxes and uncertainties for TOI-1064 from the most recent data releases for the following bandpasses; {\it Gaia} G, G$_{\rm BP}$, and G$_{\rm RP}$, 2MASS J, H, and K, and {\it WISE} W1 and W2 \citep{Skrutskie2006,Wright2010,GaiaCollaboration2021}, and use the \textsc{atlas} Catalogues \citep{Castelli2003} of model stellar spectral energy distributions. Internal to the MCMC, the computed posterior distributions of stellar angular diameters are converted to distributions of stellar radii using the {\it Gaia} EDR3 parallax \citep{GaiaCollaboration2021}, from which we obtain the stellar radius of TOI-1064 and $E(B-V)$ to be $R_{\star}=0.726\pm0.007\, R_{\odot}$ and $E(B-V)=0.056\pm0.032$, respectively. The distance to TOI-1064 is calculated in the IRFM using the {\it Gaia} EDR3 parallax with the parallax offset of \cite{Lindegren2021} applied. These stellar parameters are reported in Table~\ref{tab:stellarParam}.

By adopting $T_{\mathrm{eff}}$, [Fe/H], and $R_{\star}$ as input parameters, we infer the stellar age $t_{\star}$ and mass $M_{\star}$ through stellar evolutionary models. To make our analysis more robust, we employ two different techniques each applied to a different set of stellar isochrones and tracks. The first technique derives $t_{\star}$ and $M_{\star}$ using the isochrone placement method described in \citet{Bonfanti2015,Bonfanti2016}, which interpolates the input parameters within pre-computed grids of isochrones and tracks generated by the PARSEC\footnote{\textsl{PA}dova and T\textsl{R}ieste \textsl{S}tellar \textsl{E}volutionary \textsl{C}ode: \url{}} v1.2S code \citep{Marigo2017}. The second technique, instead, directly fits the input parameters in the CLES \citep[Code Liègeois d'Évolution Stellair,][]{Scuflaire2008} code to then retrieve $t_{\star}$ and $M_{\star}$ following a Levenberg-Marquardt minimisation scheme as described in \citet{Salmon2021}.
Once the two pairs of age and mass values are derived, first their consistency is checked through a $\chi^2$ test, and then their probability distributions are combined together to provide the final $t_{\star}$ and $M_{\star}$ values with errors at the 1-$\sigma$ level (see Table \ref{tab:stellarParam}). Specific details about the statistical derivation of both $t_{\star}$ and $M_{\star}$ may be found in \citet{Bonfanti2021}.

From the stellar radius and mass determined via the IRFM and isochrone placement techniques we obtain log\,$g$ = 4.59$\pm$0.02, that is in agreement with the value derived from the spectral analysis.

The derived stellar radius was checked with the python code \href{https://github.com/jvines/astroARIADNE}{{\tt{ARIADNE}}} (described in e.g.  
\citet[][]{Acton2020}) that fitted broadband photometry to the 
{\tt {Phoenix~v2}} \citep{Husser2013}, {\tt {BtSettl}} \citep{Allard2012}, 
\citet{Castelli2003}, and \citet{Kurucz1993b} atmospheric model grids, utilising data in the following bandpasses {\it Gaia} EDR3 G, G$_{\rm BP}$, and G$_{\rm RP}$, 2MASS J, H, and K, {\it WISE} W1 and W2, and the Johnson B and V magnitudes from APASS. We used {\tt{SME}} values for $T_\mathrm{eff}$, log\,$g$, and [Fe/H] as priors to the model. 
The final radius is computed with Bayesian Model Averaging. 
We obtain a stellar radius of $0.723 \pm 0.007$~$R_\odot$. Combining this radius with the surface gravity, we obtain a mass of $0.762 \pm 0.050$~$M_\odot$. Both values are in excellent agreement with the above derived mass and radius.

\subsection{Stellar Variability}
\label{sec:stellarvar} 

Stellar activity can contribute strong signals that are apparent in RV observations and can hinder the detection and characterisation of small exoplanets via precise measurements of planetary masses \citep{Haywood2014,Rajpaul2015,Mortier2016,Dumusque2017,Faria2020}. Therefore, there have been recent efforts made to mitigate the effect of stellar activity on RV observations of exoplanets \citep{deBeurs2020,CollierCameron2021}, that will be discussed in greater detail in Section~\ref{sec:scalpels}. 

In order to properly account for the stellar activity we need to measure the stellar rotation period, which can be done via inspection of the RV and photometric data of TOI-1064. As can be seen in the Lomb-Scargle periodograms of standard stellar activity indicators of our HARPS observations (Fig.~\ref{fig:activity_inds}), no significant peaks potentially related to the stellar rotation period are apparent. Therefore, we assess the long-baseline ASAS-SN and WASP light curves to search for photometric variability. Firstly, for both datasets we performed a nightly binning that results in RMS scatters of 28 and 16\,ppt for ASAS-SN and WASP, respectively. Subsequently, we produced Lomb-Scargle periodograms of both binned light curves and find significant peaks at 25.9 and 26.6\,d as shown in Figs.~\ref{fig:asas_sn_wasp}a) and b). To independently confirm this variability period we applied a Gaussian Process (GP) regression with a quasi-periodic kernel to both ASAS-SN and WASP light curves separately. This kernel is chosen as multiple previous studies have found that it accurately represents flux modulation from stellar activity \citep{Haywood2014,Dubber2019,Mortier2020}. Using the {\tt juliet} Python package \citep{Espinoza2019} with unconstrained priors we obtain median and 1$\sigma$ uncertainties for the rotation period of the kernel to be 27.0$\pm$4.3 and 26.6$\pm$1.0\,d, for ASAS-SN and WASP, respectively. Due to the higher S/N of the WASP observations, we adopt the value obtained from that dataset and interpret this signal to be the stellar rotation period. These values agree well with the rotation period of 26.9$\pm$1.6\,d derived from the mean HARPS value of $\log{R}_{\rm HK}^{'}$ following the empirical relations in \cite{Mamajek2008}. It is worth noting that the derived rotation periods are distinct from the average lunar synodic period of approximately 29.53\,d.

\section{Validating and Fitting the System}
\label{sec:analysis}

\subsection{High Resolution Imaging Analysis and Planet Validations}
\label{sec:aoanalysis}

\begin{figure}
  \includegraphics[width = 8.5cm]{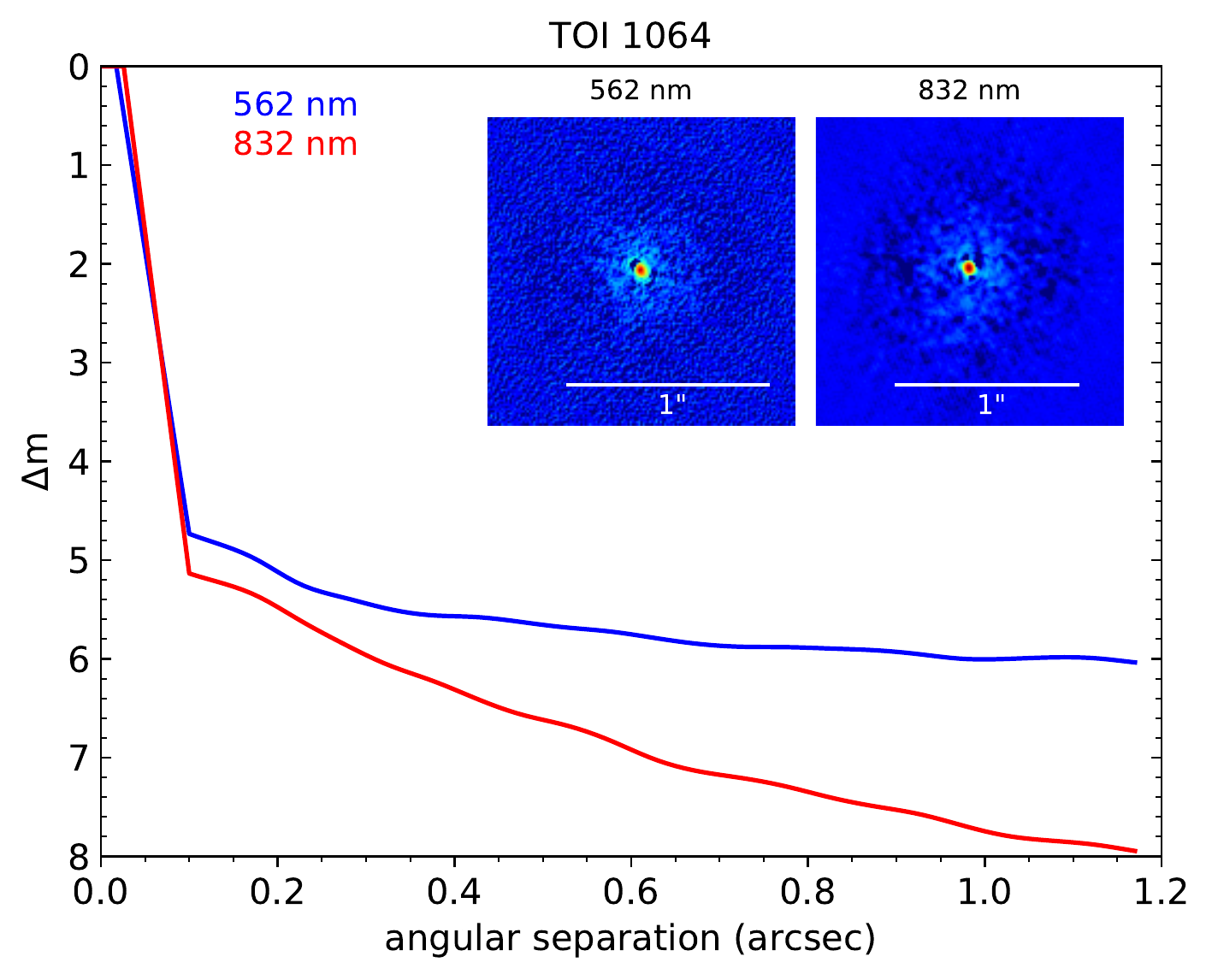}
  \caption{The 5$\sigma$ contrast curves of Gemini/Zorro speckle high-resolution images at 562\,nm (blue) and 832\,nm (red), with insets showing the central region of the images centred on TOI-1064. }
  \label{fig:gemini_ao}
\end{figure}

To establish if there is a potentially contaminating source nearby to TOI-1064 we analysed high resolution images obtained with the Gemini/Zorro instrument by determining 5$\sigma$ contrast curves in both bandpasses. Fig.~\ref{fig:gemini_ao} shows our final contrast curves and the reconstructed speckle images. We find that TOI-1064 is a single star from 20\,mas out to 1.2\arcsec with no companion brighter than 5-8 magnitudes below that of the target star beyond 100\,mas. At the distance calculated in Section~\ref{sec:hoststar}, this corresponds to the absence of a main sequence star at spatial limits of 1.36\,au to 82\,au. This isolation is supported by SOAR/HRCam observations that find no companion within 1\arcsec at a 4.5 magnitude contrast limit \citep{Ziegler2021}.

\begin{figure}
  \includegraphics[width = 8.5cm]{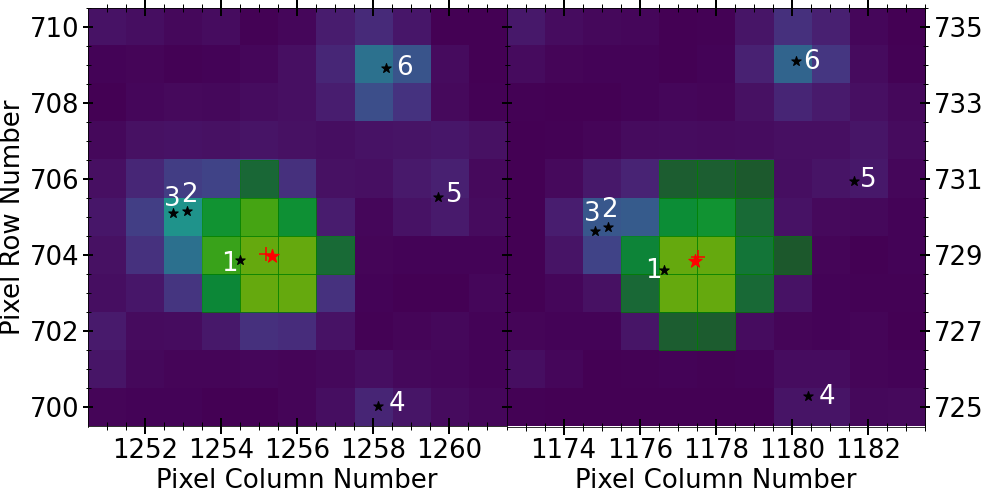}
  \includegraphics[width = 8.5cm]{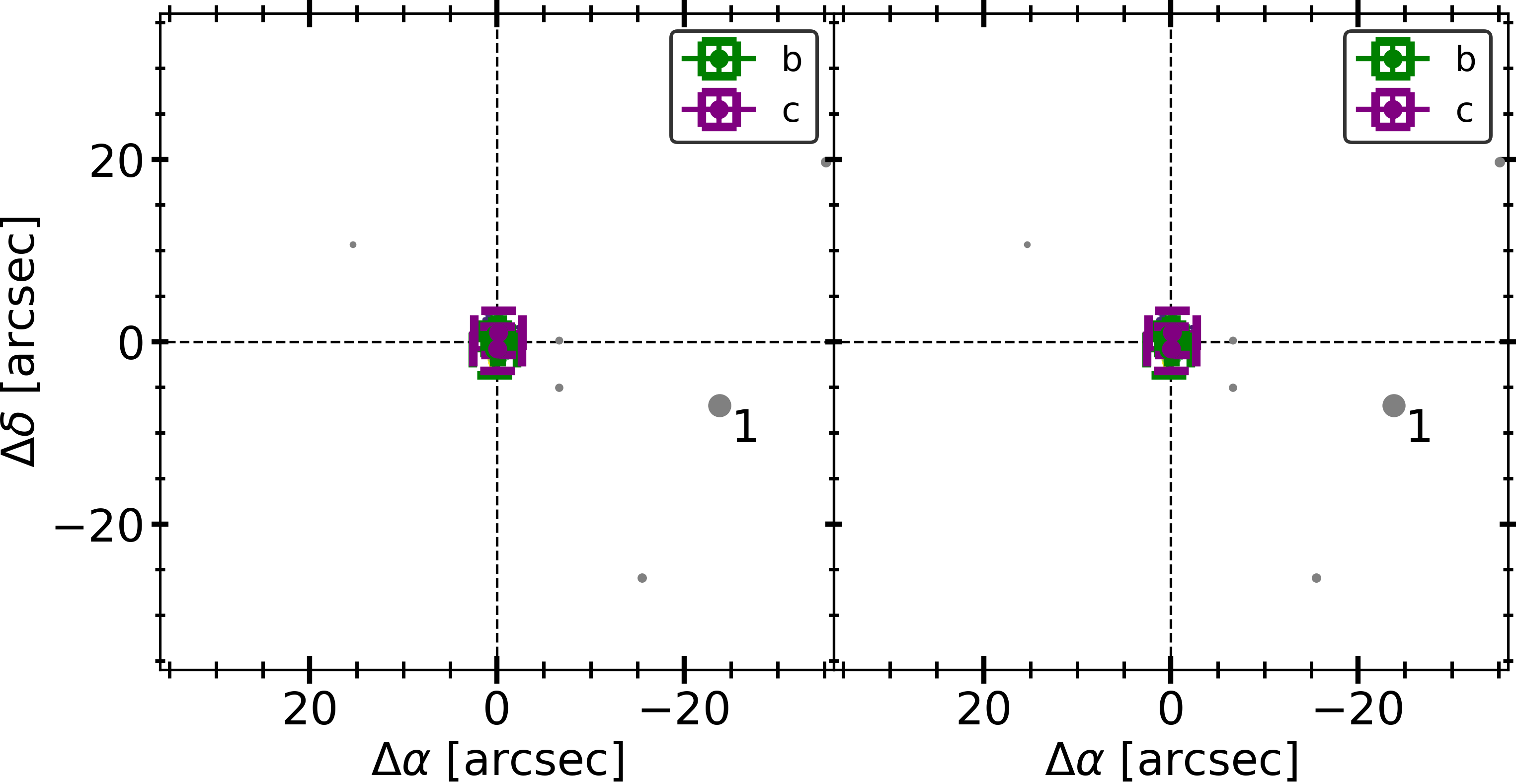}
  \caption{{\em Top panels}: Example {\it TESS} target pixel files covering 3.5$\times$3.5' for Sectors 13 ({\em left}) and 27 ({\em right}) showing TOI-1064 (red star) and nearby stars with $G < $ 15\,mag (numbered, black stars). The {\em TESS} pipeline aperture photometry mask is shown as green squares with the average centroid position given as a red plus. {\em Bottom panels}: 35$\times$35\arcsec regions for Sectors 13 ({\em left}) and 27 ({\em right}) showing all nearby stars with $G < $ 20\,mag, including source 1 seen in the {\it TESS} target pixel files. In-transit offsets to out-of-transit positions for planet b and c are shown in green and purple, respectively.}
  \label{fig:tess_cens}
\end{figure}

Therefore, we conclude that the {\it CHEOPS} photometry is not contaminated by nearby sources undetected by {\it Gaia}, whereas all other sources are corrected for by subtracting the contamination estimate described in Section~\ref{sec:cheops}. However, as the 21$\arcsec$ {\it TESS} pixels are substantially larger than those of {\it CHEOPS}, the nearby source with $\Delta G$ = +3.9\,mag seen in Fig.~\ref{fig:cheops_psf} may affect the {\it TESS} photometry. Thus, we analysed the centroid position of the {\it TESS} observations in order to assess if this source affects the photometry.

Firstly, we inspected the {\em TESS} target pixel files of both sectors to ascertain which nearby sources may affect the observations. As can be seen in the top panels of Fig.~\ref{fig:tess_cens} there is only one $G < $ 15\,mag source with the core of its PSF within the pipeline aperture photometry mask. The wings of the PSFs of additional nearby sources may also fall within the aperture photometry mask, however, as the core of the PSFs and the majority of the flux from these stars are outside of the photometric mask we conclude that these objects do not contaminate the {\it TESS} photometry of TOI-1064. To assess if the nearby object is affecting the data, we computed the average in-transit centroid positions for all transits of both planets in the system and subsequently determined the offsets between these values and the average out-of-transit centroid positions. The bottom panels of Fig.~\ref{fig:tess_cens} show that the in-transit data is obtained on target with an average offset in Sector 13 of 0.1$\pm$1.3\arcsec\ and 0.0$\pm$1.5\arcsec, and in Sector 27 of 0.2$\pm$3.6\arcsec\ and 0.2$\pm$3.6\arcsec\ for planet candidates 01 and 02, respectively. Thus, as the singular $G < $ 15\,mag object in the aperture mask is 22\arcsec\ (roughly 15$\sigma$ away) and 29\arcsec\ (roughly 8$\sigma$ away) in Sectors 13 and 27, respectively, we conclude that it does not affect the {\em TESS} photometry of TOI-1064 and that the observed transits can be attributed to the target star. This finding is consistent with the {\em TESS} SPOC difference image centroiding results that constrained the source of the transits to within 4\arcsec of the target using data from both sectors. Lastly, it should be noted that whilst there are additional nearby objects, their comparatively greater $G$-band magnitudes ($\Delta G > $ +9\,mag) mean that they do not contaminate the {\it TESS} photometry substantially.

Previous work has noted that multi-planet systems have a very low probability of being false positives \citep{Lissauer2012} and thus we may consider these candidates to be verified based on the multiplicity of the system. However, to further confirm both planet candidates we utilised the statistical validation tool, {\tt triceratops} \citep{Giacalone2021}, that uses stellar and transit parameters, transit photometry, and high-resolution speckle imaging in order to determine the False Positive Probability (FPP) of planetary candidates. This is done by first querying the TIC to calculate the flux contribution of nearby stars in a given aperture and determining which sources are bright enough to feasibly host a body with a given transit depth. The {\it TESS} and any additional light curves are subsequently fitted with transit and eclipsing binary models to determine the probability in a Bayesian manner of a transiting planet, an eclipsing binary, and an eclipsing binary on twice the orbital period around the target or nearby star in the case of no unresolved companion, around the primary or secondary star in the case of an unresolved bound companion, and around the target or background star in the case of an unresolved background star, resulting in 18 scenarios. From these probabilities, the overall FPP of a transiting planet is computed. Using our stellar mass, radius, effective temperature, and parallax values from Table~\ref{tab:stellarParam}, and all photometric data we find that both TOI-1064{\bf \,b} and{\bf \,c} have FPP values < 1\% and thus, confirm the presence of both planets.

\subsection{PSF and CCF Shape Monitoring using {\sc scalpels}}
\label{sec:scalpels} 

\subsubsection{PSF}

\begin{figure}
  \includegraphics[width = 8.5cm]{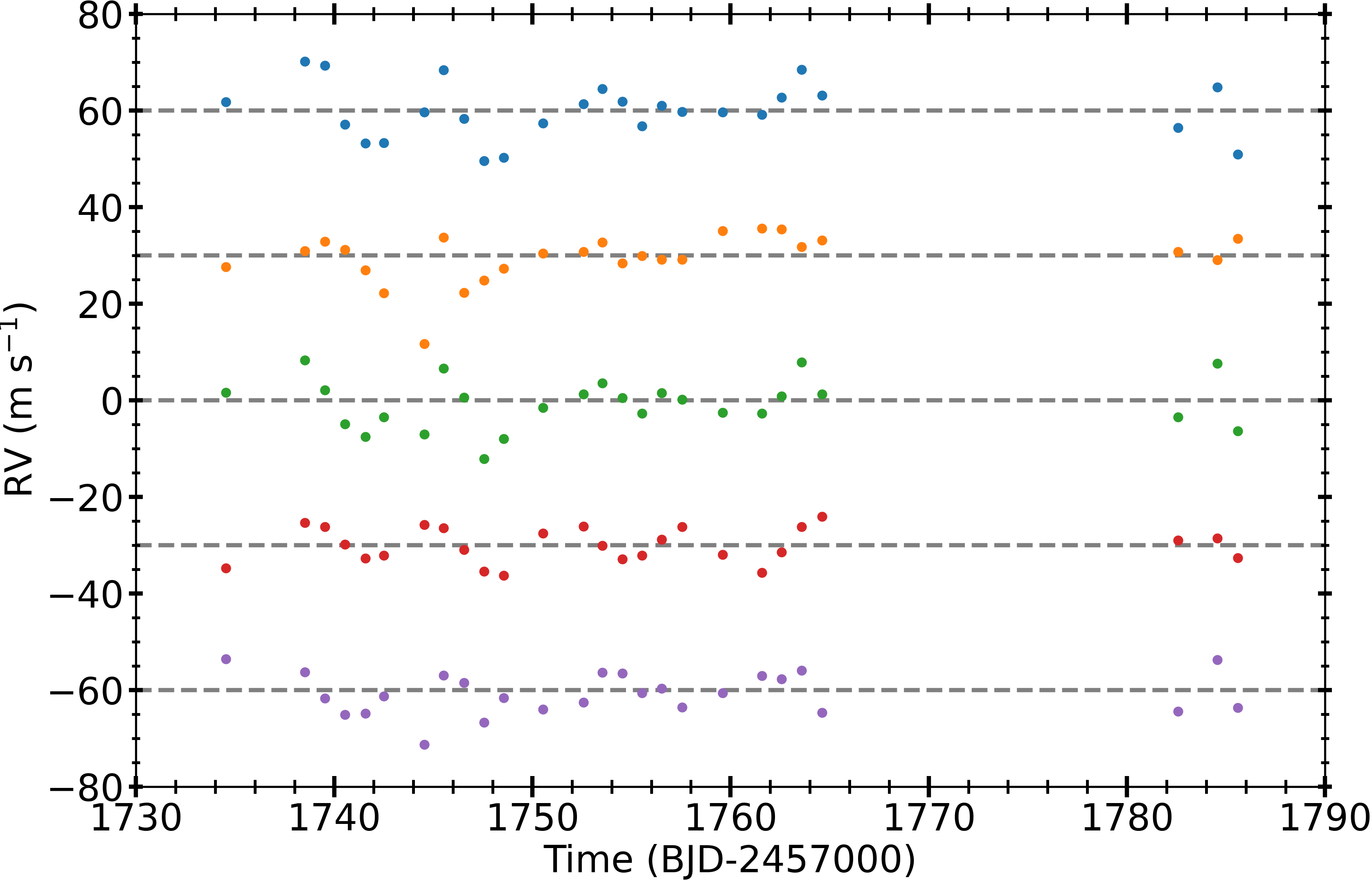}
  \caption{The HARPS time series of TOI-1064, offset for clarity, showing; the observed, median-subtracted RVs in blue, the CCF shape-based RVs in orange, the CCF shift-based RVs in green, the fit of two Keplerian orbits to the cleaned RVs in red, and the residuals to the fit in purple.}
  \label{fig:scalpels_rv}
\end{figure}

{\em CHEOPS} photometry is known to suffer from at least two types of systematic error arising from thermal effects which are found to be correlated with the output of temperature sensors in the telescope structure \citep{Morris2021}. It has been established that when the sunlight illumination pattern on the spacecraft changes, thermal flexure of the telescope structure causes subtle changes in the shape of the PSF. This in turn causes a change in the fraction of the flux entering the pupil that falls within the fixed-radius photometric aperture in the focal plane used for signal extraction. The first type of systematic is a recognised secular "ramp" effect as the telescope structure settles into a new equilibrium following a change of pointing direction (\citealt{Morris2021}, Maxted et al. {\it accepted}), and the second type of systematic is shorter-term periodic fluctuation modulated on the 98.77\,min orbital period as the spacecraft passes in and out of the Earth's shadow. Such "ramps" have been seen in space-based photometry of other telescopes \citep{Deming2006,Berta2012,Demory2015}. The details of the interplay between spacecraft illumination, thermal changes in the telescope structure and the response of the PSF shape cannot be modelled directly. We can, however, use a simple linear unsupervised machine-learning approach to establish correlations between changes in the shape of the PSF and changes in the encircled fraction of the total stellar flux.

Conceptually the problem is similar to the systematic errors produced in radial-velocity measurements by stellar activity-driven changes in the shapes of spectral lines, and the solution we adopt here is modelled on the {\sc scalpels} algorithm developed by \citet{CollierCameron2021} for separating shape-driven radial-velocity offsets from genuine stellar Doppler shifts, and is detailed in Appendix~\ref{sec:scalpels_desc}.

In this study, we applied our novel method to the {\em CHEOPS} photometry and the CCF-based {\sc scalpels} \citep{CollierCameron2021} to the HARPS data in order to model flux modulation due to PSF shape changes and RV variation due to stellar activity. For the {\it CHEOPS} light curves, we used the aforementioned method on the DRP produced fluxes (Section~\ref{sec:cheops}) for the six visits separately, with the number of principal components, $\mathbf{\theta}$, chosen by the LOOCV method to be; 24, 17, 23, 10, 26, and 42, respectively, corresponding to: 5\%, 6\%, 8\%, 4\%, 10\%, and 5\%, of the vectors produced by the principal component analysis. Subsequently, we decorrelated the {\em CHEOPS} datasets against the selected vectors using the linear regression method, masking the in-transit fluxes. The linear models produced by the regression, and used to decorrelate the light curves, are shown in the bottom panels of Fig.~\ref{fig:cheops_raw1} and Figs.~\ref{fig:cheops_raw2}-\ref{fig:cheops_raw6}. In order to be conservative in the subsequent global analysis detailed below and preserve the uncertainty of the decorrelation fits, the errors on the linear models were estimated from one thousand samples drawn from the posterior distributions of the regression coefficients and added in quadrature with the flux errors of the detrended photometry for each dataset. These decorrelated light curves yield 3\,h noise estimates of: 58.4\,ppm, 51.9\,ppm, 49.5\,ppm, 53.9\,ppm, 45.8\,ppm, and 53.0\,ppm, respectively, and were used in our joint analysis below. 

\subsubsection{CCF}

Following the stellar variability analysis (see Section~\ref{sec:stellarvar}), it was noted that the {\it TESS}-derived 12.2\,d orbital period for TOI-1064\,c is close to the 13.3\,d first harmonic of the stellar rotation period. As stellar activity signals can impede the detection of exoplanets in RV data \citep{Haywood2014,Rajpaul2015,Mortier2016,Dumusque2017}, we employed the CCF-based {\sc scalpels} \citep{CollierCameron2021} to separate any stellar activity signals from planet-induced RV variations via the re-extraction of RVs from the 26 HARPS observations (see Section~\ref{sec:harps}). For this dataset, the LOOCV approach selected three principal components (12\%) that well model the CCF shape variations and hence stellar activity. 

As reported in \cite{CollierCameron2021}, by conducting a joint fit of the stellar activity and planetary RVs, orbital signals injected into Solar RV data can be more reliably retrieved using {\sc scalpels}. Therefore, as a preliminary study of the HARPS data, we performed simultaneous detrending using the CCF-based {\sc scalpels} method and a fit two Keplerian orbits utilising the approach presented in the appendix of the aforementioned paper. Using the {\it TESS}-derived values as priors and assuming circular orbits, we detected TOI-1064\,b with a semi-amplitude, $K_{\rm b} = 4.7\pm0.5$\,m\,s$^{-1}$, and tentatively found TOI-1064\,c with a semi-amplitude, $K_{\rm c} = 1.8\pm0.6$\,m\,s$^{-1}$. The median-subtracted, raw RV time series, along with the {\sc scalpels}-determined stellar activity signal, the resulting planetary RV component, and the orbital fit and residuals, are shown in Fig.~\ref{fig:scalpels_rv}.

The {\sc scalpels} produced vectors represent variations in observed RVs due to stellar activity, and so can be used to potentially remove such variation via a linear regression. Thus, the three identified components were subsequently used to detrend the HARPS RVs in our global analysis simultaneously with the RV fitting.

\subsection{Joint Photometric and Radial Velocity Analysis}
\label{sec:joint_analysis} 

\begin{figure*}
  \includegraphics[width = \textwidth]{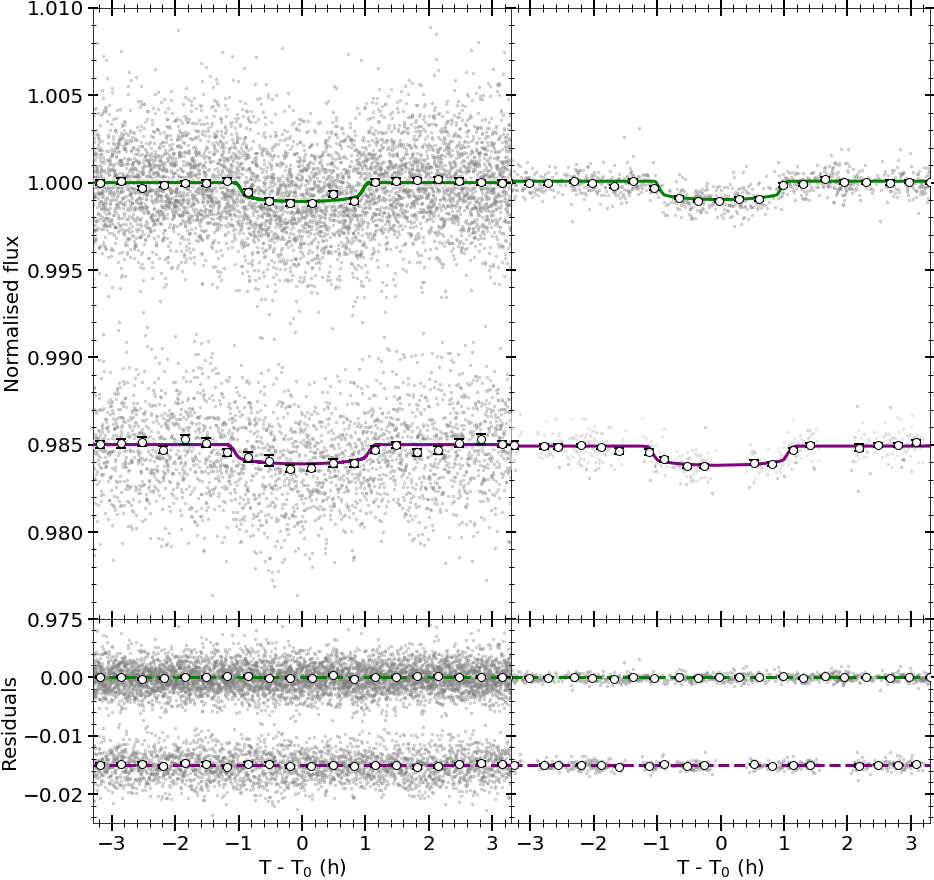}
  \caption{{\it Top panels}: Detrended {\it TESS} ({\it left}) and {\it CHEOPS} ({\it right}) photometry phase-folded to the orbital periods of planet\,b and\,c (offset by -0.015 for clarity). Unbinned data are shown in grey, with fluxes binned every 15\,min given as white circles. The best-fit transit models for TOI-1064\,b and\,c are presented in green and purple, respectively. {\it Bottom panels}: Residuals to the fits shown in the top panels with the same colouration.}
  \label{fig:tess_cheops_transits}
\end{figure*}

In order to determine the properties of TOI-1064\,b and c, we conducted a global analysis of the {\it TESS}, {\it CHEOPS}, LCOGT, NGTS, and ASTEP transit photometry, and the HARPS RV data. Prior to the joint fitting, we carried out additional preliminary checks to further assess the data and ascertain a complete model for use in the analysis. 

\subsubsection{Preliminary Analyses}

As can be seen from Fig.~\ref{fig:cheops_raw1} and Figs.~\ref{fig:cheops_raw2}-\ref{fig:cheops_raw6}, the linear models produced by the PSF-based {\sc scalpels} provide an excellent reproduction of the flux variation seen in the {\em CHEOPS} light curves, and indicate that the flux modulation typically seen on {\em CHEOPS} orbit timescales can be well accounted for via modelling of the PSF shape changes alone. To check whether additional detrended basis vectors were needed to model {\it CHEOPS} roll angle flux variation, we used the {\tt pycheops}\footnote{\url{https://github.com/pmaxted/pycheops}} Python package (Maxted et al. {\it accepted}) to evaluate the detrended data produced by the method outlined above. For each visit, we performed simultaneous transit fitting and detrending for all combinations of standard basis vectors used in the decorrelation of {\it CHEOPS} data (i.e. background, contamination, smear, x and y centroid positions, and first, second, and third-order harmonics of the roll angle), and assessed the reported Bayes Factors for the considered basis vectors. We found that for all visits, the fitting favoured not including additional detrending parameters, and thus for the joint fit no further basis vectors are used.

We also performed a preliminary analysis of the HARPS data utilising the RV fitting capabilities of {\tt radvel} within {\tt juliet} \citep{Fulton2018,Espinoza2019} in order to determine whether there are any long-term trends seen in the data that should be accounted for. Using the stellar and planetary priors outlined below, we conducted two fits with the inclusion or not of a RV slope and intercept. By comparing the nested sampling-produced Bayes evidences, we found that excluding a long-term RV trend is favoured, and therefore, it is omitted from the global analysis below.

Lastly, we conducted a stability analysis of the system to determine dynamically plausible regions of the eccentricity and argument of periastron planes for both planets, that can be used to constrain priors on these parameters in the global fit. We consider an architecture unstable if the trajectories of planets would result in intersecting orbits calculated as satisfying the following:

\begin{equation}
    \Bigg| \frac{p_1-p_2}{\sqrt{(p_1e_2)^2+(p_2e_1)^2-2p_1p_2e_1e_2\cos(\omega_1-\omega_2)}} \Bigg| < 1
\label{eq:intersect}
\end{equation}

where $p_j=a_j(1-e_j)$ for $j$ = 1,2. Using this formulation, we perform 100,000 calculations randomly sampling the $\sqrt{e}$cos$\omega$ and $\sqrt{e}$sin$\omega$ parameter space between -1 to 1 for both planets, ensuring that the eccentricities remain below one, whilst fixing the nominal values for all remaining parameters. In total, we find that 23.2\% of scenarios remain stable with the remaining architectures resulting in orbit crossing events that occur more frequently with larger values of the eccentricity and argument of periastron components. Therefore, in order to provide bounds for the eccentricity and argument of periastron priors and to avoid the subsequent global fit from returning unstable results, cut-offs for these priors are needed. In our simulations we find that bounds of $\pm$0.5 for both components on both planets result in 96.8\% of the returned orbits being stable. Thus, this constraint provides a good compromise and assurance that the fitted eccentricities and arguments of periastron yield a dynamically stable scenario and are used in the joint fit below.

\subsubsection{Global Fit}
\label{sec:globfit}

For the global analysis of the TOI-1064 system, we used the {\tt juliet} package \citep{Espinoza2019} that employs the {\tt batman} code \citep{Kreidberg2015} for transit photometry fitting and {\tt radvel} \citep{Fulton2018} for the modelling of RVs. To explore the parameter space of the fit and estimate the Bayesian posteriors, we used nested sampling algorithms provided in the {\tt dynesty} package \citep{Speagle2020}, that computes the Bayesian evidences allowing for robust model comparison. The parameterisation of the fit, and the priors used, are listed in Table~\ref{tab:priors} and outlined below:

\begin{itemize}
  \item The orbital period, $P$, and the mid-transit time, $T_{\rm 0}$ for both planets were taken as uniform priors with bounds equal to the preliminary {\it TESS} values minus and plus three times the corresponding uncertainty.
  
  \item The ($r_1$,$r_2$) parametrisation as introduced by \cite{Espinoza2018} was used, as it permits the physically plausible area of the ($b$,$p$) parameter space to be efficiently investigated through the introduced random variates $r_1$ and $r_2$, where $b$ is the transit impact parameter and $p$ is the planet-to-star radius ratio. We set uniform priors on both $r_1$ and $r_2$ of $\mathcal{U}(0,1)$ for both planets.
  
  \item The $\sqrt{e}$cos$\omega$ and $\sqrt{e}$sin$\omega$ parametrisation \citep{Eastman2013}, where $e$ is the eccentricity and $\omega$ is the argument of periastron, was used, with uniform priors on $\sqrt{e}$cos$\omega$ and $\sqrt{e}$sin$\omega$ with bounds of $\mathcal{U}(-0.5,0.5)$ taken from the preliminary analysis detailed above on both components for both planets.

  \item The RV semi-amplitude, $K$, for both planets were taken as uniform priors, $\mathcal{U}(0,20)$\,$\mathrm{ms}^{-1}$.
  
  \item The stellar density, $\rho_\star$, parametrisation was used as, when combined with the orbital periods, it provides scaled semi-major axes for both planets that are anchored to a single common value rather than setting priors on separate scaled semi-major axes. We set a normal prior on $\rho_\star$ using the value from Table~\ref{tab:stellarParam}.
  
  \item The quadratic limb-darkening coefficients parametrised in the ($q_1$,$q_2$) plane \citep{Kipping2013} were used, with the coefficients for each bandpass ({\it TESS}, {\it CHEOPS}, LCOGT PanSTARRS z$_{\rm s}$ and Bessel B, NGTS, and ASTEP Cousins R) calculated using the {\tt ld} module of {\tt pycheops} (Maxted et al. {\it accepted}). Normal priors were set for all limb-darkening coefficients centred on the determined values with a 1-$\sigma$ uncertainty of 0.1 taken.
\end{itemize}

In addition to the modelling of planetary signals in transit photometry and RV data, and {\sc scalpels} basis vectors to simulate the stellar activity in the RVs, we included further parameters to model instrumental and astrophysical noise in the form of GPs. For all transit photometry datasets, we used GPs with a Mat\'{e}rn-3/2 kernel against time in order to model long-term correlated noise, utilising the {\tt celerite} package \citep{ForemanMackey2017} within {\tt juliet}. An example of a fitted GP to the {\it TESS} fluxes can be seen in Fig.~\ref{fig:tess_lcs}. Moreover, for both photometry and RV data we fitted jitter terms in order to reflect any extra noise not previously accounted for, that is subsequently summed in quadrature with the uncertainties on the data. Lastly, we fitted the HARPS RVs with a zero-point RV offset, $\gamma_\star$.

\subsubsection{Results}
\label{sec:results}

\begin{table}
\begin{center}
\caption{Fitted and derived parameter values for TOI-1064\,b and\,c based on the joint fit to the photometric and RV data detailed in Section~\ref{sec:joint_analysis}.}
\label{tab:TOI1064_restab}
\begin{tabular}{ccc}
\hline\hline                
 Parameter (unit) & b & c  \\ 
\hline 
\multicolumn{3}{c}{\textit{Fitted parameters}} \\ 
\hline 
$P$ (d) & 6.443868$\pm$0.000025 & 12.226574$^{+0.000046}_{-0.000043}$ \\ 
$T_{0}$ (BJD-2457000) & 2036.85340$^{+0.00060}_{-0.00053}$ & 2043.51289$^{+0.00069}_{-0.00067}$ \\ 
$r_\mathrm{1}^{\rm a}$ & 0.818$^{+0.020}_{-0.015}$ & 0.835$^{+0.016}_{-0.017}$ \\ 
$r_\mathrm{2}^{\rm a}$ & 0.03267$^{+0.00044}_{-0.00041}$ & 0.03347$^{+0.00042}_{-0.00041}$ \\ 
$\sqrt{e}$cos$\omega$ & -0.01$^{+0.08}_{-0.09}$ & 0.25$^{+0.13}_{-0.14}$ \\ 
$\sqrt{e}$sin$\omega$ & -0.20$^{+0.11}_{-0.08}$ & 0.06$\pm$0.14 \\ 
$K$ ($\mathrm{m\,s}^{-1}$) & 5.62$^{+0.67}_{-0.75}$ & 0.85$^{+0.59}_{-0.67}$ \\ 
$\gamma_\star$ ($\mathrm{m\,s}^{-1}$) & \multicolumn{2}{c}{21214.5$\pm$0.7} \\ 
$\rho_\star$ ($\mathrm{kg\,m^{-3}}$) & \multicolumn{2}{c}{2711$^{+75}_{-60}$} \\ 
$\rho_\star$ ($\mathrm{\rho_\odot}$) & \multicolumn{2}{c}{1.92$^{+0.05}_{-0.04}$} \\ 
\hline 
\multicolumn{3}{c}{\textit{Derived parameters}} \\ 
\hline 
$\delta_\mathrm{tr}$ (ppm) & 1067$^{+29}_{-27}$ & 1120$^{+28}_{-27}$ \\ 
$R_\mathrm{p} / R_\star$ & 0.03267$^{+0.00044}_{-0.00041}$ & 0.03347$^{+0.00042}_{-0.00041}$ \\ 
$R_\star/a$ & 0.05488$^{+0.00097}_{-0.00092}$ & 0.03581$^{+0.00063}_{-0.00060}$  \\ 
$R_\mathrm{p}/a$ & 0.001793$^{+0.000040}_{-0.000038}$ & 0.001199$^{+0.000026}_{-0.000025}$ \\ 
$R_\mathrm{p}$ ($\mathrm{R_{\oplus}}$) & 2.587$^{+0.043}_{-0.042}$ & 2.651$^{+0.043}_{-0.042}$ \\ 
$a$ (au) & 0.06152$^{+0.00086}_{-0.00089}$ & 0.09429$^{+0.00132}_{-0.00136}$ \\ 
$t_\mathrm{14}$ (h) & 1.98$^{+0.07}_{-0.08}$ & 2.37$\pm$0.10 \\ 
$b$ & 0.728$^{+0.029}_{-0.023}$ & 0.753$^{+0.024}_{-0.026}$ \\ 
$i$ (deg) & 87.709$^{+0.083}_{-0.097}$ & 88.455$^{+0.058}_{-0.056}$  \\ 
$e$ & 0.047$^{+0.038}_{-0.030}$ & 0.088$^{+0.081}_{-0.064}$ \\ 
$\omega$ (deg) & 120$^{+49}_{-89}$ & 25$^{+22}_{-16}$ \\ 
$S_\mathrm{p}$ ($\mathrm{S_{\oplus}}$) & 62.9$^{+4.2}_{-4.0}$ & 26.8$^{+1.8}_{-1.7}$ \\ 
$T_\mathrm{eq}$ (K) & 784$\pm$13 & 634$\pm$10  \\ 
$M_\mathrm{p}$ ($\mathrm{M_{\oplus}}$) & 13.5$^{+1.7}_{-1.8}$ & 2.5$^{+1.8}_{-2.0}$ ($<$8.5)$^{\rm b}$ \\ 
$\rho_\mathrm{p}$ ($\mathrm{\rho_{\oplus}}$) & 0.78$^{+0.10}_{-0.11}$ & 0.14$^{+0.09}_{-0.11}$ ($<$0.46)$^{\rm b}$ \\ 
$g_\mathrm{p}$ ($\mathrm{m\,s}^{-2}$) & 19.7$^{+2.5}_{-2.7}$ & 3.5$^{+2.4}_{-2.8}$ ($<$11.8)$^{\rm b}$ \\ 

\hline\hline    
\end{tabular}
\end{center}

$^{\rm a}$ The $r_\mathrm{1}$ and $r_\mathrm{1}$ fitting parameters used in the {\tt juliet} package \citep{Espinoza2019} are the result of a physically plausible parametrisation of the ($b$,$p$) planet as detailed in \citet{Espinoza2018}. \\
$^{\rm b}$ 3$\sigma$ upper limit.

\end{table}

\begin{figure}
  \includegraphics[width = 8.5cm]{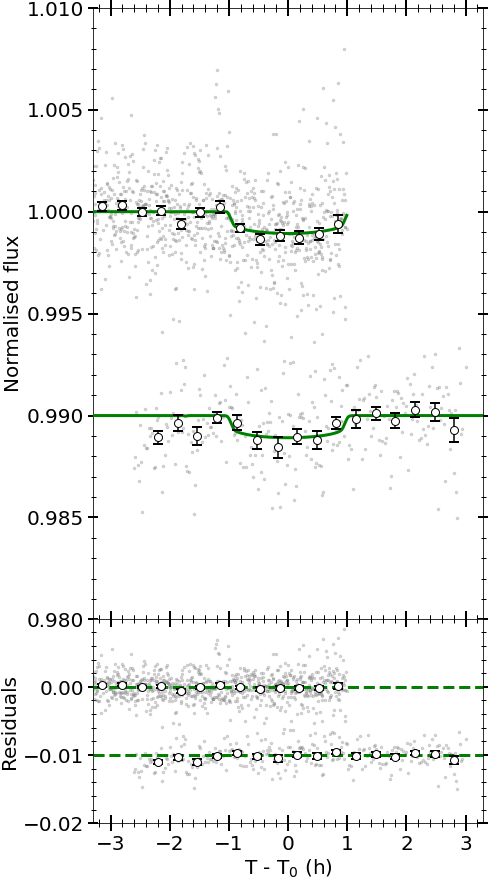}
  \caption{{\it Top panel}: Detrended LCOGT (top) and ASTEP (bottom) photometry of TOI-1064\,b  (offset by -0.01 for clarity). Unbinned data are shown in grey, with fluxes binned every 15\,min given as white circles. The best-fit transit model for TOI-1064\,b is presented in green. {\it Bottom panel}: Residuals to the fits shown in the top panel with the same colouration.}
  \label{fig:lcogt_astep_b_transits}
\end{figure}

\begin{figure}
  \includegraphics[width = 8.5cm]{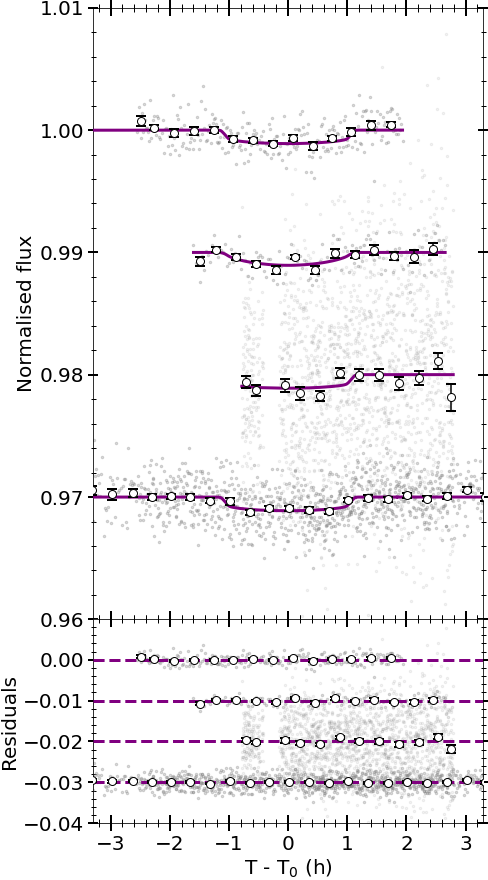}
  \caption{{\it Top panel}: Detrended LCOGT (top and upper middle), NGTS (lower middle), and ASTEP (bottom) photometry of planet\,c (offset by -0.01 for clarity). Unbinned data are shown in grey, with fluxes binned every 15\,min given as white circles. The best-fit transit model for TOI-1064\,c is presented in purple. {\it Bottom panel}: Residuals to the fits shown in the top panel with the same colouration.}
  \label{fig:lcogt_ngts_astep_c_transits}
\end{figure}

\begin{figure}
  \includegraphics[width = 8.5cm]{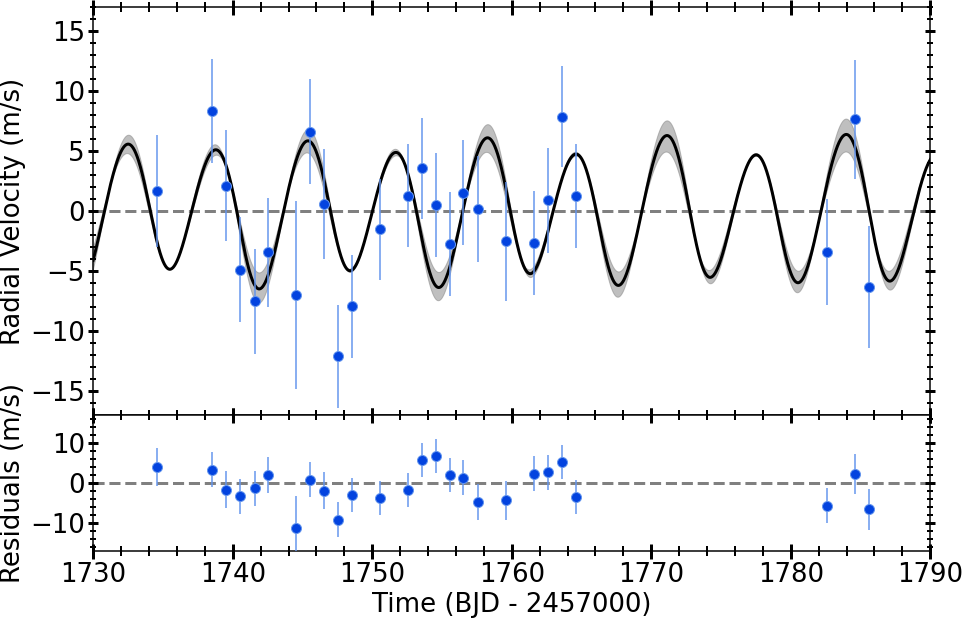}
  \includegraphics[width = 8.5cm]{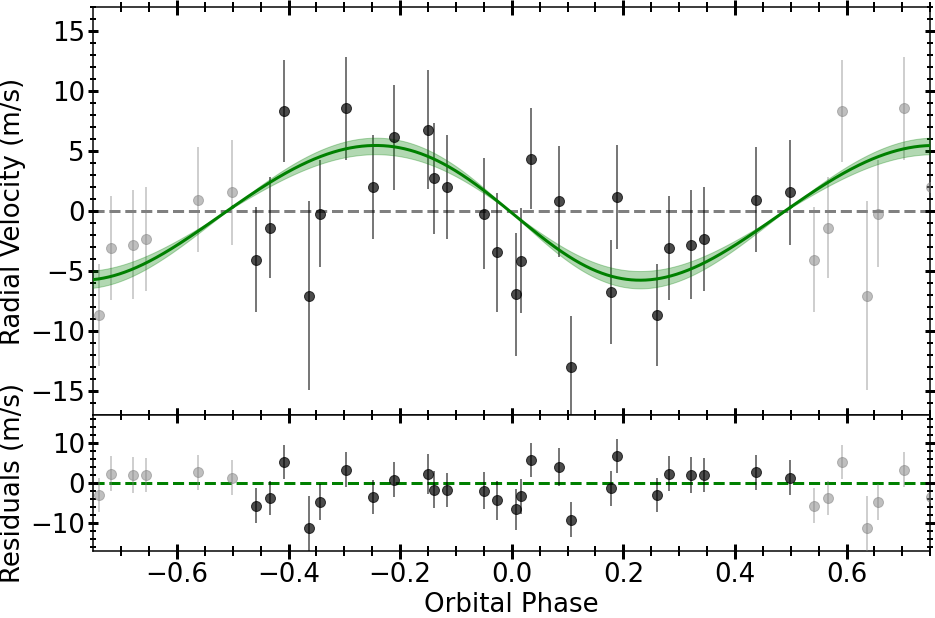}
  \includegraphics[width = 8.5cm]{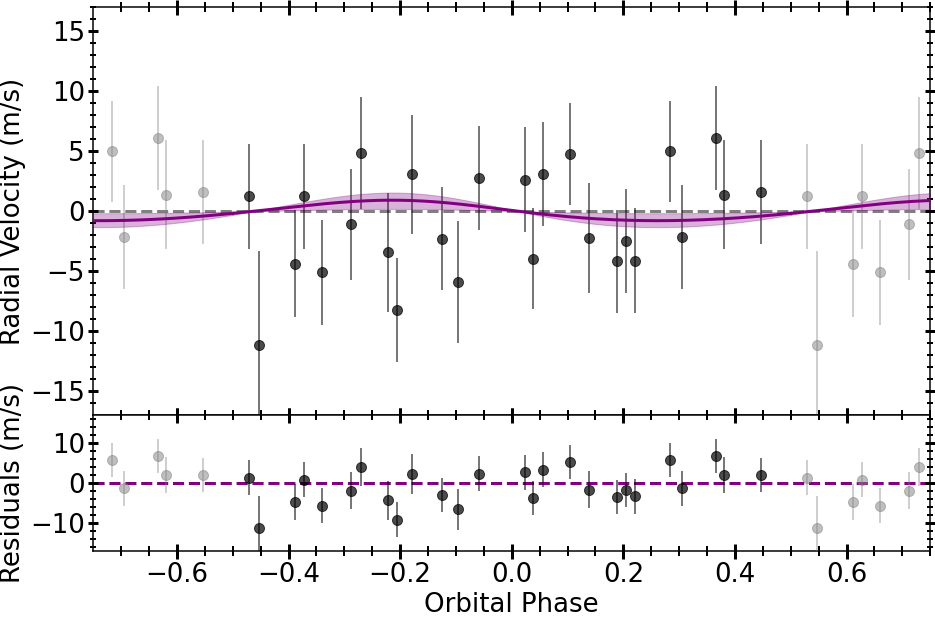}
  \caption{{\it Top panels}: The {\sc scalpels}-corrected, offset-subtracted, HARPS RV time series plotted in blue with the jitter term added in quadrature to the RV uncertainty in the error bars. The best-fit model is plotted in black with the model uncertainty shown in grey, and the residuals to the solution are given in the lower sub-panel. {\it Middle panels}: The RVs phase-folded to the orbital period of TOI-1064\,b with the best-fit model and corresponding uncertainty plotted in green and light green respectively, and the residuals to the fit given in the lower sub-panel. {\it Lower panels}: Same as the middle panels, but for planet\,c. All data shown include the jitter noise term added in quadrature with the RV error.}
  \label{fig:rv_timeseries_fits}
\end{figure}

We report the fitted results of our global analysis for both planets in the TOI-1064 system in Table~\ref{tab:TOI1064_restab}. The detrended and phase-folded {\it TESS} and {\it CHEOPS} photometry along with the fitted transit models for planets\,b and\,c are shown Fig.~\ref{fig:tess_cheops_transits}, whilst the detrended LCOGT, NGTS, and ASTEP data, and fitted transit models of both planets are reported in Figs.~\ref{fig:lcogt_astep_b_transits} and~\ref{fig:lcogt_ngts_astep_c_transits}. The {\sc scalpels}-corrected HARPS RV time series with the fit of two Keplerian orbits is presented in Fig.~\ref{fig:rv_timeseries_fits}, and the orbital period phase-folded RVs as well as the planetary models for TOI-1064\,b and\,c are shown in Fig.~\ref{fig:rv_timeseries_fits}. The posterior distributions for the main fitted parameters for both planets are given in Fig.~\ref{fig:corners_bc} with the posterior values for the fitted limb-darkening coefficients and noise terms reported in Table~\ref{tab:TOI1064_instab}.

From our analysis, we detect TOI-1064\,b and\,c in the combined {\it TESS}, {\em CHEOPS}, LCOGT, NGTS, and ASTEP transit photometry at 38.2$\sigma$ and 40.3$\sigma$, respectively. The fitted depths and derived stellar radius yield planetary radii of $R_\mathrm{p,b}$ = 2.587$^{+0.043}_{-0.042}$\,$\mathrm{R_{\oplus}}$ and $R_\mathrm{p,c}$ =  2.651$^{+0.043}_{-0.042}$\,$\mathrm{R_{\oplus}}$. We report a 8.0$\sigma$ detection of TOI-1064\,b in the HARPS RVs, and a 1.4$\sigma$ signal of TOI-1064\,c, that result in $M_\mathrm{p,b}$ = 13.5$^{+1.7}_{-1.8}$\,$\mathrm{M_{\oplus}}$ and a $M_\mathrm{p,c}$ 3$\sigma$ upper limit of 8.5\,$\mathrm{M_{\oplus}}$ (nominal value = 2.5$^{+1.8}_{-2.0}$\,$\mathrm{M_{\oplus}}$, to be discussed in Section~\ref{sec:rv_mod_comp}). 

Combining the radius and mass of TOI-1064\,b gives a bulk density of $\rho_\mathrm{p,b}$ = 0.78$^{+0.10}_{-0.11}$\,$\mathrm{\rho_{\oplus}}$ (4.28$^{+0.57}_{-0.61}$\,$\mathrm{g\,cm^{-3}}$). We determine the orbital period of planet b to be $P_{\rm b}$ = 6.443868$\pm$0.000025\,d, and the corresponding semi-major axis of $a_{\rm b}$ = 0.06152$^{+0.00086}_{-0.00089}$\,au. At this distance TOI-1064\,b receives 62.9$^{+4.2}_{-4.0}$ times Earth's insolation and has a zero Bond albedo equilibrium temperature of 784$\pm$13\,K. For TOI-1064\,c, the subtle RV signal together with the precise radius yields a $\rho_\mathrm{p,c}$ 3$\sigma$ upper limit of 2.56\,$\mathrm{g\,cm^{-3}}$ (0.46\,$\mathrm{\rho_{\oplus}}$; nominal value = 0.14$^{+0.09}_{-0.11}$\,$\mathrm{\rho_{\oplus}}$). The fitted orbital period of $P_{\rm c}$ = 12.226573$^{+0.000046}_{-0.000043}$\,d gives a semi-major axis of $a_{\rm c}$ = 0.09429$^{+0.00132}_{-0.00134}$\,au, that results in the stellar irradiance of TOI-1064\,c being 26.8$^{+1.8}_{-1.7}$ that of Earth with a zero Bond albedo equilibrium temperature of 634$\pm$10\,K.

\begin{figure*}
  \includegraphics[width = \linewidth]{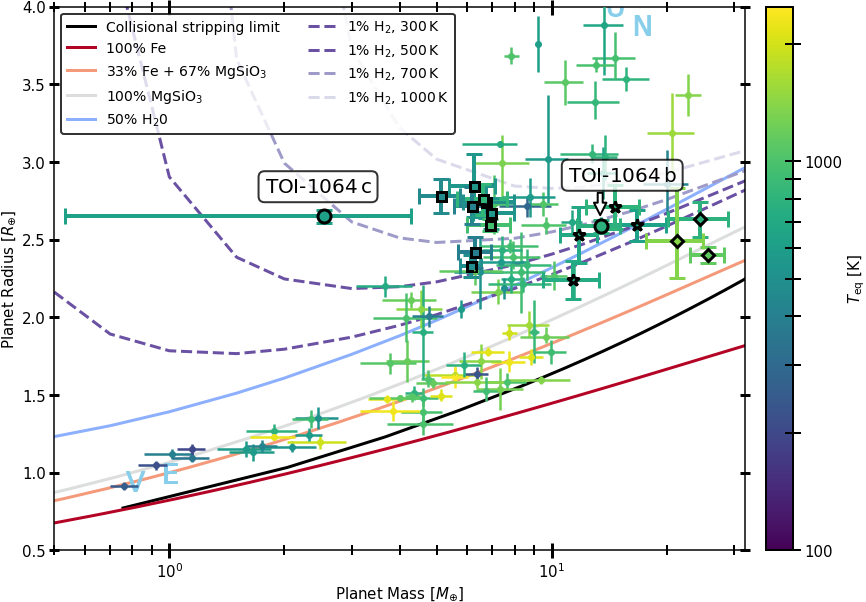}
  \caption{Mass-radius diagram for exoplanets with mass and radius measurements more precise than 20\% coloured by the zero Bond albedo equilibrium temperature with data taken from TEPCat \citep{Southworth2011}. TOI-1064\,b and\,c are highlighted via black bordered circles, with other planets discussed in Section~\ref{sec:disc} shown as stars, diamonds, and squares. The black solid line represents the upper bound of the forbidden region according to collisional mantle stripping via impact \citep{Marcus2010}. The remaining solid lines depict theoretical mass-radius curves for solid planets with various bulk compositions \citep{Zeng2013} and the dashed lines represent mass-radius curves for an Earth-like core enveloped with a 1\% mass fraction H$_2$ atmosphere with various equilibrium temperatures \citep{Zeng2019}. Venus, Earth, Uranus, and Neptune are indicated in blue.}
  \label{fig:mr_diag}
\end{figure*}

Placing TOI-1064\,b and\,c on a mass-radius diagram, as shown in Fig.~\ref{fig:mr_diag} alongside well characterised planets with masses and radii known to better than 20\%, we see that planet b is one of the smallest planets known with a mass above 10\,$\mathrm{M_{\oplus}}$. Conversely, the tentative signal of planet c places it at the lower end of the mass distribution for planets of this size. However, further RV follow-up observations are needed to confirm the mass precisely and to determine if TOI-1064\,c remains one of the lowest density sub-Neptunes known. 

Given the refined nature of the radii of the two planets, with uncertainties of 1.63\% and 1.59\%, we are able to characterise the planets further and conduct internal structure and atmospheric escape modelling of the bodies, as detailed below. Additionally, we find the timing uncertainties for individual transits of TOI-1064\,b to be between 2 and 10\,min, and between 4 and 13\,min for TOI-1064\,c.

In addition to the information about the planets in the TOI-1064 system, from our analysis we determined the density and RV of the host star. We find a stellar density of $\rho_\star$ = 1.92$^{+0.05}_{-0.04}$\,$\mathrm{\rho_\odot}$, that is in agreement with the value derived from IRFM and isochrone placement methods reported in Table~\ref{tab:stellarParam}. From the HARPS data, we obtain a $\gamma_\star$ value of 21.2\,$\mathrm{km\,s}^{-1}$ that agrees with the {\it Gaia} value of 20.7\,$\mathrm{km\,s}^{-1}$ \citep{GaiaCollaboration2021}. To assess the robustness of the fitted eccentricities we conducted two further analyses of the data. Firstly, we carry out a fit of the complete transit photometry and RV dataset setting the eccentricities of both planets to 0 whilst using the same priors as detailed above for the remaining parameters. We find a difference in log evidences between this fit and our global analysis of $\Delta {\rm ln} Z$ = 17.8 in favour of the non-zero eccentricity solution, indicating a decisive preference for this result \citep{Kass1995,Gordon2007}. Second, we fit the transit photometry using the same priors as previously set on the appropriate parameters to evaluate the contribution of the RVs. We find that the derived eccentricities from the transit photometry alone ($e_{\rm b}$ = 0.38$^{+0.05}_{-0.20}$ and $e_{\rm c}$ = 0.33$\pm$0.07) agree within 2$\sigma$ with the lower eccentricities listed in Table~\ref{tab:TOI1064_restab} that were derived from the RV and photometric data. Due to the compact nature of the system, we conclude that the lower, non-zero eccentricity solution from our global analysis is likely correct.

\begin{figure*}
    \subfloat{{\includegraphics[width = 8.5cm]{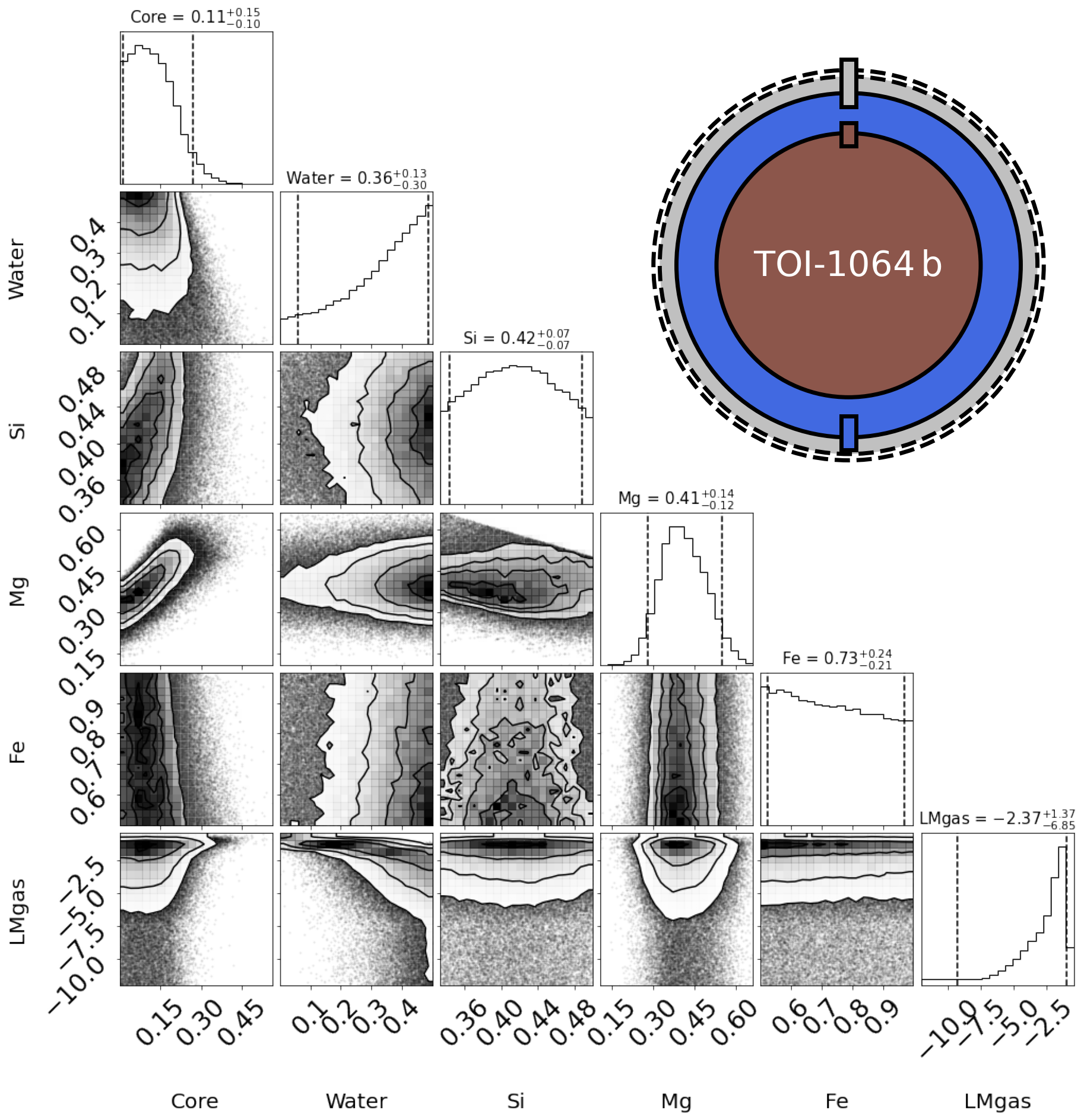} }}
    \qquad
    \subfloat{{\includegraphics[width = 8.5cm]{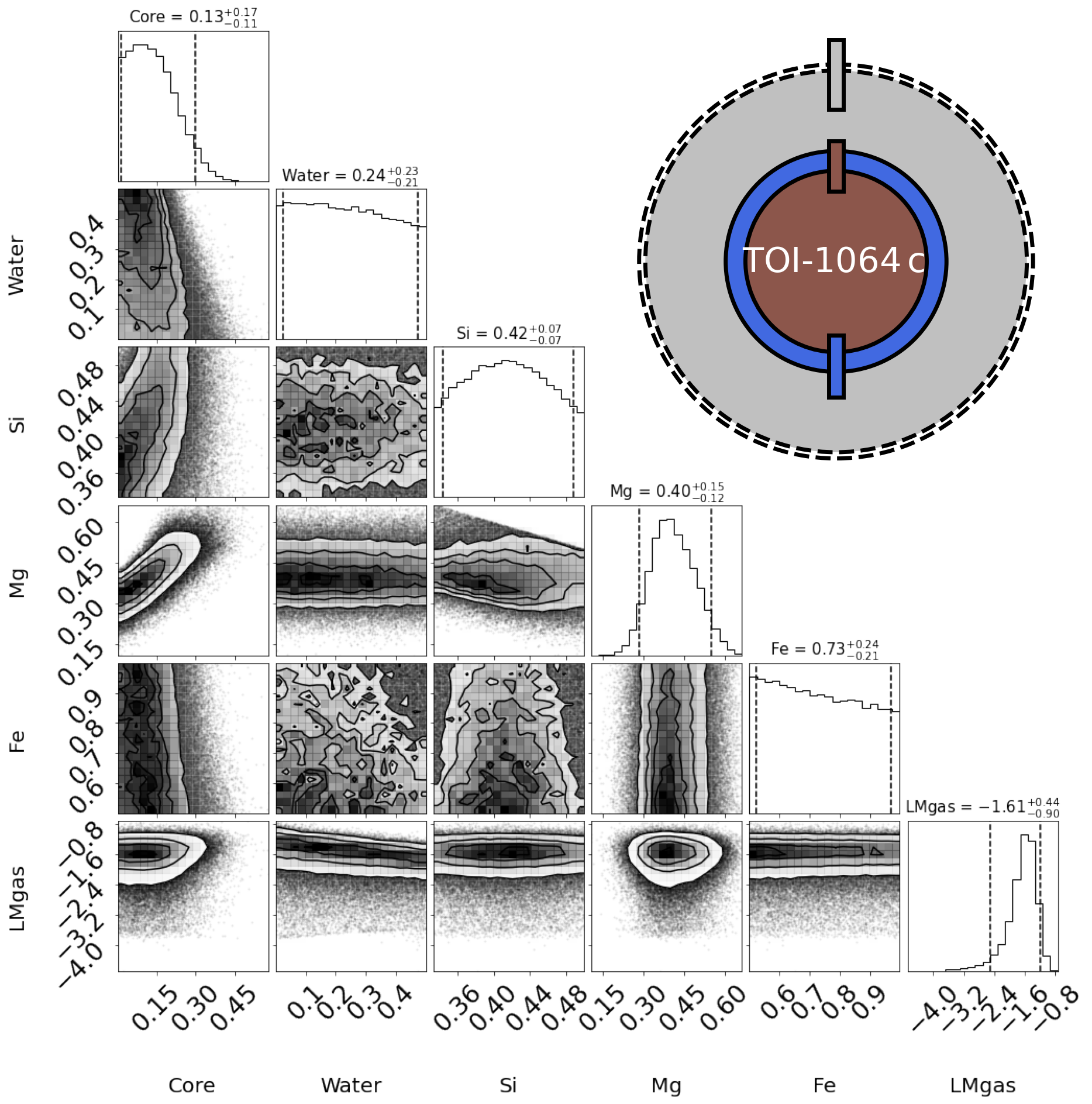} }}
  \caption{Corner plot showing the main internal structure parameters of TOI-1064\,b ({\em left}) and TOI-1064\,c ({\em right}). Shown are the mass fraction of the inner core, the mass fraction of water, the Si and Mg mole fractions in the mantle, the Fe mole fraction in the inner core, and the mass of gas (log scale). The values on top of each column are  the mean and 5\% and 95\% quantiles. Next to each corner plot, we also show illustrations of both planets representing the radius fractions of the high-Z components (inner core and mantle; brown), water layer (blue), and gas envelope (grey), corresponding to the medians of the posterior distributions, with the dashed line outer rings representing the uncertainty on the total radius.}
  \label{fig:struct_planets_bc}
\end{figure*}

\subsubsection{RV Model Comparison}
\label{sec:rv_mod_comp}

To give confidence in the subtle RV signal of TOI-1064\,c found by the global analysis, we conduct two additional fits of the complete transit photometry and RV dataset. In these supplementary analyses we set the priors of the majority of the parameters as detailed above, and either remove planet c from the RV model or fix the semi-amplitude of TOI-1064\,c to 0. By comparing the log Bayesian evidences between these analyses and our previous fit, we find $\Delta {\rm ln} Z{\rm s}$ of 55.6 and 26.1, respectively, in favour of our global analysis presented in Section~\ref{sec:globfit}. This indicates decisive preferences of using a non-massless prior to model the RV signal of TOI-1064\,c \citep{Kass1995,Gordon2007}, and provides evidence for the validity of the tenuous signal seen in the HARPS RVs.

It should be noted that the semi-amplitudes and masses for both planets derived from the global analysis are subtly different to the values retrieved by the CCF-based {\sc scalpels} algorithm. As presented in the Sections~\ref{sec:scalpels} and~\ref{sec:results}, TOI-1064\,b and\,c are more and less massive by 0.4$\sigma$ and 0.8$\sigma$, respectively, in the nested sampling joint fit. Importantly, the {\sc scalpels}-fitted semi-amplitude uncertainties are lower which allows for a more confident detection of planet c in the HARPS RVs. As these differences can have a significant effect on the bulk density, and internal structure and atmospheric escape modelling, we conducted two fits of the RV data using {\tt radvel} within {\tt juliet} \citep{Fulton2018,Espinoza2019} and the nested sampling algorithms in the {\tt dynesty} package \citep{Speagle2020} to reconcile this difference. For these analyses, we set the semi-amplitude priors to the values and uncertainties produced by either the {\sc scalpels} or our global fit, with wide uniform priors on $P$, $T_{\rm 0}$, $\sqrt{e}$cos$\omega$, and $\sqrt{e}$sin$\omega$ identical to the values set out in Section~\ref{sec:globfit}.

To assess if one model (priors taken from the fit produced by the {\sc scalpels}) is preferred over the other (priors taken from our global analysis), we consult the log Bayesian evidences reported by the nested sampling and compute an odds ratio between the two models. We find a difference in log evidences of $\Delta {\rm ln} Z$ = 0.49 and thus an odds ratio of 1.63 in favour of the model produced by our global fit. Following the standard reference levels \citep{Kass1995,Gordon2007}, this translates to a weak and non-conclusive preference. However, given the more comprehensive datasets analysed in our global fit, we adopt the values presented in Table~\ref{tab:TOI1064_restab} as the nominal values for this system.

\section{Characterisation of the System}
\label{sec:charac}

\subsection{Internal Structure}
\label{sec:is} 

We analysed the internal structure of the two planets in the TOI-1064 system using the method employed by \citet{Leleu2021} for TOI-178. The method is based on a global Bayesian model that fits the observed properties of the star (mass, radius, age, effective temperature, and the photospheric abundances [Si/Fe] and [Mg/Fe]) and planets (planet-star radius ratio, the RV semi-amplitude, and the orbital period).

In terms of the forward model, we assume a fully differentiated planet, consisting of a core composed of Fe and S, a mantle composed of Si, Mg, Fe, and O, a pure water layer, and a H and He layer. The temperature profile is adiabatic, and the equations of state (EoS) used for these calculations are taken from \citet{Hakim2018} and \citet{Fei2016} for the core materials, from \citet{Sotin2007} for the mantle materials, and \citet{Haldemann2020} for water.
The thickness of the gas envelope is computed using the semi-analytical model of \citet{Lopez2014}.

In our model, we assume the following priors: 
The logarithms of the gas-to-solid ratios in planets have uniform distributions; the mass fractions of the planetary cores, mantles, and overlying water layers have uniform positive priors except that the mass fractions of water are limited to a maximum value of 0.5. 
The bulk Si/Fe and Mg/Fe mole ratios in the planet are assumed to be equal to the values determined for the atmosphere of the star, given in Table~\ref{tab:stellarParam}.
Recent work by \citet{Adibekyan2021} has, however, found that whilst the abundances of planets and host stars are correlated, the relation may not be one-to-one. Finally, we note that the solid and gas parts of the planets are computed independently, which means that we do not include the compression effect of the planetary envelope on its core. This is justified given the small masses of the gas envelopes (see below). The results of our modelling for both planets are shown in Fig.~\ref{fig:struct_planets_bc}.

\subsection{Atmospheric Evolution}
\label{sec:ae} 

We modelled the atmospheric evolution of the two planets using a modified version of the algorithm presented by \citet{Kubyshkina2019a,Kubyshkina2019b}. This Bayesian tool requires input parameters that are both stellar ($M_{\star}$, $t_{\star}$, and the present-day rotation period $P_{\mathrm{rot,}\star}$) and planetary (semi-major axes $a$ and masses $M_p$). The stellar rotational period $P_{\mathrm{rot}}$ is assumed as a proxy for the stellar high-energy emission, which plays a significant role in controlling the atmospheric escape rate, and is modelled over time as a broken power-law with a variable exponent within the first 2\,Gyr \citep{Tu2015}, and afterwards $P_{\rm rot} \propto t^{0.566}$ following \citet{Mamajek2008}.
We translated $P_{\mathrm{rot}}$ into the stellar X-ray and extreme-UV (together XUV) luminosities using the scaling relations in \citet{Wright2011,SanzForcada2011,McDonald2019}. Lastly, the remaining input parameters include; the planetary equilibrium temperature $T_{\mathrm{eq}}$, radius $R_p$, mass $M_p$, atmospheric mass $M_{\mathrm{atm}}$, and orbital semi-major axis $a$, with the evolution of $T_{\mathrm{eq}}$ over time due to changes in stellar bolometric luminosity $L_{\mathrm{bol,}\star}$ calculated  by interpolating within grids of stellar evolutionary models \citep{Choi2016,Dotter2016}.

The main model hypotheses are that orbital semi-major axes are fixed to the present-day value and that planetary atmospheres are hydrogen dominated. The free parameters of the algorithm are the exponent of the stellar rotation period evolution power law and the initial atmospheric mass fractions of the planets $f_{\mathrm{at}}^{\mathrm{start}}$. Millions of planetary evolutionary tracks are generated within a Bayesian context employing a Markov Chain Monte Carlo scheme \citep{Cubillos2017}. After rejecting those thats do not fulfil the constraint imposed by the present-day atmospheric content derived in Section \ref{sec:is}, we obtained the posterior distributions of the free parameters. Further details about the tool may be found in \citet{Delrez2021}. The results of the joint evolution of both planet b and c are shown in Figs.~\ref{fig:evoAtmStar} and~\ref{fig:evoAtmPla}.

\begin{figure*}
  \includegraphics[width = \linewidth]{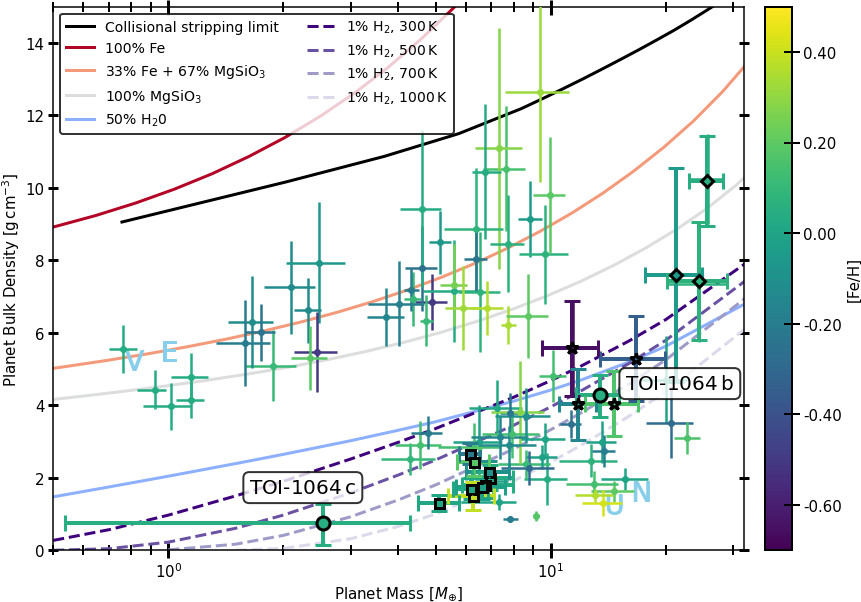}
  \caption{Mass-density diagram for exoplanets with mass and radius measurements more precise than 20\% coloured by the host star metallicity with data taken from TEPCat \citep{Southworth2011}. TOI-1064\,b and\,c are highlighted via black bordered circles, with other planets discussed in Section~\ref{sec:disc} shown as stars, diamonds, and squares. The black solid line represents the lower bound of the forbidden region according to collisional mantle stripping via impact \citep{Marcus2010}. The remaining solid lines depict theoretical mass-density curves for solid planets with various bulk compositions \citep{Zeng2013} and the dashed lines represent mass-density curves for an Earth-like core enveloped with a 1\% mass fraction H$_2$ atmosphere with various equilibrium temperatures \citep{Zeng2019}. Venus, Earth, Uranus, and Neptune are indicated in blue.}
  \label{fig:mrho_diag}
\end{figure*}

\begin{figure}
  \includegraphics[width = \linewidth]{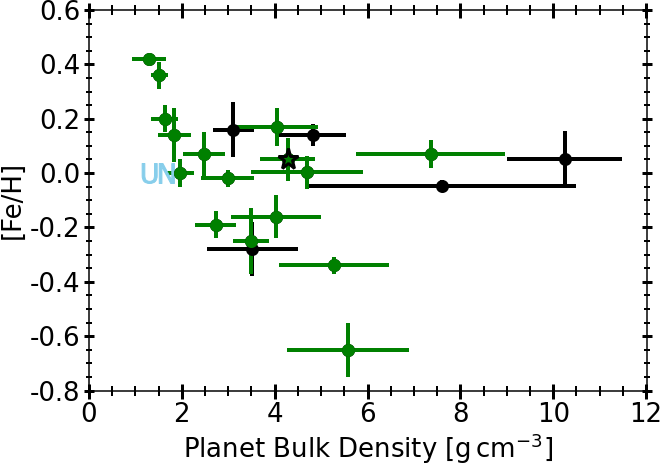}
  \caption{The bulk density versus stellar metallicity for planets with masses above 10\,$\mathrm{M_\oplus}$ and radius and mass uncertainties less than 20\%, shown in Figs.~\ref{fig:mr_diag} and~\ref{fig:mrho_diag}. Black circles indicate highly irradiated planets ($S>200\,\mathrm{S_{\oplus}}$), with green circles showing mildly irradiated bodies. TOI-1064\,b is highlighted by a black bordered star, with Uranus and Neptune indicated in blue.}
  \label{fig:rho_feh}
\end{figure}

\section{Discussion}
\label{sec:disc} 

From our analysis, we clearly detect TOI-1064\,b in the {\em TESS}, {\em CHEOPS}, LCOGT, and ASTEP photometry and the HARPS RVs. TOI-1064\,c is confidently detected in the {\em TESS}, {\em CHEOPS}, LCOGT, NGTS, and ASTEP photometry, but does not register a significant signal in the HARPS RV data. The photometric periods and HARPS RVs secure a mass of $13.5\pm2.0\,M_\oplus$ for planet b, and a 3$\sigma$ upper limit of $8.5\,M_\oplus$ for planet c. The proximity of the orbital period to the first harmonic of the stellar rotation period hinders a more confident detection in the RVs. However, with more data over an extended baseline {\sc scalpels} should be able to separate the two signals due to a more apparent shift in the CCFs, whilst periodograms may also be able to find two peaks due to a higher frequency resolution.

\subsection{Mass-Radius Comparisons and Bulk Density-Metallicity Correlation}

Figs.~\ref{fig:mr_diag} and~\ref{fig:mrho_diag} highlight that, although the radii of TOI-1064\,b and\,c are similar (in agreement with the ``peas in a pod'' scenario; \citealt{Weiss2018}), the current mass values are likely significantly different and thus, the bulk planetary densities are considerably distinct. TOI-1064\,b is one of the smallest, and therefore densest, well characterised sub-Neptunes with a mass greater 10\,$\mathrm{M_\oplus}$ known. This places planet b amongst a small family of dense (4.0$-$4.3\,$\mathrm{g\,cm^{-3}}$), warm (690$-$810\,K), and mildly irradiated (32-71\,$\mathrm{S_{\oplus}}$) sub-Neptunes orbiting around K-dwarf stars that also includes HIP 116454\,b \citep{Vanderburg2015} and Kepler-48c \citep{Steffen2013,Marcy2014}. The other planets in this parameter space orbit stars that are more metal-poor ([M/H] = $-0.16\pm0.08$) and metal-rich ([Fe/H] = $0.17\pm0.07$) than TOI-1064 ([Fe/H] = $0.05\pm0.08$), respectively. There are two further planets in this region of the mass-radius diagram that both are similarly warm and irradiated, but are more dense (5.0$-$5.6\,$\mathrm{g\,cm^{-3}}$); K2-110\,b \citep{Osborn2017} and K2-180\,b \citep{Korth2019}. Interestingly they both orbit metal-poor K-dwarfs with [Fe/H] = $-0.34\pm0.03$ and $-0.65\pm0.10$ for K2-110\,b and K2-180\,b, respectively. These four planets are highlighted in Figs.~\ref{fig:mr_diag} and~\ref{fig:mrho_diag} as black bordered stars. This may hint that metallicity could affect the bulk density of sub-Neptunes in this parameter space via differing formation conditions. 

To test this apparent correlation for massive sub-Neptunes, we select all planets with masses greater than 10\,$\mathrm{M_\oplus}$ in our well-characterised sample (uncertainties on the radii and masses less than 20\%), and compared their bulk densities against host star metallicities as seen in Fig.~\ref{fig:rho_feh}. Whilst large uncertainties on the bulk density of some exoplanets, especially at larger values, results in a scatter to the data there appears to be a negative correlation. We quantify this trend for our sample using a Bayesian correlation tool, that aims characterise the strength of the correlation between two parameters in a Bayesian framework \citep{Figueira2016}, and reports a correlation distribution that represent values akin to a Spearman's rank value. For this sample, we find the peak of the correlation posterior distribution to be -0.26, with 95\% lower and upper bounds of -0.60 and 0.11. Interestingly, if we exclude highly irradiated planets ($S>200\,\mathrm{S_{\oplus}}$) we retrieve a peak density-metallicity correlation of -0.45, with 95\% lower and upper bounds of -0.78 and -0.09. If this trend of less dense sub-Neptunes orbiting metal-rich stars is indicative of formation conditions it could suggest that metal-rich stars form planets that accrete bigger envelopes or have more massive cores and are able to retain their atmospheres \citep{Owen2018}. This correlation could also imply that planets around metal-rich stars have metal-rich atmospheres that have reduced photo-evaporation driven atmospheric mass-loss rates \citep{Owen2012}. This correlation is similar to that seen in giant planets in which more massive planets were found around metal-rich stars \citep{Guillot2006}. This formation picture could be supported by the strengthening of the correlation with the removal of highly irradiated planets that may have an observed high density due to evolution processes such as atmospheric stripping.

Thus, as there are overlapping mass-radius curves in this region, accurate characterisation of additional dense sub-Neptunes over a range of metallicities should help clarify this trend and inform whether these planets are water dominated or terrestrial with a H$_2$ atmosphere, and indicate the processes that sculpt these planets. 

We note that the similarly sized and irradiated planet HD 119130\,b is denser and orbits a more massive Solar-like star than the dense sub-Neptunes discussed above \citep{Luque2019}. Lastly, two planets (Kepler-411\,b and K2-66\,b; \citealt{Wang2014,Sun2019,Crossfield2016,Sinukoff2017}) have similar radii to TOI-1064\,b, but larger masses and substantially higher irradiation (>200\,$\mathrm{S_{\oplus}}$) meaning that these ultra-dense bodies are likely separate from the warm, dense sub-Neptunes mentioned previously that include TOI-1064\,b. However, these objects may represent the lower-mass end of a stripped planet core family that includes the more massive, but equally dense TOI-849\,b \citep{Armstrong2020}. These three planets are highlighted in Figs.~\ref{fig:mr_diag} and~\ref{fig:mrho_diag} as black bordered diamonds.

From the substantial constraint on the radius of TOI-1064\,c there exists a band of possible bulk densities from the large uncertainty on the mass. Taken at the nominal value, planet c would be one of the least dense sub-Neptunes known and could represent a planet akin to Kepler-307\,c \citep{Xie2014,JontofHutter2016}, albeit with a more extended atmosphere. If this is the case, the TOI-1064 planets would resemble TOI-178\,c and\,d both in density difference and orbital period ratio \citep{Leleu2021}. However, if the mass is confirmed to be nearer the upper limit, the low density and mild irradiation would place TOI-1064\,c closer to the outer planets of Kepler-80 (b and\,c; \citealt{Xie2014}; \citealt{MacDonald2016}) or GJ 1214\,b \citep{Charbonneau2009}, and in between planets of similar density and higher (HIP 41378\,b and TOI-125\,c; \citealt{Vanderburg2016}; \citealt{Santerne2019}; \citealt{Quinn2019}; \citealt{Nielsen2020}) and lower irradiation (Kepler-26\,b and\,c, TOI-270\,c, and LTT 3780\,c; \citealt{Steffen2012}; \citealt{JontofHutter2016}; \citealt{Gunther2019}; \citealt{VanEylen2021}; \citealt{Cloutier2020}; \citealt{Nowak2020}). These eight planets are highlighted in Figs.~\ref{fig:mr_diag} and~\ref{fig:mrho_diag} as black bordered squares.

\subsection{Internal Structure and Atmospheric Escape}

The internal structure models show that TOI-1064\,b has a small gas envelope (log\,$M_{\rm gas}$ between -9.2 and -1.0 - all given values are the 5\% or 95 \% quantiles). It has a relatively small inner core and a large mass fraction of water (comprised between 0.01 and 0.26, and 0.06 and 0.49 of the mass of the solid body, respectively), with the remainder of the mass in the Si and Mg dominant mantle. 
It should be noted that in our atmospheric escape modelling we assume the presence of an accreted primordial atmosphere. However, the internal structure modelling results highlight that the tenuous atmosphere of TOI-1064\,b means that the mass of the high-Z elements (i.e. the inner core, mantle, and volatile layers) is essentially equal to the total mass determined from the RV fitting. 
This value, and other modelled properties, could thus be similar to the properties of the high-Z core prior to the accretion of an atmosphere and may inform formation models \citep{Pollack1996,Ikoma2012}. Further high-precision photometry and internal structure modelling of additional ultra-dense sub-Neptunes (such as Kepler-411\,b and K2-66\,b; \citealt{Wang2014,Sun2019,Crossfield2016,Sinukoff2017}) may help shed light on the formation of these bodies.

In contrast, planet c likely has a substantial gas envelope (log\,$M_{\rm gas}$ between -2.5 and -1.7, corresponding to mass fractions of up to 3\%). The presence of a non-negligible gas envelope results in the mass fraction of water being unconstrained as seen in Fig.~\ref{fig:struct_planets_bc}.

The results of the atmospheric escape modelling based on fitting simultaneously both planets, respect the priors well (see Figs.~\ref{fig:evoAtmStar} and~\ref{fig:evoAtmPla}). These results suggest that the rotation of TOI-1064 evolved as observed in the majority of late-type stars of similar mass. We also find that planet b has lost (almost) the entirety of its H/He envelope, assuming it accreted one, at some time in the past, as indicated by the flat posterior on the initial atmospheric mass fraction (Fig.~\ref{fig:evoAtmPla}, left column). As a consequence of the main assumptions of the atmospheric evolution algorithm (namely that planets are assumed to have formed and spent their whole evolution at the same distance from the star), planet b is considered to have migrated to its current position while still embedded in the protoplanetary disc, as is thought to have happened to the Kepler-223 \citep{Mills2016}, TRAPPIST-1 \citep{Luger2017}, and TOI-178 planetary systems \citep{Leleu2021}. This migration scenario is favoured with respect to in-situ formation as, to form TOI-1064\,b at 0.06\,au, the protoplanetary disc mass must have been substantially enhanced compared to solar values \citep{Schlichting2014} and the presence of a substantial volatile layer, as determined by our internal structure modelling, that must have been accreted beyond the ice line. To test this migration scenario, atmospheric studies of TOI-1064\,b could be undertaken to determine if the tenuous envelope is a secondary, high molecular weight atmosphere \citep{Fortney2013}, similar to that inferred at $\pi$\,Men\,c \citep{GarciaMunoz2021}, that could be comprised of steam \citep{Marounina2020,Mousis2020}, given the large volatile fraction. Another possibility is that the thin layer is primordial in nature and is the remnant of an atmosphere that has undergone escape. This picture would support recent modelling which has suggested that highly irradiated planets with a high water mass fraction can lose their H/He envelopes and result in observed sub-Neptunes \citep{Aguichine2021}.  

Indeed, due to the overlap of multiple mass-radius curves in the parameter space around TOI-1064\,b, future work assessing the atmospheric properties of high density sub-Neptunes will help inform internal structure and atmospheric modelling in this region.

As seen in the right column of Fig.~\ref{fig:evoAtmPla}, the posterior of the initial atmospheric mass fraction of TOI-1064\,c is also flat. This distribution could suggest that planet c also migrated through the protoplanetary disc having accreted a larger primordial envelope at an exterior orbital distance prior to migrating inwards and losing part of its envelope. However, it is also possible that the large uncertainty on the planet mass means that the atmospheric mass-loss rate of the atmosphere of planet c is unconstrained and thus the initial atmospheric mass fraction cannot be modelled.

We derived the current atmospheric mass-loss rate by inserting the system parameters and the current stellar high-energy emission at the distance of the planet derived from the atmospheric evolution framework (4.3$\times$10$^{5}$\,$\mathrm{erg\,cm^{-1}\,s^{-1}}$) into the interpolation routine of \cite{Kubyshkina2018} obtaining a value of 6.11$\times$10$^{11}$\,$\mathrm{g\,s^{-1}}$. With such a high mass-loss rate, the planet is expected to lose its H/He atmosphere as estimated by the planetary structure models within just 8\,Myr, which implies that the planetary mass is likely to be higher than the best fit. With a mass of 4.3\,$\mathrm{M_{\oplus}}$, corresponding to the best fitting planetary mass plus 1$\sigma$, we obtain that the atmosphere would require 100\,Myr to escape. At the upper mass limit of 8.5\,$\mathrm{M_{\oplus}}$, instead, the atmosphere would escape within about 400\,Myr.
We remark that these calculations consider the atmospheric mass fraction computed for the best-fitting planetary mass. If the planet were more massive, then the planetary surface gravity would be higher, leading to longer atmospheric-escape time-scales. This strengthens the possibility that planet c has likely gone through, and possibly is still going through, significant migration. 
Utilising a formulation of the Jeans escape criteria \citep{Jeans1925} that includes Roche lobe effects \citep{Erkaev2007} as detailed in equations 7-9 of \citet{Fossati2017} and setting $\tilde{\lambda}^{*}$ to the recommended value of 25, we compute the minimum mass of planet c required to retain a hydrogen atmosphere to be $\sim$6.1\,$\mathrm{M_{\oplus}}$. Whilst these analyses imply that the true mass of TOI-1064\,c is near the upper mass limit, in order to accurately constrain the mass of this planet and to advance our understanding of the formation and evolution of this planetary system, further RV observations are needed.

Lastly, our atmospheric escape modelling permits us to calculate the integrated X-ray flux of the planets over the lifetime of the system. Even under the assumption that both planets in the TOI-1064 system formed at their current orbital distances, the integrated X-ray fluxes we determine (F$_{\rm x} = 1.9\times 10^{20}$ erg\,cm$^{-2}$ and F$_{\rm x} = 8.1\times 10^{19}$ erg\,cm$^{-2}$, respectively) are lower than amount of flux required to strip a sub-Neptune sized planet around a K-dwarf via photoevaporation alone, \citep{McDonald2019}. Therefore, as we think that migration may have occurred, hence lowering the received X-ray flux, and that our internal structure modelling finds that planet b has a tenuous atmosphere, we conclude that core-powered mass-loss also plays an important role in sculpting exoplanet atmospheres.

\subsection{Planetary Architecture and Formation Mechanisms}

The planets of the TOI-1064 system have a transit-chord-length ratio, $\xi=1.03\pm$0.06, indicative of coplanar orbits \citep{Steffen2010,Fabrycky2014} that is common for multi-planet systems \citep{Fang2012}. The eccentricities for both planets are also typical of compact, multi-planet systems \citep{Limbach2015,VanEylen2015,VanEylen2019,Mills2019}. The coplanar and roughly circular orbits support formation at larger semi-major axes and subsequent migration through a disc, either proto-planetary or planetesimal \citep{Terquem2007}. Currently, the planets are in orbits that are short or ``narrow'' of the 2:1 MMR (1.90:1) that is somewhat rare \citep{Fabrycky2014} as planets have been observed to occur more frequently beyond MMRs in so-called ``pile-ups'' \citep{Batygin2013,Silburt2015,Winn2015}. Whilst a precise mass value for planet c still needs to be obtained, our 3$\sigma$ upper limit of 8.5\,$\mathrm{M_{\oplus}}$ yields one of the largest mass ratios among pairs of small planets with similar radii. Thus, TOI-1064 likely joins other ``different twin'' systems that have a significant mass, and therefore density, disparity such as Kepler-107 \citep{Bonomo2019}, KOI-1599 \citep{Panichi2019}, and TOI-125 \citep{Nielsen2020}, although TOI-1064 would be only the second system with such planets near a MMR. If this mass imbalance is verified, the migration scenario for TOI-1064 may support the energy optimisation modelling of ``peas in a pod'' systems presented in \citet{Adams2020}, which proposes that compact systems of similar-mass planets can form in-situ, whereas non-comparable mass planets should form in the outer planetary systems where the critical mass (eq.~(22) in \citealt{Adams2020}) is small and there can be large mass disparities. 

Subsequent to the discovery of the radius valley \citep{Fulton2017,Owen2017}, many studies have aimed at understanding the underlying physical processes with observationally well-characterised planets playing a key-role \citep{Cloutier2020,VanEylen2021}. As noted above, planet b has a thin gaseous envelope likely having undergone atmospheric loss in the past. Using the quantification of the radius valley in period-radius space from \cite{VanEylen2018}, we find that TOI-1064\,b lies on the upper edge of the valley that is supported by our observations and internal structure modelling.

Recent work using archival data has shown that there appears to be a gap in the period-mass distribution for small, short period planets \citep{Armstrong2019}. This valley is reported to decrease from 18.9\,$\mathrm{M_\oplus}$ at a rate of -0.9\,$\mathrm{M_\oplus}$\,d$^{-1}$and extend over a width of 4.0\,$\mathrm{M_\oplus}$. With a mass of 13.5$^{+1.7}_{-1.8}$\,$\mathrm{M_{\oplus}}$, TOI-1064\,b falls within this valley. Although the underlying process is currently unknown, the presence of planet b in the gap can help refine the gradient and width of the valley, and under the assumption of the migration of TOI-1064\,b, it could point towards a zero-torque location mechanism for shaping the gap, although other processes have been proposed \citep{Lyra2010,Bitsch2015,Morbidelli2016}.

\section{Summary}
\label{sec:concs} 

In this study, we present the multi-planet TOI-1064 system identified by {\it TESS}. We report the discovery of the pair of sub-Neptunes; TOI-1064\,b and\,c, and characterise both planets using high-precision space-based photometry from {\em TESS} and {\em CHEOPS}, and ground-based photometry from LCOGT, NGTS, and ASTEP, and HARPS RVs. We also present a novel method for detrending photometric data via the monitoring of PSF shape changes and successfully apply it to the {\it CHEOPS} photometry. 

From our analyses find that TOI-1064 is a bright K-dwarf with a stellar rotation period of $\sim$26\,d, that hosts two warm sub-Neptunes, TOI-1064\,b and\,c, on orbital periods of $\sim$6.44\,d and $\sim$12.23\,d. Interestingly, although this pair of sub-Neptunes has similar radii, the masses are substantially different resulting in a density disparity. Placing planets\,b and\,c on a mass-radius diagram, we find that TOI-1064\,b is one of the smallest planets with a mass greater than 10\,$\mathrm{M_{\oplus}}$. It thus joins a small group of dense, warm and mildly irradiated sub-Neptunes around old K-dwarfs. TOI-1064\,c, on the other hand, is potentially one of the least-dense small planets known, however further RV observations are needed to confirm this.

Upon analysis of well-characterised, massive sub-Neptunes, including TOI-1064\,b, we find a negative correlation between bulk density and host star metallicity. This trend is strengthened when highly-irradiated planets are removed from the sample, potentially indicating that this correlation is a result of the formation of planets rather than subsequent evolutionary processes.

Our internal structure analysis shows that TOI-1064\,b has a tenuous atmosphere and a large volatile component by mass fraction. The mass fractions of the component layers and the total refractory mass of TOI-1064\,b may help shed light on the formation mechanisms for such dense bodies. Whilst there are similarities in the mass fractions for the high-Z components of both planets, TOI-1064\,c likely has an extended gaseous envelope. Our atmospheric escape evolution modelling shows that TOI-1064\,b likely lost its primordial atmosphere after migration through the protoplanetary disc. TOI-1064\,c may have also lost some of its envelope subsequent to migration, however understanding the origin and evolution of this planetary system requires first the acquisition of additional RV data to better measure the mass of planet c.

Future atmospheric studies may shed light on whether the atmosphere of TOI-1064\,b is primordial or secondary in nature, and may potentially constrain formation pathways of TOI-1064\,c after the acquisition of further RV data.

\section*{Acknowledgements}

The authors would like to thank the anonymous referee for helpful comments that improved the manuscript. CHEOPS is an ESA mission in partnership with Switzerland with important contributions to the payload and the ground segment from Austria, Belgium, France, Germany, Hungary, Italy, Portugal, Spain, Sweden, and the United Kingdom. The CHEOPS Consortium would like to gratefully acknowledge the support received by all the agencies, offices, universities, and industries involved. Their flexibility and willingness to explore new approaches were essential to the success of this mission. Funding for the TESS mission is provided by NASA's Science Mission Directorate. We acknowledge the use of public TESS data from pipelines at the TESS Science Office and at the TESS Science Processing Operations Center. This research has made use of the Exoplanet Follow-up Observation Program website, which is operated by the California Institute of Technology, under contract with the National Aeronautics and Space Administration under the Exoplanet Exploration Program. Resources supporting this work were provided by the NASA High-End Computing (HEC) Program through the NASA Advanced Supercomputing (NAS) Division at Ames Research Center for the production of the SPOC data products. Part of this work was done using data taken by KESPRINT, an international consortium devoted to the characterization and research of exoplanets discovered with space-based missions (http://www.kesprint.science/). This work makes use of observations from the LCOGT network. Part of the LCOGT telescope time was granted by NOIRLab through the Mid-Scale Innovations Program (MSIP). MSIP is funded by NSF. Based on data collected under the NGTS project at the ESO La Silla Paranal Observatory. The NGTS facility is operated by the consortium institutes with support from the UK Science and Technology Facilities Council (STFC) projects ST/M001962/1 and  ST/S002642/1. This work has made use of data from the European Space Agency (ESA) mission {\it Gaia} (https://www.cosmos.esa.int/gaia), processed by the {\it Gaia} Data Processing and Analysis Consortium (DPAC, https://www.cosmos.esa.int/web/gaia/dpac/consortium). Funding for the DPAC has been provided by national institutions, in particular the institutions participating in the {\it Gaia} Multilateral Agreement. This work makes use of observations from the ASTEP telescope. ASTEP benefited from the support of the French and Italian polar agencies IPEV and PNRA in the framework of the Concordia station program and from Idex UCAJEDI (ANR-15-IDEX-01). Some of the observations in the paper made use of the High-Resolution Imaging instrument Zorro. Zorro was funded by the NASA Exoplanet Exploration Program and built at the NASA Ames Research Center by Steve B. Howell, Nic Scott, Elliott P. Horch, and Emmett Quigley. Zorro was mounted on the Gemini South telescope of the international Gemini Observatory, a program of NSF’s OIR Lab, which is managed by the Association of Universities for Research in Astronomy (AURA) under a cooperative agreement with the National Science Foundation on behalf of the Gemini partnership: the National Science Foundation (United States), National Research Council (Canada), Agencia Nacional de Investigación y Desarrollo (Chile), Ministerio de Ciencia, Tecnología e Innovación (Argentina), Ministério da Ciência, Tecnologia, Inovações e Comunicações (Brazil), and Korea Astronomy and Space Science Institute (Republic of Korea). 
T.G.W., A.C.C., and K.H. acknowledge support from STFC consolidated grant numbers ST/R000824/1 and ST/V000861/1, and UKSA grant ST/R003203/1. 
Y.A. and M.J.H. acknowledge the support of the Swiss National Fund under grant 200020\_172746. 
D.G. and L.M.S. gratefully acknowledge financial support from the CRT foundation under Grant No. 2018.2323 “Gaseous or rocky? Unveiling the nature of small worlds''. 
D.G., M.F., X.B., S.C., and J.L. acknowledge their roles as ESA-appointed CHEOPS science team members. 
C.M.P., M.F., J.K., and A.J.M. gratefully acknowledge the support of the Swedish National Space Agency (SNSA.; DNR 65/19, 174/18, 2020-00104, and Career grant 120/19C). 
A. De. and D.E. acknowledge support from the European Research Council (ERC) under the European Union’s Horizon 2020 research and innovation programme (project {\sc Four Aces}; grant agreement No 724427). 
A.De., A. Le., and H.O. acknowledge support from the Swiss National Centre for Competence in Research “PlanetS” and the Swiss National Science Foundation (SNSF). 
S.H. gratefully acknowledges CNES funding through the grant 837319. 
S.E.S. have received funding from the European Research Council (ERC) under the European Union’s Horizon 2020 research and innovation programme (grant agreement No 833925, project STAREX). 
S.G.S. acknowledge support from FCT through FCT contract nr. CEECIND/00826/2018 and POPH/FSE (EC).
We acknowledge support from the Spanish Ministry of Science and Innovation and the European Regional Development Fund through grants ESP2016-80435-C2-1-R, ESP2016-80435-C2-2-R, PGC2018-098153-B-C33, PGC2018-098153-B-C31, ESP2017-87676-C5-1-R, MDM-2017-0737 Unidad de Excelencia Maria de Maeztu-Centro de Astrobiologí­a (INTA-CSIC), as well as the support of the Generalitat de Catalunya/CERCA programme. The MOC activities have been supported by the ESA contract No. 4000124370. 
S.C.C.B. and V.A. acknowledge support from FCT through FCT contracts nr. IF/01312/2014/CP1215/CT0004 and IF/00650/2015/CP1273/CT0001, respectively.
A.Br. was supported by the SNSA. 
This project was supported by the CNES. 
L.D. is an F.R.S.-FNRS Postdoctoral Researcher. The Belgian participation to CHEOPS has been supported by the Belgian Federal Science Policy Office (BELSPO) in the framework of the PRODEX Program, and by the University of Liège through an ARC grant for Concerted Research Actions financed by the Wallonia-Brussels Federation. 
This work was supported by FCT - Fundação para a Ciência e a Tecnologia through national funds and by FEDER through COMPETE2020 - Programa Operacional Competitividade e Internacionalizacão by these grants: UID/FIS/04434/2019, UIDB/04434/2020, UIDP/04434/2020, PTDC/FIS-AST/32113/2017 \& POCI-01-0145-FEDER- 032113, PTDC/FIS-AST/28953/2017 \& POCI-01-0145-FEDER-028953, PTDC/FIS-AST/28987/2017 \& POCI-01-0145-FEDER-028987, O.D.S.D. is supported in the form of work contract (DL 57/2016/CP1364/CT0004) funded by national funds through FCT. 
B.-O.D. acknowledges support from the Swiss National Science Foundation (PP00P2-190080). 
D. D. acknowledges support from the TESS Guest Investigator Program grant 80NSSC19K1727 and NASA Exoplanet Research Program grant 18-2XRP18\_2-0136. 
M.G. is an F.R.S.-FNRS Senior Research Associate. 
K.G.I. is the ESA CHEOPS Project Scientist and is responsible for the ESA CHEOPS Guest Observers Programme. She does not participate in, or contribute to, the definition of the Guaranteed Time Programme of the CHEOPS mission through which observations described in this paper have been taken, nor to any aspect of target selection for the programme. 
G.L. acknowledges support by CARIPARO Foundation, according to the agreement CARIPARO-Università degli Studi di Padova (Pratica n. 2018/0098). 
This work was granted access to the HPC resources of MesoPSL financed by the Region Ile de France and the project Equip@Meso (reference ANR-10-EQPX-29-01) of the programme Investissements d'Avenir supervised by the Agence Nationale pour la Recherche.
M.L. acknowledges support from the Swiss National Science Foundation under grant no. PCEFP2\_194576. 
P.F.L.M. acknowledges support from STFC research grant number ST/M001040/1. 
L.D.N. thanks the Swiss National Science Foundation for support under Early Postdoc. Mobility grant P2GEP2\_200044. 
This work was also partially supported by a grant from the Simons Foundation (PI Queloz, grant number 327127). 
I.R. cknowledges support from the Spanish Ministry of Science and Innovation and the European Regional Development Fund through grant PGC2018-098153-B- C33, as well as the support of the Generalitat de Catalunya/CERCA programme. 
This project has been supported by the Hungarian National Research, Development and Innovation Office (NKFIH) grant K-125015, the MTA-ELTE Lend\"ulet Milky Way Research Group and the City of Szombathely under Agreement No. 67.177-21/2016. 
This research received funding from the European Research Council (ERC) under the European Union's Horizon 2020 research and innovation programme (grant agreement n$^\circ$ 803193/BEBOP), and from the Science and Technology Facilities Council (STFC.; grant n$^\circ$ ST/S00193X/1). 
V.V.G. is an F.R.S-FNRS Research Associate.

\section*{Data Availability}

The data underlying this article will be made available in the {\it CHEOPS} mission archive (https://cheops.unige.ch/archive\_browser/). This paper includes data collected by the {\it TESS} mission, which is publicly available from the Mikulski Archive for Space Telescopes (MAST) at the Space Telescope Science Institure (STScI) (https://mast.stsci.edu).



\bibliographystyle{mnras}
\bibliography{1064} 

\bigskip

\noindent
\hrulefill \\
$^{1}$ Centre for Exoplanet Science, SUPA School of Physics and Astronomy, University of St Andrews, North Haugh, St Andrews KY16 9SS, UK\\
$^{2}$ Dipartimento di Fisica, Universita degli Studi di Torino, via Pietro Giuria 1, I-10125, Torino, Italy\\
$^{3}$ Thüringer Landessternwarte Tautenburg, Sternwarte 5, D-07778 Tautenburg, Germany\\
$^{4}$ Physikalisches Institut, University of Bern, Gesellsschaftstrasse 6, 3012 Bern, Switzerland\\
$^{5}$ Space Research Institute, Austrian Academy of Sciences, Schmiedlstrasse 6, A-8042 Graz, Austria\\
$^{6}$ Department of Space, Earth and Environment, Chalmers University of Technology, Onsala Space Observatory, 43992 Onsala, Sweden\\
$^{7}$ Leiden Observatory, University of Leiden, PO Box 9513, 2300 RA Leiden, The Netherlands\\
$^{8}$ Department of Space, Earth and Environment, Astronomy and Plasma Physics, Chalmers University of Technology, 412 96 Gothenburg, Sweden\\
$^{9}$ Center for Space and Habitability, Gesellsschaftstrasse 6, 3012 Bern, Switzerland\\
$^{10}$ Observatoire Astronomique de l'Université de Genève, Chemin Pegasi 51, Versoix, Switzerland\\
$^{11}$ Department of Astronomy, Stockholm University, AlbaNova University Center, 10691 Stockholm, Sweden\\
$^{12}$ Department of Astronomy, Stockholm University, SE-106 91 Stockholm, Sweden\\
$^{13}$ Aix Marseille Univ, CNRS, CNES, LAM, Marseille, France\\
$^{14}$ Division Technique INSU, BP 330, 83507 La Seyne cedex, France\\
$^{15}$ Instituto de Astrofisica e Ciencias do Espaco, Universidade do Porto, CAUP, Rua das Estrelas, 4150-762 Porto, Portugal\\
$^{16}$ Laboratoire d'Astrophysique de Marseille, 38 Rue Frédéric Joliot Curie, 13013 Marseille, France\\
$^{17}$ Lund Observatory, Dept. of Astronomy and Theoretical Physics, Lund University, Box 43, 22100 Lund, Sweden\\
$^{18}$ Université Côte d’Azur, Observatoire de la Côte d’Azur, CNRS, Laboratoire Lagrange, CS 34229, 06304 Nice Cedex 4, France\\
$^{19}$ Instituto de Astrofisica de Canarias, 38200 La Laguna, Tenerife, Spain\\
$^{20}$ Departamento de Astrofisica, Universidad de La Laguna, 38206 La Laguna, Tenerife, Spain\\
$^{21}$ Institut de Ciencies de l'Espai (ICE, CSIC), Campus UAB, Can Magrans s/n, 08193 Bellaterra, Spain\\
$^{22}$ Institut d'Estudis Espacials de Catalunya (IEEC), 08034 Barcelona, Spain\\
$^{23}$ European Space Agency (ESA), European Space Research and Technology Centre (ESTEC), Keplerlaan 1, 2201 AZ Noordwijk, The Netherlands\\
$^{24}$ Admatis, 5. Kandó Kálmán Street, 3534 Miskolc, Hungary\\
$^{25}$ Depto. de Astrofisica, Centro de Astrobiologia (CSIC-INTA), ESAC campus, 28692 Villanueva de la Cañada (Madrid), Spain\\
$^{26}$ Departamento de Fisica e Astronomia, Faculdade de Ciencias, Universidade do Porto, Rua do Campo Alegre, 4169-007 Porto, Portugal\\
$^{27}$ Université Grenoble Alpes, CNRS, IPAG, 38000 Grenoble, France\\
$^{28}$ Department of Physics, University of Warwick, Gibbet Hill Road, Coventry CV4 7AL, United Kingdom\\
$^{29}$ Department of Physics and Astronomy, University of Leicester, Leicester LE1 7RH, United Kingdom\\
$^{30}$ Concordia Station, Dome C, Antarctica\\
$^{31}$ Institute of Planetary Research, German Aerospace Center (DLR), Rutherfordstrasse 2, 12489 Berlin, Germany\\
$^{32}$ Université de Paris, Institut de physique du globe de Paris, CNRS, F-75005 Paris, France\\
$^{33}$ NASA Exoplanet Science Institute, Caltech/IPAC, 1200 E. California Ave., Pasadena, CA, 91125, USA\\
$^{34}$ Center for Astrophysics \textbar \ Harvard \& Smithsonian, 60 Garden Street, Cambridge, MA, 02138, USA\\
$^{35}$ Center for Planetary Systems Habitability and McDonald Observatory, The University of Texas at Austin, Austin, TX, 78712, USA\\
$^{36}$ NASA Goddard Space Flight Center, Greenbelt, MD, 20771, USA\\
$^{37}$ Centre for Mathematical Sciences, Lund University, Box 118, SE-22100 Lund, Sweden\\
$^{38}$ Astrobiology Research Unit, Université de Liège, Allée du 6 Août 19C, B-4000 Liège, Belgium\\
$^{39}$ Space sciences, Technologies and Astrophysics Research (STAR) Institute, Université de Liège, Allée du 6 Août 19C, 4000 Liège, Belgium\\
$^{40}$ Department of Physics and Astronomy, University of New Mexico, 1919 Lomas Blvd NE, Albuquerque, NM, 87131, USA\\
$^{41}$ School of Physics \& Astronomy, University of Birmingham, Edgbaston, Birmingham, B15 2TT, United Kingdom\\
$^{42}$ Department of Astronomy and Tsinghua Centre for Astrophysics, Tsinghua University, Beijing, 100084, China\\
$^{43}$ NASA Ames Research Center, Moffett Field, CA, 94035, USA\\
$^{44}$ Rhenish Institute for Enviromental Research, dep. of Planetary Research, at the University of Cologne, Aachener Strasse 209, 50931 Cologne, Germany\\
$^{45}$ Department of Astrophysics, University of Vienna, Türkenschanzstrasse 17, 1180 Vienna, Austria\\
$^{46}$ Science and Operations Department - Science Division (SCI-SC), Directorate of Science, European Space Agency (ESA), European Space Research and Technology Centre (ESTEC), Keplerlaan 1, 2201-AZ Noordwijk, The Netherlands\\
$^{47}$ Department of Physics \& Astronomy, Swarthmore College, Swarthmore PA, 19081, USA\\
$^{48}$ Konkoly Observatory, Research Centre for Astronomy and Earth Sciences, 1121 Budapest, Konkoly Thege Miklós út 15-17, Hungary\\
$^{49}$ Dipartimento di Fisica e Astronomia "Galileo Galilei", Universita degli Studi di Padova, Vicolo dell'Osservatorio 3, 35122 Padova, Italy\\
$^{50}$ INAF, Osservatorio Astronomico di Padova, Vicolo dell'Osservatorio 5, 35122 Padova, Italy\\
$^{51}$ IMCCE, UMR8028 CNRS, Observatoire de Paris, PSL Univ., Sorbonne Univ., 77 av. Denfert-Rochereau, 75014 Paris, France\\
$^{52}$ Institut d'astrophysique de Paris, UMR7095 CNRS, Université Pierre \& Marie Curie, 98bis blvd. Arago, 75014 Paris, France\\
$^{53}$ Department of Physics and Kavli Institute for Astrophysics and Space Research, Massachusetts Institute of Technology, Cambridge, MA, 02139, USA\\
$^{54}$ Department of Astronomy, University of Tokyo, 7-3-1 Hongo, Bunkyo-ku, Tokyo 113-0033, Japan\\
$^{55}$ Instituto de Astrof\'isica de Andaluc\'ia (IAA-CSIC), Glorieta de la Astronom\'ia s/n, 18008 Granada, Spain\\
$^{56}$ Astrophysics Group, Keele University, Staffordshire, ST5 5BG, United Kingdom\\
$^{57}$ Astronomy Department, University of California, Berkeley, CA, 94720, USA\\
$^{58}$ Space Telescope Science Institute, 3700 San Martin Drive, Baltimore, MD, 21218, USA\\
$^{59}$ Department of Physics, University of Oxford, OX13RH, Oxford, UK\\
$^{60}$ Mullard Space Science Laboratory, University College London, Holmbury St Mary, Dorking, Surrey, RH5 6NT, UK\\
$^{61}$ INAF, Osservatorio Astrofisico di Catania, Via S. Sofia 78, 95123 Catania, Italy\\
$^{62}$ Institute of Optical Sensor Systems, German Aerospace Center (DLR), Rutherfordstrasse 2, 12489 Berlin, Germany\\
$^{63}$ Cavendish Laboratory, JJ Thomson Avenue, Cambridge CB3 0HE, UK\\
$^{64}$ Center for Astronomy and Astrophysics, Technical University Berlin, Hardenberstrasse 36, 10623 Berlin, Germany\\
$^{65}$ Institut für Geologische Wissenschaften, Freie UniversitÃ¤t Berlin, 12249 Berlin, Germany\\
$^{66}$ Astronomy Department and Van Vleck Observatory, Wesleyan University, Middletown, CT, 06459, USA\\
$^{67}$ Patashnick Voorheesville Observatory, Voorheesville, NY, 12186, USA\\
$^{68}$ Department of Earth, Atmospheric and Planetary Sciences, Massachusetts Institute of Technology, Cambridge, MA, 02139, USA\\
$^{69}$ Department of Aeronautics and Astronautics, MIT, 77 Massachusetts Avenue, Cambridge, MA, 02139, USA\\
$^{70}$ Hazelwood Observatory, Australia\\
$^{71}$ ELTE Eötvös Loránd University, Gothard Astrophysical Observatory, 9700 Szombathely, Szent Imre h. u. 112, Hungary\\
$^{72}$ MTA-ELTE Exoplanet Research Group, 9700 Szombathely, Szent Imre h. u. 112, Hungary\\
$^{73}$ MTA-ELTE Lendület Milky Way Research Group, Hungary\\
$^{74}$ Institute of Astronomy, University of Cambridge, Madingley Road, Cambridge, CB3 0HA, United Kingdom\\
$^{75}$ Department of Astrophysical Sciences, Princeton University, 4 Ivy Ln., Princeton, NJ, 08544, USA\\



\appendix

\section{A Description of the PSF-Based {\sc scalpels} Method}
\label{sec:scalpels_desc}

Here we describe the PSF-based {\sc scalpels} method developed to model PSF shape changes in {\it CHEOPS} observations. We make use of the fact that the autocorrelation function of the PSF is, to a high degree of approximation, invariant to the small shifts imposed by spacecraft pointing jitter. The PSF is a positive definite function of position on the CCD. We apply a simple autocorrelation procedure, imposing a sequence of shifts by integer-pixel offsets in $x$ and $y$ covering the image, and co-multiplying the shifted and un-shifted images. The resulting autocorrelation function is unwrapped into a 1-dimensional array and normalised by its own mean which is given as $\mathbf{A}(\delta\mathbf{r}_i)$ where $(\delta\mathbf{r}_i)$ is the aforementioned shift.

From this point on, the procedure is as described by \citet{CollierCameron2021}, i.e. we compute the singular-value decomposition (SVD) of the time sequence of ACFs computed from the images following the standard procedure:

\begin{equation}
\mathbf{A}(\delta\mathbf{r}_i,t_j) = \left<A(\delta\mathbf{r}_i,t_j)\right> + \mathbf{U}(t_j) \cdot \rm{diag}(\mathbf{S})\cdot \mathbf{V}(\delta\mathbf{r}_i).
\label{eq:acfSVD}
\end{equation}

where ${\bf U}$, ${\bf S}$, and ${\bf V}$ are the components of the SVD. The diagonal matrix $\mathbf{S}$ lists the singular values (eigenvalues) of the principal components in decreasing order. The columns of $\mathbf{U}(t_j)$ define an orthonormal basis in the time domain with each column comprising the time-domain coefficients of the corresponding eigenvector $\mathbf{V}(\delta\mathbf{r}_i)$. 

We mask out images affected by cosmic rays, saturation and other unwanted chance effects by identifying strong outliers in the time-domain coefficients $\mathbf{U}$ of the eigenfunctions $\mathbf{V}$. We use leave-one-out cross-validation (LOOCV; \citealt{Celisse2014}) to separate the subset of the columns of $\mathbf{U}$ whose variations are driven by genuine PSF changes, from those that simply describe noise. The LOOCV method systematically leaves single data points (or in this study, rows of the autocorrelation function) out of the analysis and aims to reconstruct the missing data. If this prediction accurately matches the removed data we consider this row to represent PSF changes, whereas if the reconstruction poorly reproduces the data it is interpreted as noise.

The response of the photometric data time-series $\mathbf{d}_{\rm obs}$ to the set of normal modes of PSF shape variation described by the columns of $\mathbf{U}$ is found by applying the projection operator $\mathbf{U}\cdot\mathbf{U}^T$ to the light curve. Subtracting this response from the light curve is equivalent to projecting the light curve into the orthogonal complement of $\mathbf{U}$, via the linear operator $P_\perp\equiv(\mathbf{I}-\mathbf{U}\cdot\mathbf{U}^T)$.

This procedure is only valid if the target flux is already corrected for intrinsic stellar variability, because the target variations may have a non-zero projection into the null space which would distort the aperture correction. In reality we are interested in modelling the intrinsic variations of the target with a time-dependent model light curve $\mathbf{m}(\mathbf{\theta})$, where $\mathbf{\theta}$ is the vector of parameters defining the light curve model.

The likelihood of the data, $\mathbf{d}_{\rm obs}$, given the light curve model parameters, $\mathbf{\theta}$, is found by subtracting the model from the data and projecting into the orthogonal complement of the null space:
\begin{eqnarray}
    \ln\mathcal{L} &= 
    &-\frac{1}{2}\left[P_\perp\cdot(\mathbf{d}_{\rm obs}-\mathbf{m}(\mathbf{\theta}))\right]^T\cdot{\bf \Sigma}^{-1}\cdot\left[P_\perp\cdot(\mathbf{d}_{\rm obs}-\mathbf{m}(\mathbf{\theta}))\right]\nonumber\\
    & &-\frac{1}{2}\ln|\mathbf{\Sigma}|-\frac{N}{2}\ln(2\pi)\nonumber\\
    &= 
    &-\frac{1}{2}\left[P_\perp\cdot(\mathbf{d}_{\rm obs}-\mathbf{m}(\mathbf{\theta}))\right]^T\cdot{\bf \Sigma}^{-1}\cdot\left[P_\perp\cdot(\mathbf{d}_{\rm obs}-\mathbf{m}(\mathbf{\theta}))\right]\nonumber\\
    & & -\frac{1}{2}\left(\sum_{j=1}^{N}\ln\mathbf{\Sigma}_{jj}\right)-\frac{N}{2}\ln(2\pi),
    \label{eq:likelihood}    
\end{eqnarray}
where the covariance matrices are $\mathbf{\Sigma}_{jj}$ and thus $\sigma_j^2={\rm var}(d_j)$ are the variances of the individual photometric measurements $d_j$.

Used with MCMC sampling, this allows us to determine the posterior probability distribution of the model parameters $\mathbf{\theta}$, while simultaneously correcting for aperture losses arising from PSF shape changes. For the optimal set of model parameters $\mathbf{\theta_*}$, the magnitude correction to the light curve is 
\begin{equation}
   \mathbf{d}_\parallel = P_\parallel\cdot(\mathbf{d}_{\rm obs}-\mathbf{m}(\mathbf{\theta_*})),
\end{equation}
where $P_\parallel = \mathbf{U}\cdot\mathbf{U}^T$ is the projection operator into the null space.
From this approach, the corrected flux light curve is then $\mathbf{f}_{\rm corr} = \mathbf{d}_{\rm obs}-\mathbf{d}_\parallel$.

\section{Raw {\it CHEOPS} Lightcurves}

\begin{figure}
  \includegraphics[width = 8.5cm]{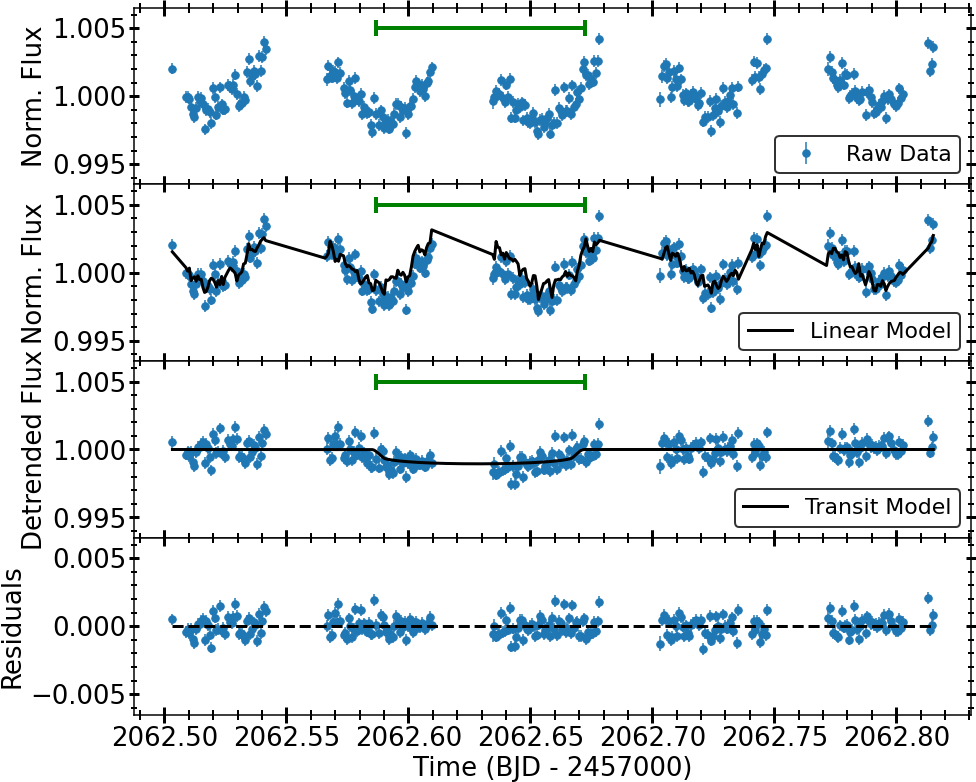}
  \caption{The normalised light curve of the second {\em CHEOPS} visit of TOI-1064 with the location of the transit of planet b (green) shown. {\em Top panel}: The raw DRP-produced fluxes. {\em Second panel}: The DRP fluxes with the fitted linear model produced from the measured PSF shape changes, as detailed in Appendix~\ref{sec:scalpels_desc}. {\em Third panel}: The detrended fluxes with the nominal transit model from the global analysis (Section~\ref{sec:joint_analysis}). {\em Bottom panel}: Residuals to the fit.}
  \label{fig:cheops_raw2}
\end{figure}

\begin{figure}
  \includegraphics[width = 8.5cm]{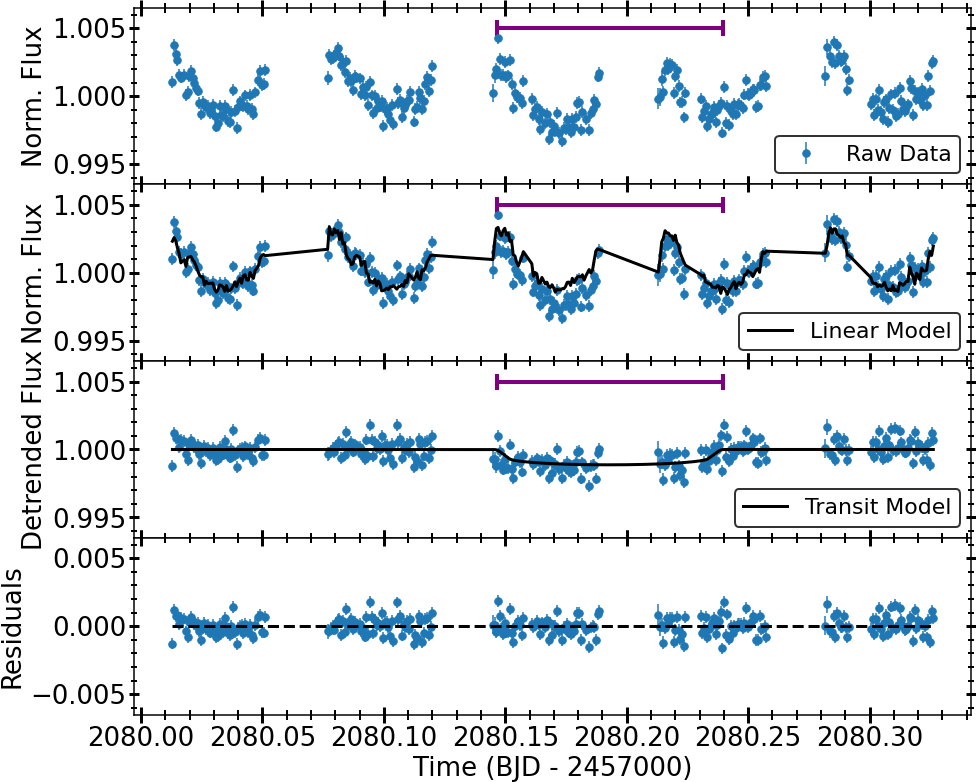}
  \caption{Same as Fig.~\ref{fig:cheops_raw2}, but for the third {\em CHEOPS} visit with the location of the transit of planet c (purple) shown.}
  \label{fig:cheops_raw3}
\end{figure}

\begin{figure}
  \includegraphics[width = 8.5cm]{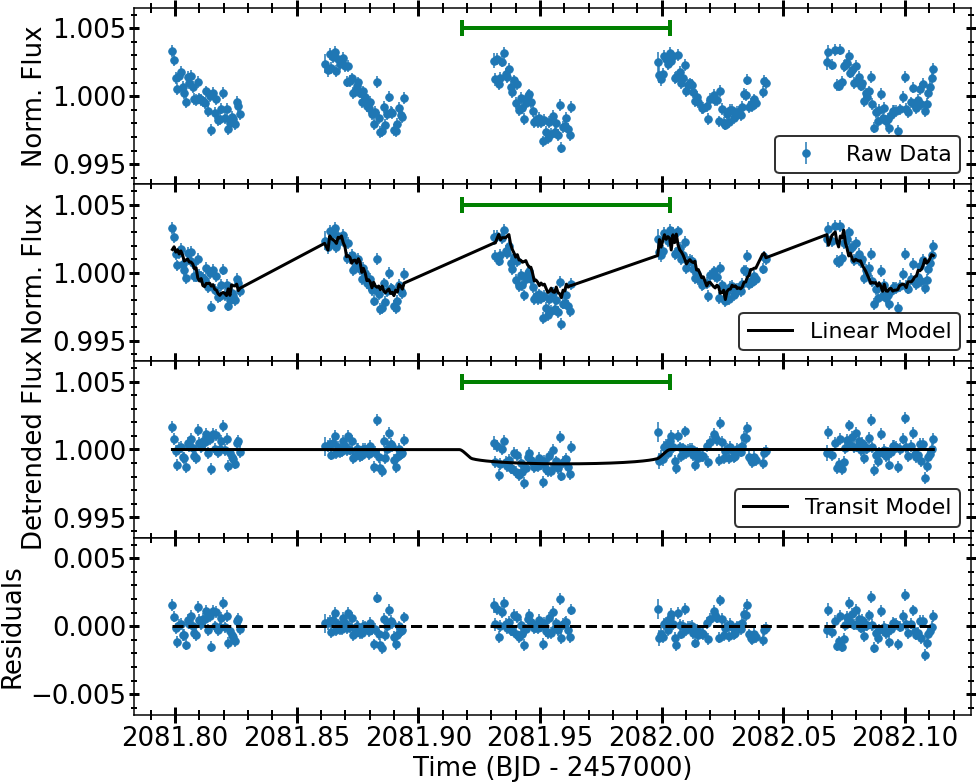}
  \caption{Same as Fig.~\ref{fig:cheops_raw2}, but for the fourth {\em CHEOPS} visit with the location of the transit of planet b (green) shown.}
  \label{fig:cheops_raw4}
\end{figure}

\begin{figure}
  \includegraphics[width = 8.5cm]{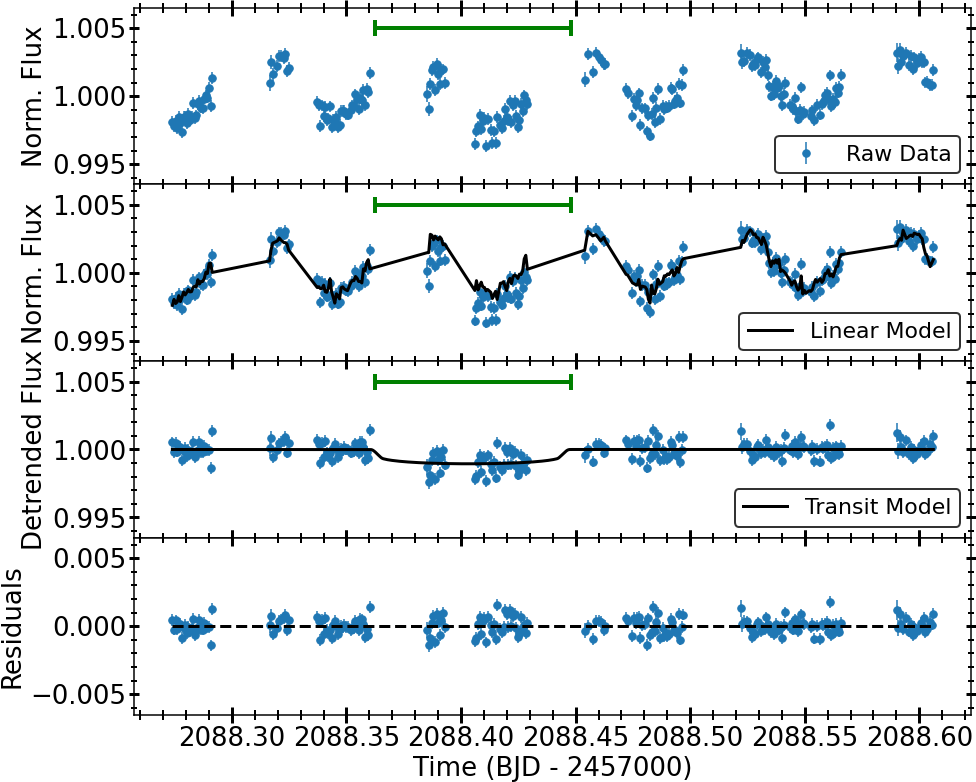}
  \caption{Same as Fig.~\ref{fig:cheops_raw2}, but for the fifth {\em CHEOPS} visit with the location of the transit of planet b (green) shown.}
  \label{fig:cheops_raw5}
\end{figure}

\begin{figure}
  \includegraphics[width = 8.5cm]{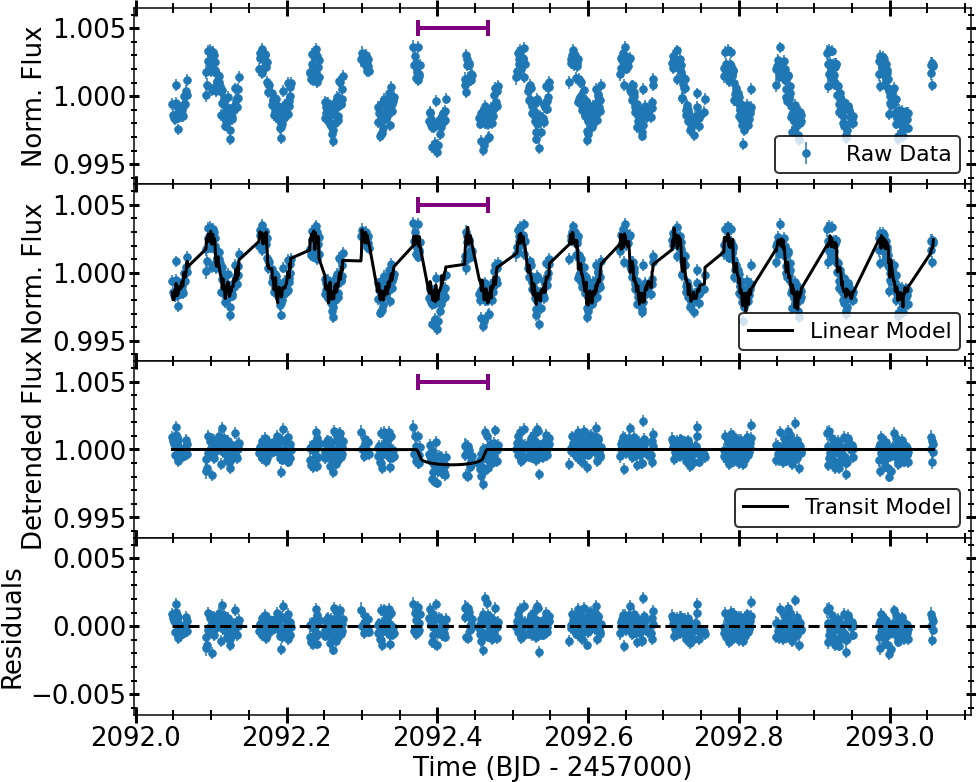}
  \caption{Same as Fig.~\ref{fig:cheops_raw2}, but for the sixth {\em CHEOPS} visit with the location of the transit of planet c (purple) shown.}
  \label{fig:cheops_raw6}
\end{figure}

In addition to the raw data and fitted linear model of the first {\em CHEOPS} visit of TOI-1064 seen in Fig.~\ref{fig:cheops_raw1}, here for completeness we present the raw lightcurves and linear models for the remaining five {\em CHEOPS} observations (Figs.~\ref{fig:cheops_raw2}-\ref{fig:cheops_raw6}). The raw data plotted in the top panels are the {\em CHEOPS} DRP-produced photometry outlined in Section~\ref{sec:cheops}, with the linear models in the bottom panels constructed following the principal component analysis of the auto-correlation function of the {\em CHEOPS} PSF as described in Appendix~\ref{sec:scalpels_desc}.

\section{Periodogram analysis of the HARPS measurements}

\begin{figure}
  \includegraphics[width = 8.5cm]{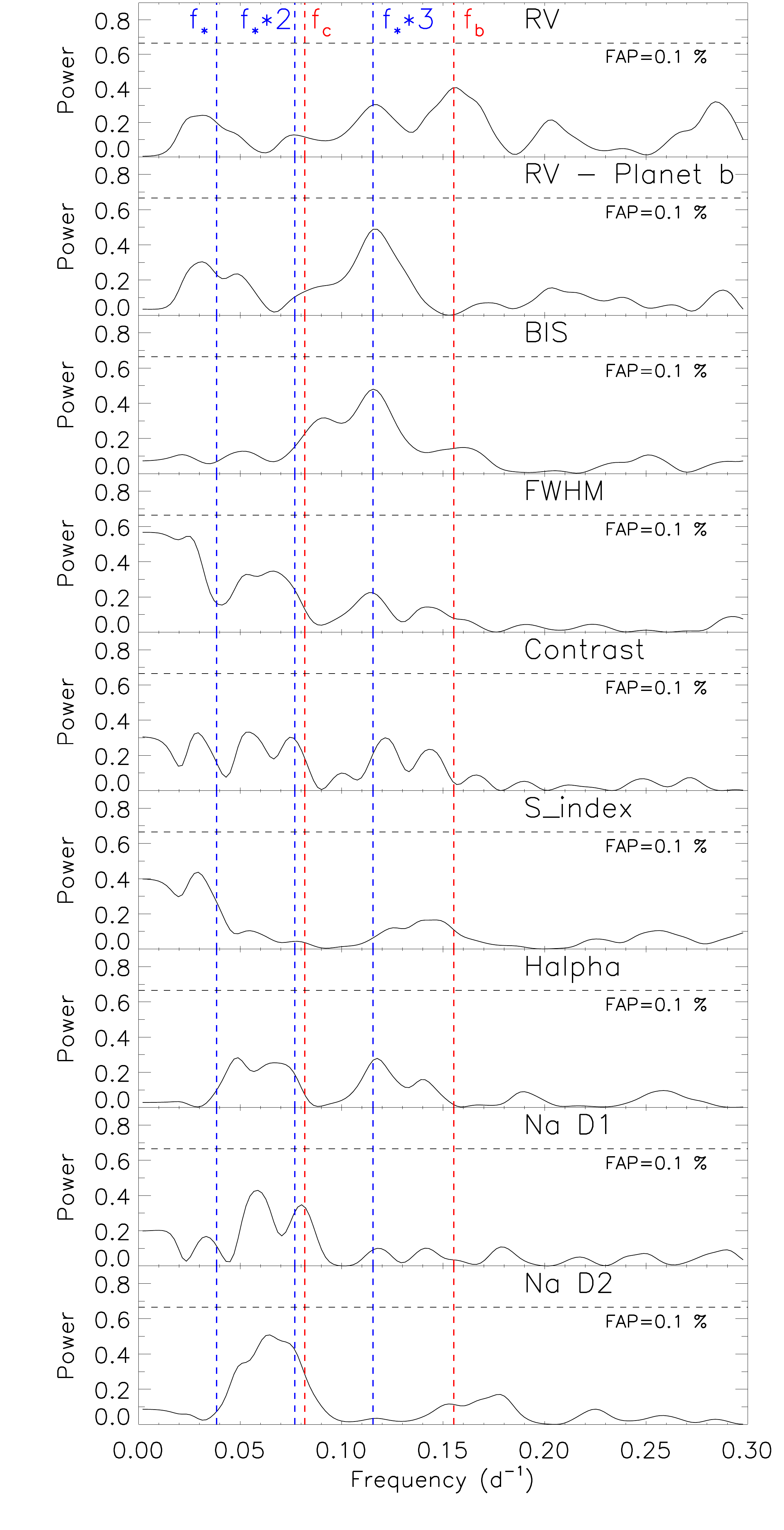}
  \caption{Generalized Lomb-Scargle periodograms of the HARPS RVs and spectral activity indicators from the HARPS DRS and \texttt{TERRA}. The horizontal dashed line marks the 0.1\% FAP level. The vertical dashed red lines mark the orbital frequencies of the two transiting planets (f$_{\rm b}$ = 0.155\,d$^{-1}$ and f$_{\rm c}$ = 0.082\,d$^{-1}$). The vertical dashed blue lines mark the frequencies of the stellar signal at $\sim$0.038\,d$^{-1}$ and its first and second harmonics. {\em Upper panel}: HARPS DRS RVs. {\em Second panel}: RV residuals following the subtraction of the signal of TOI-1064\,b. {\em Remaining panels}: the activity indicators of TOI-1064; the BIS, the FWHM, and the contrast of the CCF, the Ca\,{\sc ii} H\,\&\,K lines activity indicator (S-index), H$\alpha$, Na D1, and Na D2.}
  \label{fig:activity_inds}
\end{figure}

We performed a frequency analysis of the HARPS RV measurements and spectral diagnostics in order to search for the Doppler reflex motion induced by the two transiting planets TOI-1064\,b and TOI-1064\,c, and look for possible additional signals that might be associated to stellar activity or other orbiting bodies in the system.

Figure~\ref{fig:activity_inds} displays the generalised Lomb-Scargle periodograms \citep[GLS;][]{Zechmeister2009} of the HARPS DRS RVs and activity indicators extracted with DRS and with TERRA. The horizontal black dashed lines mark the power corresponding to the 0.1\% false alarm probability (FAP). We estimated the FAP following the bootstrap method described in \cite{Murdoch1993}, i.e., by computing the GLS periodogram of 10$^{6}$ mock time-series obtained by randomly shuffling the RV measurements and their uncertainties whilst keeping the time stamps fixed. We defined the FAP as the fraction of those periodograms whose highest power exceeds the observed power of f$_{\rm b}$ in the periodogram of the original HARPS data over the frequency range 0-0.3\,d$^{-1}$.

The GLS periodogram of the HARPS DRS RVs (Fig.~\ref{fig:activity_inds}, upper panel) shows its highest peak at f$_{\rm b}$ = 0.155\,d$^{-1}$, i.e., the orbital frequency of the inner transiting planet initially detected in the {\it TESS} light curve (right dashed red line). Although this peak is not significant (FAP\,$>$\,0.1\%) within the frequency range 0.0-0.3\,d$^{-1}$, the occurrence of this peak at a known frequency provides strong evidence that this signal is related to planet b. We estimated the FAP for a peak at the orbital frequency of TOI-1064\,b and found the probability that the periodogram of white/red noise can display a peak at the orbital frequency of the inner planet, with a power higher than the observed power, is only 0.06\%, making the peak at f$_{\rm b}$ significant. 
The peak at f$_{\rm b}$ has no counterpart in the periodograms of the activity indicators, suggesting that the signal detected in Doppler data is induced by an orbiting companion and confirms the presence of the inner transiting planet seen in both {\em TESS} and {\em CHEOPS} data with a period of about 6.44\,d. 

We subtracted the Doppler signal of TOI-1064\,b from the HARPS RVs by fitting a circular-orbit model, fixing period and time of first transit as derived from the analysis of the light curves. The periodogram of the RV residuals (Fig.~\ref{fig:activity_inds}, second panel) displays a non-significant peak at $\sim$8.7\,d (FAP\,$>$\,0.1\%). This non-significant peak is also observed in the periodogram of the CCF BIS, FWHM, contrast, and H$\alpha$. However, observing the same peak in different unrelated activity periodograms provides evidence that it could be real. The rotation period of the star - as inferred from the WASP light curve - is about 26.6\,d. The peak at 8.7\,d would then be near the second harmonic of this period. Active regions separated by about 120 degrees in longitude could account for this signal. In addition, the FWHM, the contrast, and the S-index show powers at frequencies lower than 0.04\,d$^{-1}$ (25\,d), that might be due to activity coupled to stellar rotation. Unfortunately, although seen in {\em TESS}, {\em CHEOPS}, LCOGT, NGTS, and ASTEP photometry, the outer planet at 12.23\,d is undetected in these HARPS RVs periodograms. Therefore, the use of the {\sc scalpels} algorithm to potentially separate the stellar and planetary signals is needed, see Section~\ref{sec:scalpels}.

\section{Joint Photometry and RV Fit Priors and Posterior Distributions}

Here we present a table of the priors used in the global analysis of the TOI-1064 system, and corner plots of the posterior distributions of the main fitted transit and RV parameters for planets\,b and\,c, and the stellar density. The corner plots show a slight correlation between $r_1$ and $\sqrt{e}$sin$\omega$ that might indicate a tension between the fitting the ingresses and egresses of the transit photometry to obtain the impact parameters and two Keplerian orbits to obtain the eccentricities of the planets. This could manifest as the bimodality around 0 for the $\sqrt{e}$sin$\omega$ component of TOI-1064\,c.

\begin{figure*}
    \subfloat[\centering TOI-1064\,b]{{\includegraphics[width = 8.5cm]{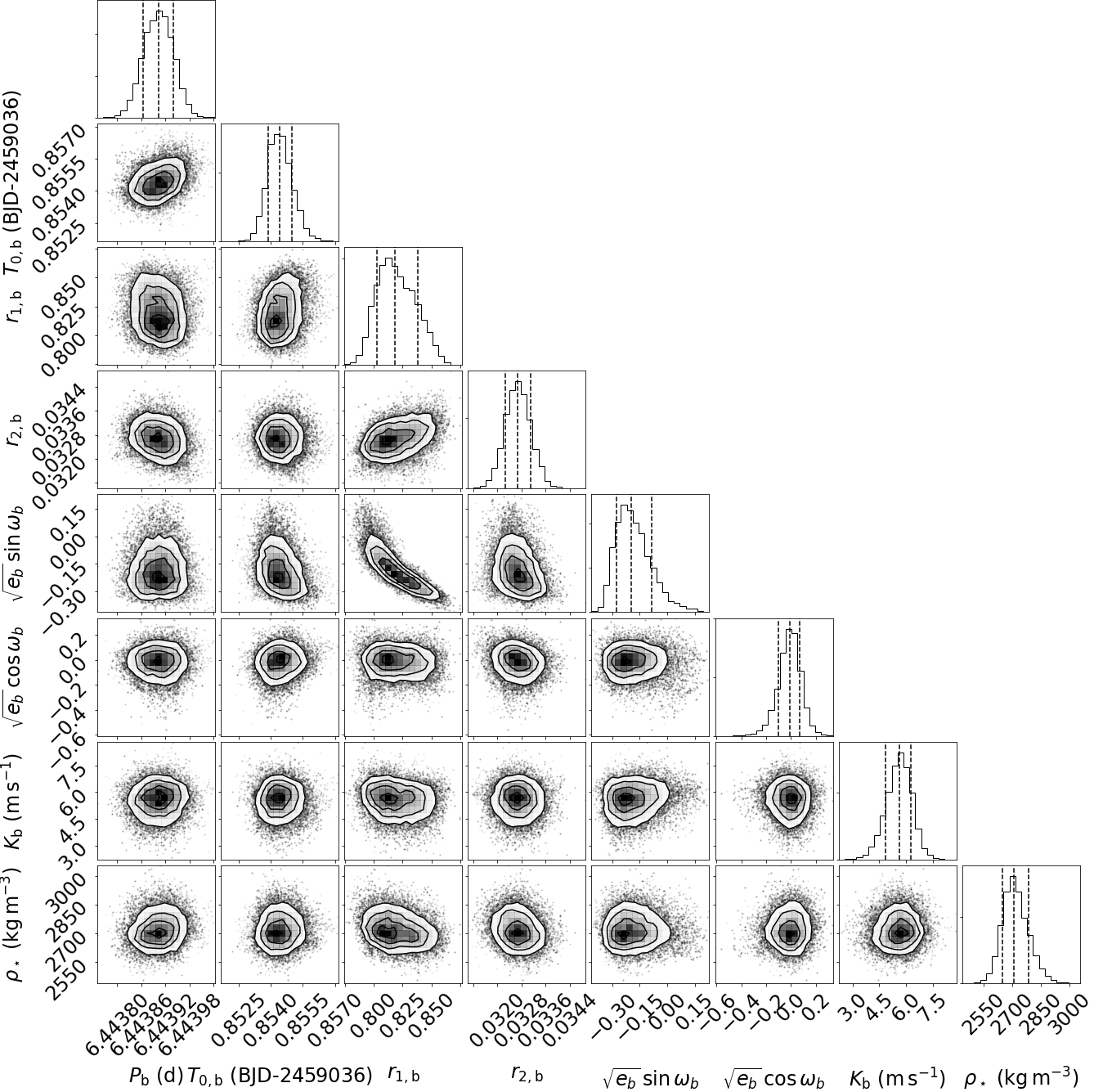} }}
    \qquad
    \subfloat[\centering TOI-1064\,c]{{\includegraphics[width = 8.5cm]{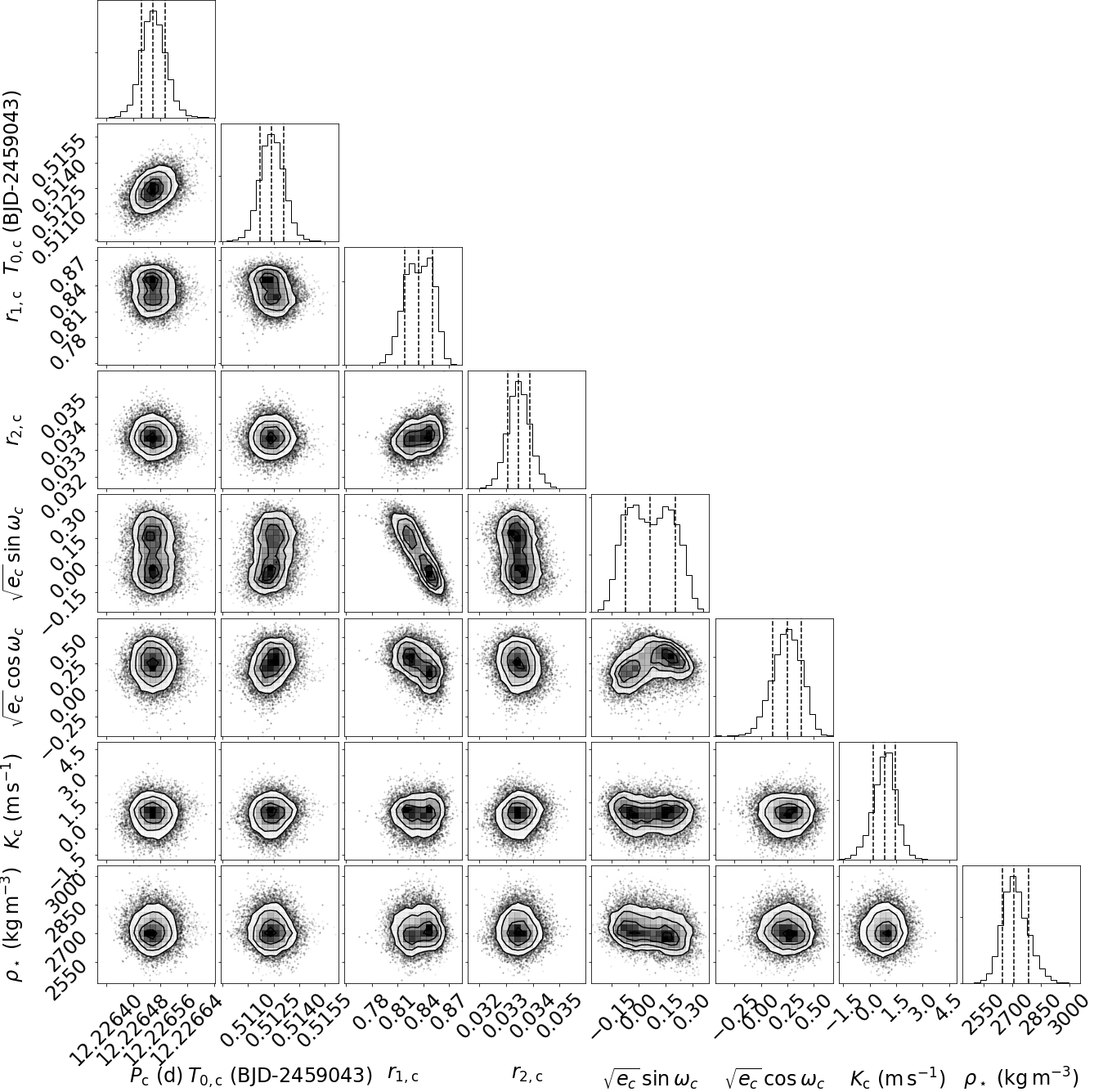} }}
  \caption{The posterior distributions of the fitted transit and RV parameters for a) planet\,b and b) planet\,c, including; the orbital period ($P$), the mid-transit time ($T_{0}$), the (r$_1$,r$_2$) parametrisation of the impact parameter ($b$) and planet-to-star radius ratio ($p$) plane, the $\sqrt{e}$cos$\omega$ and $\sqrt{e}$sin$\omega$ parametrisation where $e$ is the eccentricity and $\omega$ is the argument of periastron, the semi-amplitude ($K$), and the stellar density ($\rho_\star$).}
  \label{fig:corners_bc}
\end{figure*}

\begin{table*}
    \centering
    \caption{Priors used in the joint photometry and RV fit. Uniform priors are represented by $\mathcal{U}(a, b)$, with lower and upper bounds of $a$ and $b$, whereas $\mathcal{N}(\mu,\sigma)$ indicates a Normal (Gaussian) prior with mean, $\mu$, and standard deviation, $\sigma$.}
    \label{tab:priors}
    \begin{tabular}{ccc}
\hline\hline                
 Parameter (unit) & \multicolumn{2}{c}{Prior} \\ 
\hline 
 & b & c \\ 
\hline 
$P$ (d) & $\mathcal{U}(6.437785,6.446653)$ & $\mathcal{U}(12.215945,12.258251)$ \\ 
$T_{0}$ (BJD-2457000) & $\mathcal{U}(2036.849036,2036.865344)$ & $\mathcal{U}(2043.489928,2043.519832)$ \\ 
$r_\mathrm{1}$ & $\mathcal{U}(0,1)$ & $\mathcal{U}(0,1)$ \\ 
$r_\mathrm{2}$ & $\mathcal{U}(0,1)$ & $\mathcal{U}(0,1)$ \\ 
$\sqrt{e}$cos$\omega$ & $\mathcal{U}(-0.5,0.5)$ & $\mathcal{U}(-0.5,0.5)$ \\ 
$\sqrt{e}$sin$\omega$ & $\mathcal{U}(-0.5,0.5)$ & $\mathcal{U}(-0.5,0.5)$ \\ 
$K$ ($\mathrm{m\,s}^{-1}$) & $\mathcal{U}(0,20)$ & $\mathcal{U}(0,20)$ \\ 
$\gamma_\star$ ($\mathrm{m\,s}^{-1}$) & \multicolumn{2}{c}{$\mathcal{U}(21186.8,21243.0)$} \\ 
$\rho_\star$ ($\mathrm{kg\,m^{-3}}$) & \multicolumn{2}{c}{$\mathcal{N}(2760,140)$} \\ 
\hline
$q_{1,{\mathrm{TESS}}}$ & \multicolumn{2}{c}{$\mathcal{N}(0.32,0.1)$} \\
$q_{2,{\mathrm{TESS}}}$ & \multicolumn{2}{c}{$\mathcal{N}(0.53,0.1)$} \\
$q_{1,{\mathrm{CHEOPS}}}$ & \multicolumn{2}{c}{$\mathcal{N}(0.33,0.1)$} \\
$q_{2,{\mathrm{CHEOPS}}}$ & \multicolumn{2}{c}{$\mathcal{N}(0.44,0.1)$} \\
$q_{1,{\mathrm{LCOGT_{zs}}}}$ & \multicolumn{2}{c}{$\mathcal{N}(0.34,0.1)$} \\
$q_{2,{\mathrm{LCOGT_{zs}}}}$ & \multicolumn{2}{c}{$\mathcal{N}(0.60,0.1)$} \\
$q_{1,{\mathrm{LCOGT_{B}}}}$ & \multicolumn{2}{c}{$\mathcal{N}(0.36,0.1)$} \\
$q_{2,{\mathrm{LCOGT_{B}}}}$ & \multicolumn{2}{c}{$\mathcal{N}(0.26,0.1)$} \\
$q_{1,{\mathrm{NGTS}}}$ & \multicolumn{2}{c}{$\mathcal{N}(0.32,0.1)$} \\
$q_{2,{\mathrm{NGTS}}}$ & \multicolumn{2}{c}{$\mathcal{N}(0.47,0.1)$} \\
$q_{1,{\mathrm{ASTEP_{Rc}}}}$ & \multicolumn{2}{c}{$\mathcal{N}(0.30,0.1)$} \\
$q_{2,{\mathrm{ASTEP_{Rc}}}}$ & \multicolumn{2}{c}{$\mathcal{N}(0.43,0.1)$} \\
\hline\hline
    \end{tabular}
\end{table*}

\begin{table}
\begin{center}
\caption{Fitted light curve and RV parameter posterior values for the joint fit to the data detailed in Section~\ref{sec:joint_analysis}.}
\label{tab:TOI1064_instab}
\begin{tabular}{cc}
\hline\hline                
\multicolumn{2}{c}{\textit{Fitted light curve and RV parameters}} \\ 
\hline 

$q_{1,{\mathrm{TESS}}}$ & 0.336$^{+0.070}_{-0.068}$ \\
$q_{2,{\mathrm{TESS}}}$ & 0.464$^{+0.076}_{-0.068}$ \\
$\sigma_{{\mathrm{jitter,TESS}}}$ & 0.0007652$^{+0.0000094}_{-0.0000095}$ \\
$q_{1,{\mathrm{CHEOPS}}}$ & 0.265$^{+0.047}_{-0.050}$ \\
$q_{2,{\mathrm{CHEOPS}}}$ & 0.521$^{+0.066}_{-0.058}$ \\
$\sigma_{{\mathrm{jitter,CHEOPS}}}$ & 0.0000012$^{+0.0001557}_{-0.0000012}$ \\
$q_{1,{\mathrm{LCOGT_{zs}}}}$ & 0.343$^{+0.081}_{-0.075}$ \\
$q_{2,{\mathrm{LCOGT_{zs}}}}$ & 0.592$^{+0.075}_{-0.079}$ \\
$\sigma_{{\mathrm{jitter,LCOGT_{zs}}}}$ & 0.00099960$^{+0.00000027}_{-0.00000039}$ \\
$q_{1,{\mathrm{LCOGT_{B}}}}$ & 0.934$^{+0.275}_{-0.222}$ \\
$q_{2,{\mathrm{LCOGT_{B}}}}$ & 0.721$^{+0.289}_{-0.284}$ \\
$\sigma_{{\mathrm{jitter,LCOGT_{B}}}}$ & 0.000711$^{+0.000073}_{-0.000069}$ \\
$q_{1,{\mathrm{NGTS}}}$ & 0.300$^{+0.072}_{-0.074}$ \\
$q_{2,{\mathrm{NGTS}}}$ & 0.505$^{+0.074}_{-0.070}$ \\
$\sigma_{{\mathrm{jitter,NGTS}}}$ & 0.000012$^{+0.000094}_{-0.000011}$ \\
$q_{1,{\mathrm{ASTEP_{Rc}}}}$ & 0.352$^{+0.062}_{-0.053}$ \\
$q_{2,{\mathrm{ASTEP_{Rc}}}}$ & 0.428$^{+0.074}_{-0.075}$ \\
$\sigma_{{\mathrm{jitter,ASTEP_{Rc}}}}$ & 0.000851$^{+0.000042}_{-0.000043}$ \\
$\sigma_{{\mathrm{jitter,HARPS}}}$ ($\mathrm{m\,s}^{-1}$) & 4.01$^{+0.58}_{-0.49}$ \\

\hline\hline    
\end{tabular}
\end{center}

\end{table}

\section{Atmospheric evolution posterior distributions}
Here we present the posterior distribution plots of results of our atmospheric evolution modelling of the TOI-1064 system.

\begin{figure*}
    \centering
    \includegraphics[width=\linewidth]{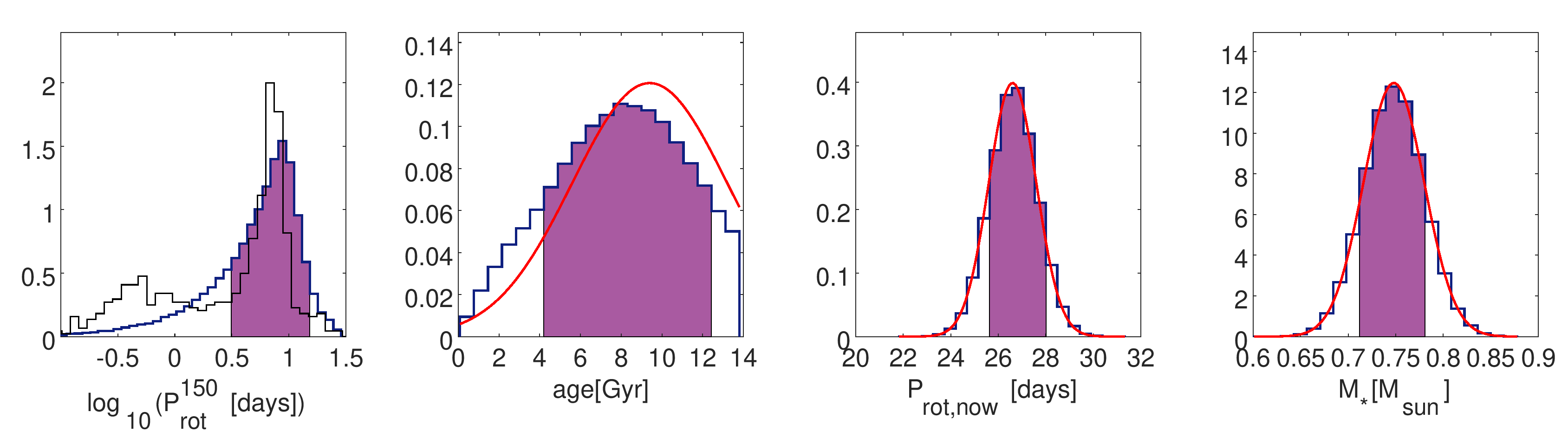}
    \caption{Posterior distributions as inferred from the atmospheric evolution code described in Section \ref{sec:ae}. Purple areas highlight the highest probability density (HPD) intervals at the 68\% level. Red lines are the normal priors imposed on the parameters, while the background histogram plotted with the thin black line in the left-most panel is the distribution of the stellar rotation period observed in open clusters of 150 Myr \citep{Johnstone2015} considering only stars whose masses deviate from $M_{\star}$ by less than 0.1 $M_{\odot}$.}
    \label{fig:evoAtmStar}
\end{figure*}
\begin{figure*}
    \centering
    \includegraphics[width=\textwidth]{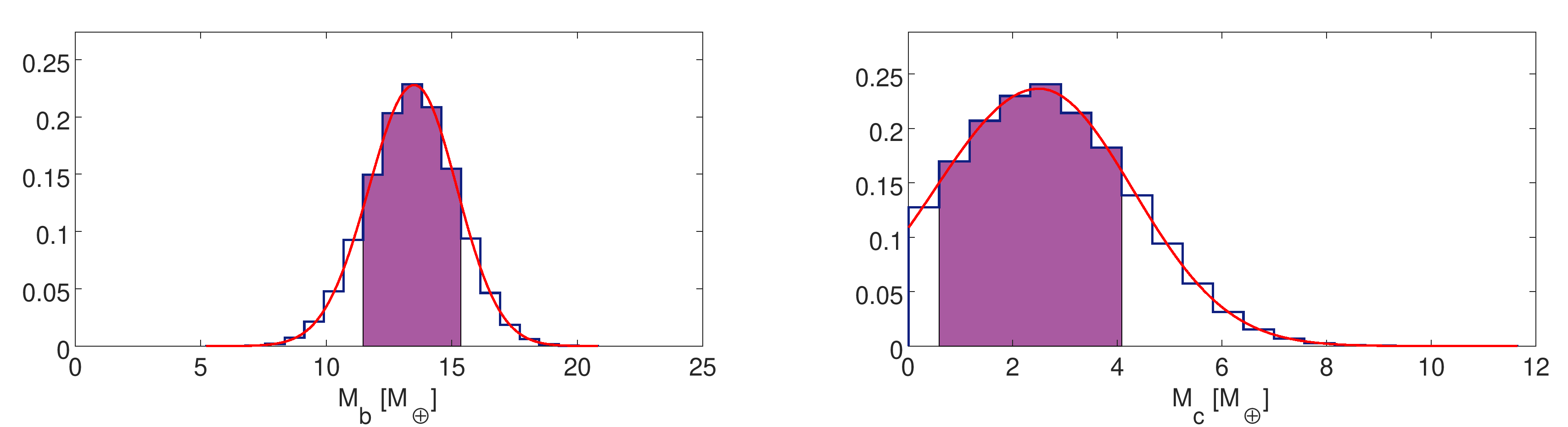} \\
    \includegraphics[width=\textwidth]{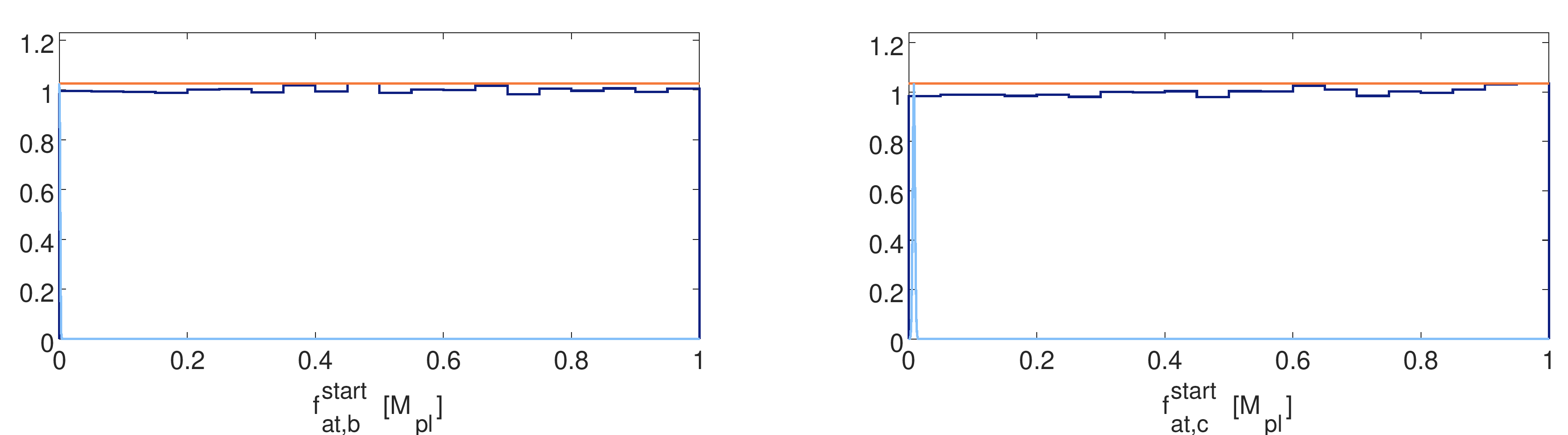}
    \caption{Posterior distributions of the planetary masses (\textit{upper panels}) and initial atmospheric mass fractions (\textit{bottom panels}). Purple areas highlight the highest probability density (HPD) intervals at the 68\% level. The red lines represent the priors imposed on the parameters, which are uniform in the case of $f_{\mathrm{at}}^{\mathrm{start}}$ (free parameters), while the light blue lines, spiking near f $ = $ 0, in the bottom panels represent the present-day atmospheric mass fractions. The atmospheric content is basically negligible for planet b, while it is barely visible, though non-negligible, for planet c.}
    \label{fig:evoAtmPla}
\end{figure*}


\bsp	
\label{lastpage}
\end{document}